\def\version{November 25, 2014}

\documentclass[12pt]{article}
\def\macrosPb{} 
\def\macrosHarxiv{} 
\usepackage[psamsfonts]{amsfonts}
\usepackage{amsmath,amssymb,amsthm}
\usepackage[dvips]{graphicx}
\usepackage{appendix}
\usepackage{bbm} 
\usepackage{amsbsy}
\usepackage{enumerate}
\usepackage{cite}

\ifdefined\macrosPa
  \usepackage[textwidth=465pt,textheight=650pt,centering]{geometry} 
\else\ifdefined\macrosPb
  \usepackage[textwidth=500pt,textheight=650pt,centering]{geometry} 
\fi\fi

\ifdefined\macrosS
  \makeatletter
  
  \def\boldsymbol{\pmb}
  \makeatother

  \usepackage{mathptmx}
  \DeclareMathAlphabet{\mathcal}{OMS}{cmsy}{m}{n}
\fi

\def\UseSection{
        \numberwithin{equation}{section}
	\theoremstyle{plain}
        \newtheorem{theorem}    {Theorem}[section]
        \DefineTheorems 
}

\def\DefineTheorems{
	
	\newtheorem{lemma}      [theorem] {Lemma}
	
	\newtheorem{prop}       [theorem] {Proposition}
	
	\newtheorem{cor}        [theorem] {Corollary}

	\theoremstyle{definition}
	\newtheorem{defn}       [theorem] {Definition}

	\newtheorem{rk} 	[theorem] {Remark}
	\theoremstyle{definition}

}

\newcommand{\bt}   {\begin{theorem}}
\newcommand{\et}   {\end  {theorem}}
\newcommand{\bl}   {\begin{lemma}}
\newcommand{\el}   {\end  {lemma}}
\newcommand{\bp}   {\begin{prop}}
\newcommand{\ep}   {\end  {prop}}
\newcommand{\bc}   {\begin{cor}}
\newcommand{\ec}   {\end  {cor}}
\newcommand{\bd}   {\begin{defn}}
\newcommand{\ed}   {\end  {defn}}

\newcommand{\ba}   {\begin{array}}
\newcommand{\ea}   {\end  {array}}
\newcommand{\be}   {\begin{enumerate}}
\newcommand{\ee}   {\end  {enumerate}}
\newcommand{\bi}   {\begin{itemize}}
\newcommand{\ei}   {\end  {itemize}}

\def\eq#1\en{\begin{equation}#1\end{equation}}  
\def\eqsplit#1\ensplit{
	\begin{equation}\begin{split}#1\end{split}\end{equation}
	}
\def\eqalign#1\enalign{
	\begin{align}#1\end{align}
	}
\def\eqmul#1\enmul{
	\begin{multline}#1\end{multline}
	}
\newcommand{\eqarrstar} {\begin{eqnarray*}} 
\newcommand{\enarrstar} {\end{eqnarray*}} 
\newcommand{\eqarray}   {\begin{eqnarray}} 
\newcommand{\enarray}   {\end{eqnarray}} 
\newcommand{\nnb}	{\nonumber \\} 

\newcommand{\lbeq}[1]  {\label{e:#1}}
\newcommand{\refeq}[1] {\eqref{e:#1}}    
\newcommand{\lbfg}[1]  {\label{fg: #1}}
\newcommand{\reffg}[1] {\ref{fg: #1}}

\makeatletter
\newcommand{\labelcounter}[2]{{%
	\stepcounter{#1}
	\protected@write\@auxout{}%
	{\string\newlabel{#2}{{\csname the#1\endcsname}{\thepage}}}%
	{\ref{#2}}
	}}
\makeatother
\newcommand{\Cbold} {{\mathbb C}}  
  
\newcommand{\Ebold} {{\mathbb E}}

\newcommand{\Nbold} {{\mathbb N}}

\newcommand{\Rbold} {{\mathbb R}}

\newcommand{\Zbold} {{\mathbb Z}}

\newcommand{\Acal}   {\mathcal{A}} 
\newcommand{\Bcal}   {\mathcal{B}} 
\newcommand{\Ccal}   {\mathcal{C}} 
\newcommand{\Dcal}   {\mathcal{D}} 
 
\newcommand{\Fcal}   {\mathcal{F}} 
\newcommand{\Gcal}   {\mathcal{G}} 
 
\newcommand{\Ical}   {\mathcal{I}} 
 
\newcommand{\Kcal}   {\mathcal{K}} 
\newcommand{\Lcal}   {\mathcal{L}} 
 
\newcommand{\Ncal}   {\mathcal{N}} 
 
\newcommand{\Pcal}   {\mathcal{P}}
\newcommand{\Qcal}   {\mathcal{Q}}

\newcommand{\Scal}   {\mathcal{S}} 
 
\newcommand{\Ucal}   {\mathcal{U}} 
\newcommand{\Vcal}   {\mathcal{V}} 
\newcommand{\Wcal}   {\mathcal{W}} 
\newcommand{\Xcal}   {\mathcal{X}}
\newcommand{\Ycal}   {\mathcal{Y}}

\newcommand{\Ihat} {{\hat{I} }}

\newcommand{\Vhat} {{\hat{V} }}

\newcommand{\Zd}    {{ {\Zbold}^d }}

\newcommand{\spose}[1] {{\hbox to 0pt{#1\hss}} }
\newcommand{\ltapprox} {\mathrel{\spose{\lower 3pt\hbox{$\mathchar"218$}}
 \raise 2.0pt\hbox{$\mathchar"13C$}}}
\newcommand{\gtapprox} {\mathrel{\spose{\lower 3pt\hbox{$\mathchar"218$}}
 \raise 2.0pt\hbox{$\mathchar"13E$}}}

\UseSection   
\usepackage[usenames]{color}

\definecolor{at}{rgb}{0.0, 0.5, 0.0} 

\newcommand{\LT}{{\rm Loc}  }

\newcommand{\DV}{\Dcal}

\newcommand{\rhogen}{\tilde{\rho}}
\newcommand{\rD}{r}

\renewcommand{\to} {\rightarrow}
\renewcommand{\qed}{\hfill\rule{2mm}{2mm}\bigskip}

\newcommand{\R}{\Rbold}

\newcommand{\N}{\Nbold}
\newcommand{\C}{\mathbb{C}}
\newcommand{\volume}{\mathbb{V}}
\newcommand{\Lambdabold}{\boldsymbol{\Lambda}}

\newcommand{\1}{\mathbbm{1}}

\newcommand{\nn}{\nonumber}

\newcommand{\psib}{\bar\psi}

\newcommand{\jm}{j_\Omega}

\newcommand{\Ex}{\mathbb{E}}

\newcommand{\Econst}{\alpha_{\Ebold}}

\newcommand{\EGconst}{\alpha}

\newcommand{\IstabC}{\alpha_{I}}
\newcommand{\ItilstabC}{\alpha_{I}}

\newcommand{\chicCov}{{\chi}}

\newcommand{\Qcalnabla}{\Qcal}

\newcommand{\Gtilp}{\gamma}

\newcommand{\pair}[1]{\langle #1 \rangle}

\newcommand{\Phipol}{\Pi}

\newcommand{\diam}[1]{\textrm{diam}(#1)}

\newcommand{\units}{\Ucal}

\newcommand{\cgam}{\gamma}

\newcommand{\Itilde}{\tilde{I}}

\newcommand{\Ttimes}{T}

\newcommand{\hldg}{h_{\rm lead}}
\newcommand{\cldg}{c_{\rm lead}}

\newcommand{\pt}{{\rm pt}}
\newcommand{\Ipt}{I_{\rm pt}}
\newcommand{\Ipttil}{\tilde{I}_{\rm pt}}

\newcommand{\Vpt}{V_{\rm pt}}

\newcommand{\ypt}{y_{\mathrm{pt}}}
\newcommand{\lambdapt}{\lambda_{\mathrm{pt}}}

\newcommand{\lambdaa}{\lambda^{\pp}}
\newcommand{\lambdab}{\lambda^{\qq}}

\newcommand{\qa}{q^{\pp}}
\newcommand{\qb}{q^{\qq}}

\newcommand{\dq}{\delta q}

\newcommand{\h}{\mathfrak{h}}

\newcommand{\gbar}{\bar{g}}

\newcommand{\ggen}{\tilde{g}}

\newcommand{\chigen}{\tilde{\chi}}
\newcommand{\mgen}{\tilde{m}}
\newcommand{\Iint}{\mathbb{I}}
\newcommand{\Igen}{\tilde{\mathbb{I}}}

\newcommand{\domRG}{\mathbb{D}}

\newcommand{\domr}{R}

\newcommand{\pp}{a}
\newcommand{\qq}{b}

\newcommand{\sigmab}{\bar{\sigma}}

\newcommand{\ddp}[2]{\frac{\partial #1}{\partial #2}}

\newcommand{\epdV}{\bar{\epsilon}}

\newcommand{\phib}{\bar\phi}

\newcommand{\zldg}{z_{\rm lead}}
\newcommand{\zh}{z_{h}}

\newcommand{\McalnowM}{M}

\newcommand{\hred}{h_{\rm red}}
\newcommand{\hrem}{h_{\rm rem}}

\newcommand{\Kin}   {K_{\mathrm{in}}}
\newcommand{\Uin}   {U_{\mathrm{in}}}
\newcommand{\Kout}  {K_{\mathrm{out}}}
\newcommand{\Kspace}{\Kcal}
\newcommand{\CKspace}{\Ccal\Kspace}
\newcommand{\BKspace}{\Bcal\Kspace}

\newcommand{\Iin}   {I_{\mathrm{in}}}

\newcommand{\amain} {a}
\newcommand{\ain}   {a_{\mathrm{in}}}
\newcommand{\aout}  {a_{\mathrm{out}}}

\newcommand{\abig}  {\tilde{a}}

\newcommand{\rball} {r}

\newcommand{\RG}{\text{RG}}

\ifdefined\macrosH
  \usepackage{xr-hyper}
  \usepackage{hyperref}
  \hypersetup{hypertexnames=false}
  \hypersetup{colorlinks,citecolor=blue,linkcolor=blue}  

\else\ifdefined\macrosHarxiv
  \usepackage{xr-hyper}
  \usepackage{hyperref}
  \hypersetup{hypertexnames=false}

\else
  \newcommand{\texorpdfstring}[2]{#1}
  \usepackage{xr}
\fi\fi

\title  {
       A renormalisation group method. \\
       V. A single renormalisation group step
        }

\author{
David C. Brydges\thanks{Department of Mathematics,
University of British Columbia,
Vancouver, BC, Canada V6T 1Z2.
E-mail: {\tt db5d@math.ubc.ca}, {\tt slade@math.ubc.ca}.}\;
 and Gordon Slade$^*$}

\date\version

\newcommand{\ratio}{\omega}
\newcommand{\rhoFcal}{\epdV}
\renewcommand{\rhogen}{\rho}
\newcommand{\qcpl}{q}
\renewcommand{\dq}{\delta \qcpl}

\begin{document}

\maketitle

\begin{abstract}
This paper is the fifth in a series devoted to the development of a
rigorous renormalisation group method applicable to lattice field
theories containing boson and/or fermion fields, and comprises the
core of the method.  In the renormalisation group method, increasingly
large scales are studied in a progressive manner, with an interaction
parametrised by a field polynomial which evolves with the scale under
the renormalisation group map.  In our context, the progressive
analysis is performed via a finite-range covariance decomposition.
Perturbative calculations are used to track the flow of the coupling
constants of the evolving polynomial, but on their own perturbative
calculations are insufficient to control error terms and to obtain
mathematically rigorous results.  In this paper, we define an
additional non-perturbative coordinate, which together with the flow
of coupling constants defines the complete evolution of the
renormalisation group map.  We specify conditions under which the
non-perturbative coordinate is contractive under a single
renormalisation group step.  Our framework is essentially
combinatorial, but its implementation relies on analytic results
developed earlier in the series of papers.  The results of this paper
are applied elsewhere to analyse the critical behaviour of the
4-dimensional continuous-time weakly self-avoiding walk and of the
4-dimensional $n$-component $|\varphi|^4$ model.  In particular, the
existence of a logarithmic correction to mean-field scaling for the
susceptibility can be proved for both models, together with other
facts about critical exponents and critical behaviour.
\end{abstract}

\section{Introduction and main results}

\subsection{Background}
\label{sec:intro}

This paper is the fifth in a series devoted to the development of a
rigorous renormalisation group method applicable to lattice field
theories containing boson and/or fermion fields, and comprises the
core of the method.  Its immediate goal is to prepare for the
application in \cite{BBS-saw4-log,BBS-saw4} to a specific
supersymmetric field theory that is used to analyse the critical
behaviour of the continuous-time weakly self-avoiding walk, and in
particular to prove the existence of a logarithmic correction to the
susceptibility in dimension~4.  However, our approach is more general
and applies more broadly including to the critical behaviour of the
$4$-dimensional $n$-component $|\varphi^4|$ model \cite{BBS-phi4-log}.

In the renormalisation group method, a multi-scale analysis is
performed in which increasingly large scales are studied in a
progressive manner, with an interaction parametrised by a field
polynomial which evolves with the scale under renormalisation group
transformations \cite{WK74}.  In our context, progressive integration
is performed via a finite-range covariance decomposition
\cite{Baue13a,BGM04}.  Perturbative calculations are used to track the
flow of the coefficients, or \emph{coupling constants}, of the
evolving polynomial, but on their own perturbative calculations are
insufficient to control error terms and to obtain mathematically
rigorous results.  In this paper, we employ another coordinate called
$K$, in addition to the interaction polynomial $V$, for tracking the
evolution under renormalisation group transformation.  With this
additional coordinate, we provide a framework that allows the error
terms to be rigorously controlled.  Our framework is essentially
combinatorial, but its implementation relies on analytic results
developed in earlier papers.  An important feature of our method is
that it respects supersymmetry, when this is present in the underlying
model. Euclidean invariance is not manifest since our method relies on
subdivisions of space into hypercubes. The use of such subdivisions
has been universal in nonperturbative work on the renormalisation
group, but recently \cite{RW14} a manifestly Euclidean invariant
method has been invented.

Some aspects of our approach, whose roots go back to \cite{BY90}, were
presented in \cite{Bryd09}.  We draw on the approach of
\cite{BEI92,BI03d} for \emph{hierarchical} models, but in a much
extended and generalised form that applies to $\Zd$.  The idea of
using a covariance decomposition to implement renormalisation goes
back to \cite{BCGNOPS78,BCGNOPS80}.  Recent uses of the
renormalisation group that bear some relation to our approach can be
found in \cite{ACG13,AKM12,Falc12,Falc13,MS08}.

Different approaches to the renormalisation group include the block
spin method used in \cite{GK83,GK84,GK85,GK86}, the phase space
expansion method used in \cite{FMRS87}, and the approach of Ba{\l}aban
(see e.g., \cite{Bala83}, and \cite{Dimo13} for a recent overview).
These various methods are distinguished from each other according to
how they combine perturbation theory with estimates on large
deviations connected with large fields. Balaban's method is
particularly powerful because it also applies to strong coupling
problems where the action has degenerate minima.  The books and major
reviews \cite{BG95,Bryd09,FKT02,Mast08,Riva91,Salm99} give varied
perspectives on renormalisation.

This paper is the culmination of the developments presented in
parts~I--IV \cite{BS-rg-norm,BS-rg-loc,BBS-rg-pt,BS-rg-IE} of the
series and it relies on results from all four parts.  A full assembly
of parts~I-V (and using also the result of \cite{BBS-rg-flow}), is
given for the 4-dimensional weakly self-avoiding walk in
\cite{BBS-saw4-log,BBS-saw4}, and for the 4-dimensional $|\varphi|^4$
model in \cite{BBS-phi4-log}.  To put the present paper in
perspective, we briefly summarise the other papers in the series as
they pertain to this one.
\begin{enumerate}
\item In part~I \cite{BS-rg-norm}, we present elements of the theory
of Gaussian integration involving both boson and fermion fields, and
develop norms and norm estimates for performing analysis with such
Gaussian integrals.  A renormalisation group step involves performing
a Gaussian integral whose covariance is given by a generic term in the
finite-range decomposition of an original covariance.  In the present
paper, we show how to obtain effective control on such an integration,
so that error terms do not accumulate upon repeated integration.
\item In part~II \cite{BS-rg-loc}, we define and analyse the
localisation operator $\LT$, which extracts from a functional of the
fields a polynomial that captures the components of the functional
which are relevant and marginal for the dynamical system defined by
the renormalisation group.  These are the components which must be
accurately tracked, and this tracking leads to the flow of the
coupling constants.  In the present paper, we prove that the operator
$\LT$ achieves its purpose in the sense that the non-perturbative coordinate
is contractive under the renormalisation group map.
It is this contraction that prevents
error terms from building up under successive renormalisation group
steps.
\item In part~III \cite{BBS-rg-pt}, we present a general description
of perturbation theory, in which the polynomial $V_j$ at scale $j$ is
replaced after a single Gaussian integral by a new polynomial $\Vpt$.
The polynomial $\Vpt$ is accurate to second order in the coupling
constants but does not take into account error terms that have the
potential to accumulate in repeated renormalisation group steps.  In
the present paper we show how to employ $\Vpt$ while preventing errors
from accumulating.
\item In part~IV \cite{BS-rg-IE}, we prove nonperturbative estimates
for the specific supersymmetric field theory studied in
\cite{BBS-saw4-log,BBS-saw4}.  The results include stability estimates
for the interaction, proof of accuracy of the perturbative
calculations of part~III, estimates on Gaussian expectations, and a
crucial contraction estimate which implements the achievements of the
operator $\LT$.  The estimates of part~IV provide an essential input
for the present paper.
\item As an application and d\'enoument, in
\cite{BBS-saw4-log,BBS-saw4} we obtain a statement of infrared
asymptotic freedom for the 4-dimensional weakly self-avoiding walk,
and use it to prove the existence of a logarithmic correction to
mean-field scaling for the sucsceptibility and $|x|^{-2}$ decay for
the critical two-point function.  The analysis of
\cite{BBS-saw4-log,BBS-saw4} combines the results of parts~I--V with
the main result of \cite{BBS-rg-flow} to analyse the
infinite-dimensional dynamical system arising from repeated
application of the renormalisation group.  A further application to
the 4-dimensional $n$-component $\varphi^4$ model is given in
\cite{BBS-phi4-log}.
\end{enumerate}

Throughout the paper, we concentrate on the case of dimension $d=4$.
Before stating our main results in Section~\ref{sec:mr}, we first
introduce the language and concepts needed for their formulation, as
well as the norms used in their statement.


\subsection{Polymers and local algebras of forms}
\label{sec:poly}

Let $L\ge 3$ and $N \ge 1$ be integers, and let $\Lambda = \Zd /
(L^N\Zd)$ for fixed dimension $d>0$.  We write $|\cdot|$ for the
$\ell_\infty$ distance on both $\Zd$ and the torus $\Lambda$.  Since
$N$ and $\Lambda$ are determined by each other we make $\Lambda$ the
primary object and write $N = N (\Lambda)$.  Our results concern the
renormalisation group in both finite volume $\Lambda$ and the infinite
volume $\Zd$. To cover both cases we use the symbol $\volume$ whose
values are $\Lambda$ or $\Zd$, and we set $N (\volume)=\infty$ for
$\volume =\Lambda$.  To allow for the study of the two-point function,
two particular points $\pp,\qq$ are fixed in $\Zd$. We assume
$\pp,\qq$ have distinct images in $\Lambda $, under the projection $x
\mapsto x \mod (L^N\Zd)$, and their images are also called $\pp,\qq$
so we can refer to the two distinguished points in $\volume$.  They
are called \emph{observable} points. The following definition is basic
to our setup.

\begin{defn}
\label{def:blocks} (a) \emph{Blocks.}  For each $j\in \N_{0}$ the
lattice $\Zd$ is paved in a natural way by disjoint $d$-dimensional
cubes of side $L^j$.  The cube that contains the origin at the corner
has the form
\begin{equation}
    \{x\in \Lambda:  |x|_{\infty} < L^{j}\},
\end{equation}
and all other cubes are translates of this one by vectors in
$L^{j} \Zd$.  Similarly, for $j=0,1,\ldots,N (\Lambda)$, the torus
$\Lambda$ is paved in a natural way by $L^{N-j}$ disjoint
$d$-dimensional cubes of side $L^j$.  We call these cubes
$j$-\emph{blocks}, or \emph{blocks} for short and let $\Bcal_{j} =
\Bcal_j(\volume)$ denote the set of $j$-blocks. The integer $j$ is
called a \emph{scale}.
\\
(b) \emph{Polymers.}  A union of blocks in $\Bcal_j$ is called a {\em
polymer (at scale $j$)}, and the set of polymers at scale $j$ is
denoted $\Pcal_j=\Pcal_{j} (\volume)$.  The empty union is included:
$\varnothing \in \Pcal_j$.  For $X \in \Pcal_j$, $\Bcal_{j} (X)$ denotes
the set of blocks $B \in \Bcal_{j}$ with $B\subset X$.  The size
$|X|_j$ of $X\in {\cal P}_j$ is the number of $j$-blocks in $X$, i.e.,
$|X|_j$ is the cardinality of $\Bcal_{j} (X)$.  We define $\Pcal_{*}=\sqcup_{j}\Pcal_{j}
(\Zd)$.  In particular, an element $X$ of $\Pcal_{*}$ has a scale $j (X)$.
\\
(c) \emph{Connectivity.}  A nonempty subset $X\subset \Lambda $ is
said to be \emph{connected} if for any two points $x, x'\in X$ there
exists points $ x_i \in X$ ($i=0,1,\dotsc ,n$) with
$|x_{{i+1}}-x_{i}|_\infty =1$, $x_{0} = x$ and $x_{n}=x'$.  The set
of connected polymers in $\Pcal_j$ is denoted $\Ccal_j= \Ccal_{j}
(\volume)$.  The null set $\varnothing$ is not in $\Ccal_j$.  We say
that two polymers $X,Y$ \emph{do not touch} if $\min\{|x-y|_\infty :
x \in X, y \in Y\} >1$. A polymer can be decomposed into connected
components that do not touch; we write ${\rm Comp}(X)$ for the set of
connected components of $X$.
\end{defn}

The basic setting for our analysis is detailed in
\cite[Section~\ref{IE-sec:study}]{BS-rg-IE}, and we maintain the same
setting and notation here, but now allow infinite volume as well as
finite volume.  In brief, we have a complex boson field $\phi :
\Lambda \to \C$ with its complex conjugate $\bar\phi$, a pair of
conjugate fermion fields $\psi,\bar\psi$, and a \emph{constant}
complex observable boson field $\sigma \in \C$ with its complex
conjugate $\bar\sigma$.  The fermion field is given in terms of the
1-forms $d\phi_x$ by $\psi_x = \frac{1}{\sqrt{2\pi i}} d\phi_x$ and
$\bar\psi_x = \frac{1}{\sqrt{2\pi i}} d\bar\phi_x$, where we fix some
square root of $2\pi i$.  We work with an algebra $\Ncal$ which is
defined in terms of a direct sum decomposition
\begin{equation}
\label{e:Ncaldecomp}
    \Ncal = \Ncal^\varnothing \oplus \Ncal^a \oplus \Ncal^b \oplus \Ncal^{ab}.
\end{equation}
Elements of $\Ncal^\varnothing$ are given by finite linear
combinations of products of an even number of fermion fields with
coefficients that are complex-valued functions of the boson fields.
This restriction to forms of even degree results in a commutative
algebra.  Elements of $\Ncal^a, \Ncal^b , \Ncal^{ab}$ are respectively
given by elements of $\Ncal^\varnothing$ multiplied by $\sigma$, by
$\bar\sigma$, and by $\sigma\bar\sigma$.  For example, $\phi_x
\bar\phi_y \psi_x \bar\psi_x \in \Ncal^\varnothing$, and $\sigma
\bar\phi_x \in \Ncal^a$.  There are canonical projections $\pi_\alpha:
\Ncal \to \Ncal^\alpha$ for $\alpha \in \{\varnothing, a, b, ab\}$.
We use the abbreviation $\pi_*=1-\pi_\varnothing =
\pi_a+\pi_b+\pi_{ab}$.  The algebra $\Ncal$ is discussed further
around \cite[\eqref{loc-e:1Ncaldecomp}]{BS-rg-loc}.  There $\Ncal$ is
written $\Ncal/\Ical$, but to simplify the notation we write $\Ncal$
here instead.  The quotient space notation reflects our policy of
writing arbitrary functions of $\sigma,\bar\sigma$ and identifying any
such function with the sum of the constant, $\sigma$, $\bar\sigma$ and
$\sigma\bar\sigma$ terms in its formal power series expansion in
$\sigma,\bar\sigma$.  The parameter $p_\Ncal$ which appears in its
definition is a measure of the smoothness of elements of $\Ncal$ (see
\cite[Section~\ref{norm-sec:Ncal}]{BS-rg-norm}); its precise value is
unimportant as long as it is fixed with $p_\Ncal \ge 10$ (the value
``10'' is required for Lemma~\ref{lem:KKK} below).  Constants in
estimates are permitted to depend on $p_\Ncal$, and this is
unimportant.

In \cite[\eqref{norm-e:Fxyphi},\,\eqref{norm-e:NXdef}]{BS-rg-norm},
$\Ncal (X)$ is defined to be the algebra of differential forms that
depend only on fields with spatial labels in $X$, where $X$ is a
subset of $\Lambda$.  In this paper the argument $X$ of $\Ncal (X)$ is
a subset of $\volume$, which is $\Lambda$ or $\Zd$, and $\Ncal (X)$
consists of differential forms \emph{of even degree} generated by
monomials in $\psi , \psib$ with spatial labels in $X$, so that $\Ncal
(X)$ is \emph{commutative}.  We also define the commutative algebra
\begin{equation}\label{def:NcalZd}
    \Ncal (\volume)
    =
    \bigcup_{\text{$X$ finite, $X\subset \volume$}}\Ncal (X)
.
\end{equation}
For $\volume =\Lambda$ or $\volume =\Zd$ we write $\Ncal=\Ncal
(\volume)$.  Note that $\Ncal (X)$ is a subalgebra of $\Ncal (Y)$ when
$X$ is a subset of $Y$.

In the notation of \cite[Section~\ref{norm-sec:seq}]{BS-rg-norm}, for
$X \subset \Lambda$, an element of $\Ncal (X)$ has the form
\begin{equation}
    \label{e:K}
    F
=
    \sum_{y \in \vec\Lambdabold^*} \frac{1}{y!} F_y \psi^y.
\end{equation}
The sum is over sequences $y=(x,\bar x)$, with each of
$x=(x_1,\ldots,x_p)$ and $\bar x = (\bar x_1,\ldots, \bar x_q)$ a
sequence in $X$, with $\psi^y = \psi_{x_1}\cdots \psi_{x_p}
\bar\psi_{\bar x_1} \cdots \bar\psi_{\bar x_q}$, and with $y!=p!q!$.
The coefficient $F_{y}$ is a complex valued function of $(\phi
,\sigma)$ in $\C^{\volume}\times \C$ such that $F_{y} (\phi',\sigma) =
F_{y} (\phi,\sigma)$ when $\phi'\rvert_{X}=\phi\rvert_{X}$.  The
coefficients $F_{y}$ are zero when the sequence $y$ has odd length.
As a function of $\sigma$, $F_{y}$ has the form $\alpha+\beta\sigma +
\gamma\sigmab + \delta\sigma \sigmab$, but $\beta=\delta=0$ when $X$
does not contain $\pp$ and $\gamma=\delta=0$ when $X$ does not contain
$\qq$. To understand this, one should regard $\sigma$ as associated to
the point $\pp$, and $\sigmab$ to the point $\qq$, and then the
conditions say that an element $F$ of $\Ncal (X)$ depends only on
fields in $X$.

Let $\units$ denote the set of $2d$ nearest neighbours of the origin
in $\Zd$.  For $e\in\units$, we define the finite difference operator
$\nabla^e \phi_x = \phi_{x+e}-\phi_x$, and the Laplacian $\Delta_\Zd =
-\frac{1}{2}\sum_{e \in \units}\nabla^{-e} \nabla^{e}$.  Important
examples of forms are:
\begin{equation}
    \label{e:addDelta}
    \tau_x = \phi_x \bar\phi_x + \psi_x \bar\psi_x,
    \quad
 \tau_{\nabla \nabla,x}  =
   \frac 12
   \sum_{e \in \units}
   \left(
   (\nabla^e \phi)_x (\nabla^e \bar\phi)_x +
   (\nabla^e \psi)_x (\nabla^e \bar\psi)_x
   \right)
    ,
\end{equation}
\begin{equation}
       \tau_{\Delta,x} = \frac 12 \left(
    (-\Delta \phi)_{x} \bar{\phi}_{y} +
    \phi_{x} (-\Delta \bar{\phi})_{y} +
    (-\Delta \psi)_{x} \bar{\psi}_{y} +
    \psi_{x} (-\Delta \bar{\psi})_{y}
    \right)
  .
\end{equation}
Let $\Qcalnabla$ denote the vector space of polynomials of the form
\begin{gather}
    \label{e:V}
    V
=
    V_{\varnothing} + V_{\pp} + V_{\qq} + V_{\pp\qq},
\end{gather}
where
\begin{gather}
    V_{\varnothing}
    =
    g \tau^{2} + \nu \tau +
    z \tau_{\Delta} +
    y \tau_{\nabla \nabla},
\quad
    V_{\pp}
=
    \lambda_{\pp} \sigma \bar{\phi},
\quad
    V_{\qq}
=
    \lambda_{\qq}\bar{\sigma} \phi,
\quad
    V_{\pp \qq}
=
    q_{\pp\qq} \bar{\sigma}\sigma
    \label{e:Vx}
,
\end{gather}
\begin{align}
\label{e:lambda-defs}
&
    \lambda_{\pp}
    =
    - \lambdaa \,\1_{\pp},
\quad\quad
    \lambda_{\qq}
    =
    - \lambdab \,\1_{\qq},
\quad\quad
    q_{\pp\qq}
    =
    -\frac{1}{2} (\qa\1_{\pp} + \qb\1_{\qq})
    ,
\end{align}
$g,\nu,y,z,\lambdaa,\lambdab,\qa,\qb \in \C$, and the indicator
functions are defined by the Kronecker delta $\1_{a,x}=\delta_{a,x}$.
For $X \subset \Lambda$, we write
\begin{equation}
\label{e:VXdef}
    V(X)=\sum_{x\in X}V_x.
\end{equation}
Elements $V$ of $\Qcal$ are polynomials with eight independent
coefficients, so $\Qcal$ is isomorphic to $\C^{8}$ and this
identification is sometimes useful.  The polynomial $V$ has symmetries
which are inherited by the field theory to be defined below in terms
of $V$.  To discuss these symmetries, an \emph{automorphism}
$E:\Lambda \rightarrow \Lambda$ is an injective map from $\Lambda$ to
$\Lambda$ under which nearest-neighbour points are mapped to
nearest-neighbour points under both the map and its inverse.
Translations and reflections that preserve $\Lambda$ are examples of
automorphisms.  The action of an automorphism $E:\Lambda \rightarrow
\Lambda$ as a map from $\Ncal(\Lambda)$ to itself is defined in
\cite[\eqref{loc-e:E-action-on-N}]{BS-rg-loc}.  The polynomial
$V_\varnothing$ is Euclidean covariant, in the sense that for any
automorphism $E$, $E (V_{\varnothing, x}) = V_{\varnothing, Ex}$.
Also, $V_x$ is gauge invariant and $V_{\varnothing, x}$ is
supersymmetric, where these two terms are defined for elements of
$\Ncal$ in \cite[Section~\ref{pt-sec:supersymmetry}]{BBS-rg-pt}.

\subsection{Covariance decomposition}
\label{sec:cd}

Given $m^2>0$, let $C=(-\Delta_\Lambda +m^2)^{-1}$.  As explained in
more detail in \cite[Section~\ref{IE-sec:cd}]{BS-rg-IE}, the
covariance $C$ has a finite-range decomposition $C=C_1+\cdots
C_{N-1}+C_{N,N}$ \cite{Baue13a,BGM04}.  The expectation $\Ex_C$
denotes the combined bosonic-fermionic Gaussian integration on
$\Ncal$, with covariance $C$, defined in
\cite[Section~\ref{norm-sec:Grass}]{BS-rg-norm}.  The expectation can
be performed successively, using
\begin{equation}
\lbeq{progexp}
    \Ex_C \theta = \Ex_{N} \theta \circ \Ex_{N-1} \theta \circ
    \cdots \circ \Ex_{1}\theta,
\end{equation}
where $\Ex_j$ is the expectation corresponding to the $j^{\rm th}$ covariance, and
$\theta$ denotes a type of convolution.
More precisely, we define the
map $\theta : \Ncal(\volume) \to \Ncal(\volume\sqcup
\volume')$ by making the replacement
in an element of $\Ncal$ of $\phi$ by $\phi+\xi$,
$\bar\phi$ by $\bar\phi+\bar\xi$, $\psi$ by $\psi+\eta$, and
$\bar\psi$ by $\bar\psi+\bar\xi$.
In applying $\Ex_{j+1}\theta$,  the fields $\xi,\bar\xi,\eta,\bar\eta$ are integrated out by
$\Ex_j$, with $\phi, \bar\phi, \psi, \bar\psi$ kept fixed.
The expectation $\Ex_C$ can be obtained as the special case of
\refeq{progexp} resulting from setting $\xi=\bar\xi=\eta=\bar\eta=0$ in
$\Ex_N\theta$.

We assume that the covariance decomposition obeys the estimates listed and
discussed in
\cite[Section~\ref{IE-sec:frp}]{BS-rg-IE}.
In particular, for \cite[\eqref{IE-e:scaling-estimate-Omega}]{BS-rg-IE},
we restrict $m^2$ to lie in a small interval $[0,\delta]$ when considering $C_j$
with $j<N$, but make the further restriction $m^2 \in [\delta L^{-2(N-1)},\delta]$
for $C_{N,N}$.
The covariances obey the \emph{finite-range property} that $C_{j} (x,y)=0$ for $|x-y|\ge
\frac{1}{2}L^{j}$, for each scale $j$.  These properties are established for
the covariance decomposition of \cite{Baue13a} in \cite{BBS-rg-pt}.

In analogy with ordinary Gaussian random variables, there is an
independence consequence of the finite-range property, called the
\emph{factorisation property} of the expectation.  The latter states
that if $X_1,\ldots, X_n \in \Pcal_{j+1} (\Lambda)$ do not touch each
other, and if $F_m(X_m) \in \Ncal(X_m)$ for each $m$, then
\begin{equation}
\label{e:Efaczz}
    \Ex_{j+1} \theta \prod_{m=1}^n F_m(X_m)
    =
    \prod_{m=1}^n \Ex_{j+1} \theta F_m(X_m).
\end{equation}
This factorisation property is a consequence of
\cite[Proposition~\ref{norm-prop:factorisationE}]{BS-rg-norm}.
It plays an important role.

\subsection{Perturbative and non-perturbative coordinates}

As in \cite[\eqref{IE-e:Fsoptb}]{BS-rg-IE}, the interaction is
defined, for $V\in \Qcal$, $B \in \Bcal_j$ and $X \in \Pcal_j$, by
\begin{equation}
\label{e:IVB}
    I_j(V,B)
    =
    e^{-V(B)}
    \left( 1+W_j(V,B) \right) ,
    \quad
    \quad
    I_j(V,X)
    =
    \prod_{B \in \Bcal_{j}(X)}
    I_j(V,B),
\end{equation}
where $W_{j}$ is a certain non-local polynomial in the fields, which
is an explicit quadratic function of $V$ discussed in detail in
\cite[Section~\ref{IE-sec:formint}]{BS-rg-IE}.  In the present paper,
we rely on properties of $I$ proved in \cite{BS-rg-IE} and the
specifics of its definition play a minor role.

Recall the function $\Vpt : \Qcalnabla \to \Qcalnabla$
defined in \cite[\eqref{pt-e:Vptdef}]{BBS-rg-pt} and explained
in \cite[Section~\ref{pt-sec:WPjobs}]{BBS-rg-pt}.
In \cite[Proposition~\ref{pt-prop:I-action}]{BBS-rg-pt}, we show that
\begin{equation}
\lbeq{EIapprox}
    \Ex_{j+1} \theta I_j(V,\Lambda)
    \approx
    I_{j+1}(\Vpt,\Lambda),
\end{equation}
where the approximation is accurate up to and including second order,
as formal power series in the coupling constants.
Under this approximate perturbative calculation, the effect of a single
expectation is captured by the map $V \mapsto \Vpt$, and we refer to $V$
as the \emph{perturbative coordinate}.  We introduce a \emph{non-perturbative
coordinate} $K$ which accurately tracks all the errors in the
approximation \refeq{EIapprox}.  For this, the following definition is needed.

\begin{defn}
\label{def:circleproduct} \emph{Circle product.}
Given $F,G :{\cal P}_j \to \Ncal$, we define
$F\circ G :{\cal P}_j \to \Ncal$ by
\begin{equation}
\label{e:circdef}
    (F \circ G)(X) = \sum_{Y\in \Pcal_j (X)}
    F(Y) G(X \setminus Y) \quad\quad (X \in \Pcal_j).
\end{equation}
This \emph{circle product} is
commutative and associative.
\end{defn}

The circle product depends on $j$ but this is left implicit in
the notation.
All functions $F:\Pcal_j \to \Ncal$ that we consider are required to obey $F(\varnothing)=1$.
The sum in \eqref{e:circdef} includes the degenerate terms $Y =
\varnothing, X$
(in particular, $(F \circ G)(\varnothing) = F(\varnothing) G(\varnothing)=1$).
The identity element
for the circle product is $\1_\varnothing$, defined by setting
$\1_\varnothing(X)=1$ if $X=\varnothing$ and $\1_\varnothing(X)=0$ otherwise.
From \refeq{progexp}, we obtain
\begin{equation}
\lbeq{progcirc}
    \Ex_C \theta I_0(V,\Lambda) = \Ex_C\theta(I_0 \circ \1_\varnothing)(\Lambda)
    = \Ex_N \theta \circ \Ex_{N-1}\theta \circ \cdots \circ \Ex_1 \theta (I_0 \circ \1_\varnothing)(\Lambda).
\end{equation}

Let $\Qcal^{(0)}$ be the subspace of $\Qcal$ with $y=\qa=\qb=0$.  Let
$j<N (\volume)$, let $\qcpl_{j}\in\C$, let $V_{j} \in
\Qcalnabla^{(0)}$, and let $K_{j}:\Pcal_j \to\Ncal$.  The
\emph{renormalisation group map} $\text{RG} = \text{RG}_{j}$ is a
description of the action of $\Ex_{j+1}\theta$ as a map $\text{RG}:
(\qcpl_j,V_j,K_j) \mapsto (\qcpl_{j+1},V_{j+1},K_{j+1})$, with
$\qcpl_{j+1}\in\C$, $V_{j+1} \in \Qcalnabla^{(0)}$, and
$K_{j+1}:\Pcal_{j+1}\to \Ncal$, such that
\begin{equation}
    \label{e:rgmapdef}
    e^{\qcpl_j \sigma\bar\sigma}\,\Ex_{j+1} \theta \big(I_{j} (V_j)\circ K_j\big)(\Lambda)
    =
    e^{\qcpl_{j+1}\sigma\bar\sigma}\big(I_{j+1} (V_{j+1})\circ K_{j+1}\big)(\Lambda)
.
\end{equation}
This allows \refeq{progcirc} to be evaluated iteratively.
In particular,
the flow of $\qcpl$ under repeated applications of the renormalisation group map
turns out to be
central to the proof obtaining the decay of the critical two-point function
of the continuous-time weakly self-avoiding walk in \cite{BBS-saw4}.
By dividing \refeq{rgmapdef} by $e^{\qcpl_j \sigma\bar\sigma}$ and setting
$\delta q_{j+1}=q_{j+1}-q_j$, we obtain the equivalent equation
\begin{equation}
    \label{e:rgmapdef0}
    \Ex_{j+1} \theta \big(I_{j} (V_j)\circ K_j\big)(\Lambda)
    =
    e^{\delta\qcpl_{j+1}\sigma\bar\sigma}\big(I_{j+1} (V_{j+1})\circ K_{j+1}\big)(\Lambda)
.
\end{equation}
Thus we can regard $\text{RG}$ as the map
\begin{equation}
\lbeq{rgq0}
    \text{RG}_j: (V_j,K_j) \mapsto (\delta\qcpl_{j+1},V_{j+1},K_{j+1})
\end{equation}

The existence of a
map obeying \refeq{rgmapdef} is easy: there are $\qcpl_{j+1},K_{j+1}$ that solve this
equation for any choice of $V_{j+1}$, and they are not unique. An example
is given in Section~\ref{sec:example} below.  It is \emph{much} harder to choose
the map $\text{RG}$ and a Banach space in which $K_{j+1}$ does
not grow in norm under iteration of the renormalisation group map,
and the main achievement of the present paper is
to exhibit such a choice.

\subsection{Simplified construction of \texorpdfstring{$K_1$}{K1}}
\label{sec:example}

For illustrative purposes, we
now provide an example of a simplified construction of $K_1$
from $(V_0,K_0)=(V_0,\1_\varnothing)$.
The idea in this section is used in
Section~\ref{sec:polymers} below, but the complete construction of RG
requires a better (but less simple)
choice of $K_{+}$ than the one in the example.

The
following elementary lemma, which relates the circle product and
binomial expansion, is useful here and also later.  It uses notation
discussed in more detail around \refeq{B-extension}.  Namely,
given $F:\Bcal_j \to \Ncal$ and $X \in \Pcal_j$,
we write $F^X=F(X)=\prod_{B \in \Bcal_j(X)}F(B)$.

\begin{lemma}
\label{lem:bin}
For $F_1,F_2:\Bcal_j \to \Ncal$ and $X \in \Pcal_j$,
\begin{equation}
    (F_1+F_2)^X
    = (F_1\circ F_2)(X).
\end{equation}
\end{lemma}

\begin{proof}
By \eqref{e:B-extension}, followed by expansion of the product and
application of \eqref{e:circdef}, we find that
\begin{equation}
    (F_1+F_2)^X = \prod_{B \in \Bcal_j(X)} (F_1+F_2)(B)
    =
     \sum_{Y \in \Pcal_j(X)} F_1^Y F_2^{X \backslash Y}
     = (F_1\circ F_2)(X),
\end{equation}
and the proof is complete.
\end{proof}

\begin{figure}[t]
\begin{center}
  \input{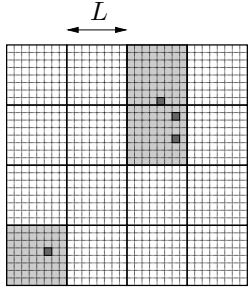}
  \caption{The four small dark squares represent a polymer in $\mathcal{P}_0$,
  and the three larger shaded squares represent its closure in $\mathcal{P}_1$.}
  \label{fig:polymers01}
\end{center}
\end{figure}

We also need the following definition, which is depicted in Figure~\ref{fig:polymers01}.

\begin{defn}
\label{def:blocks1}
The \emph{closure}
$\overline X$ of $X \in \Pcal_j$ is the smallest $Y \in \Pcal_{j+1}$ such that $X \subset
Y$.  Given $U \in \Pcal_{j+1}$, we write
\begin{gather}
    \overline \Pcal_{j}(U)=\{X\in \Pcal_j \mid \overline X = U\}
.
\end{gather}
\end{defn}

The following proposition provides an example of a construction of
$K_1$ from the pair $I_0$ and $K_0=\1_\varnothing$, for
\emph{arbitrary} choice of $V_0,V_1$ each with $q_{\pp\qq}=0$.

\begin{prop}
For any $V_0,V_1\in \Qcal$, each with $q_{\pp\qq}=0$,
\begin{equation}
    \Ex_1(I_0(V_0) \circ \1_\varnothing)(\Lambda) = (I_1(V_1) \circ \tilde K_1)(\Lambda),
\end{equation}
where
\begin{equation}\label{e:K1Definition}
    \tilde{K}_1(U) = \sum_{X\in\overline{\Pcal}_0 (U)}
    I_{1}^{U \setminus X} \Ex_{C_1}  \delta I_{1}^X
\end{equation}
with $\delta I_{1}^X =\prod_{x\in X}(\theta I_0(x)-I_{1}(x))$.
\end{prop}

\begin{proof}
For $X \in \Pcal_0$, let $\delta I_{1}^X =\prod_{x\in X}(\theta
I(x)-I_{1}(x))$; this depends on
$\phi_{1},\bar\phi_1,\psi_1,\bar\psi_1$ via $I_{1}$, as well as on the
fields $\phi_0= \phi_{1} + \xi_1, \bar\phi_0= \bar\phi_{1} +
\bar\xi_1, \psi_0 = \psi_1+\eta_1, \bar\psi_0 = \bar\psi_1+\bar\eta_1$
via $\theta I$.  The integration implied by $\Ex_{1} \theta$
integrates out only the fluctuation fields $\xi_1, \bar\xi_1, \eta_1,
\bar\eta_1$, leaving dependence on the scale-$1$ fields only.  Thus we
obtain, using Lemma~\ref{lem:bin} for the third equality,
\begin{align}
  \Ex_{1} \theta (I_0 \circ \1_\varnothing)(\Lambda)
  &=
    \Ex_{1} \theta I_0(\Lambda)
  = \Ex_{1} (I_{1} +\delta I_{1} )^\Lambda
  \nnb &
  = \Ex_{1} (I_{1} \circ \delta I_{1} )(\Lambda)
  = (I_{1} \circ \Ex_{1} \delta I_{1} )(\Lambda)
  .
\label{e:Z1}
\end{align}
The above circle products are at scale $0$.
Using \refeq{K1Definition} for the last equality, we  obtain
\begin{gather}
    \big(I_{1}\circ\Ex_{1} \theta I\big)(\Lambda)
    =
    \sum_{X\in\Pcal_0} I_{1}^{\Lambda\setminus X} \Ex_{1}
    \delta I_{1}^{X}
    =
    \sum_{U \in \Pcal_1}
    \sum_{X\in \overline{\Pcal}_0 (U)} I_{1}^{\Lambda\setminus X} \Ex_{1}
    \delta I_{1}^{X}
    \nnb
    \label{e:Z1Final}
    = \sum_{U \in \Pcal_1} I_{1}^{\Lambda\setminus U} \tilde{K}_1(U)
    =
    (I_{1} \circ \tilde K_{1})(\Lambda),
\end{gather}
where the circle product on the right-hand side is at scale $1$.
This completes the proof.
\end{proof}

An important fact is that $\tilde K_1$ has
a certain component factorisation property.  For example, if $U\in \Pcal_1$ has
connected components $U_1,U_2$, then with the help of
Figure~\ref{fig:polymers01} it is straightforward to check that the
factorisation property \refeq{Efaczz}
of the expectation
implies that $\tilde{K} (U) = \tilde{K} (U_{1})\tilde{K} (U_{2})$.
We make a formal definition of the component factorisation property in the next section.

\subsection{Setting for non-perturbative coordinate}
\label{sec:coordinates}

We now define the basic setting for
the non-perturbative coordinate $K:\Pcal_j \to \Ncal$, including the spaces
$\Ccal\Kcal_j$ and $\Kcal_j$.

We say that a function $K:\Pcal_{j} \rightarrow \Ncal$ is
\emph{Euclidean covariant} if $E (K (X)) = K (EX)$ for all polymers $X
\in \Pcal_{j}$ and all automorphisms $E$ of $\volume$.  We say that
$K$ is \emph{gauge invariant} (\emph{supersymmetric}) if $K(X)$ is
gauge invariant (supersymmetric) for all $X$ in $\Pcal_{j}$; these two
terms are defined for elements of $\Ncal$ in
\cite[Section~\ref{pt-sec:supersymmetry}]{BBS-rg-pt}.  We say that $K$
\emph{has zero constant part} if the result of setting $\phi=0$ and
$\psi=0$ in $K(X)$ is zero for all non-empty polymers $X$.  We need
the following two definitions.

\begin{defn}
\label{def:blocks2}
\emph{Small sets.}  A polymer $X\in \Pcal_{*}$ is said to be a
\emph{small set} if $|X|_{j (X)} \le 2^{d}$ and $X \in \Ccal_{j (X)}$.
Let $\Scal_{j}$ be the set of all small sets in $\Pcal_{j}$.  The
\emph{small set neighbourhood} of $X \in \Pcal_{*}$ is defined by
\begin{equation}
\label{e:ssn}
    X^{\Box}
=
    \bigcup_{Y\in \Scal_{j (X)}:X\cap Y \not =\varnothing } Y.
\end{equation}
\end{defn}

For the next definition, we define the \emph{coalescence scale}
$j_{ab}$ by
\begin{equation}
   \label{e:Phi-def-jc}
    j_{\pp \qq}
    =
    \big\lfloor
   \log_{L} (2 |\pp - \qq|)
   \big\rfloor
   .
\end{equation}

\begin{defn}\label{def:Kspace}
For $j \le N (\volume)$ with $j<\infty$, let $\Ccal\Kspace_{j} =
\Ccal\Kspace_{j} (\volume)$ denote the complex vector space of functions
$K : \Ccal_j (\volume) \to \Ncal (\volume)$ with the properties:
\begin{itemize}
\item Field Locality: For all $X\in \Ccal_j(\volume)$, $K (X) \in
\Ncal (X^{\Box})$.  Also, (i) $\pi_{a} K (X)=0$ unless $a\in X$, (ii) $\pi_{b}
K (X)=0$ unless $b\in X$, and (iii)
$\pi_{ab} K (X)=0$ unless $a\in X$
and $b\in X^\Box$ or vice versa, and
$\pi_{ab} K(X)=0$ if $X\in \Scal_j$ and $j<j_{ab}$.
\item Symmetry: (i) $K$ is gauge invariant; (ii) $\pi_\varnothing K$
is supersymmetric and has no constant part;
(iii) $\pi_{\varnothing}K$ is Euclidean covariant.
\end{itemize}
Let $\Kspace_{j} = \Kspace_{j} (\volume)$ be the complex vector space
of functions $K : \Pcal_j (\volume) \to \Ncal (\volume)$ which have
the properties listed above and in addition
\begin{itemize}
\item Component Factorisation: for all polymers $X$, $K (X) = \prod_{Y
\in {\rm Comp}( X)}K (Y)$.
\end{itemize}
\end{defn}

\bigskip
Every element of $\Kspace_{j}$ determines an element of $\CKspace_{j}$
by restriction to connected sets, and every element of $\CKspace_{j}$
determines an element of $\Kspace_{j}$ by the factorisation condition.
The same symbol is used for both elements related by this
correspondence.  Under this correspondence, $\1_{\varnothing} \in
\Kspace_{j}$ becomes $0 \in \CKspace_{j}$, because
the empty set is not a connected set.

Let $\BKspace_{j}=\BKspace_{j} (\volume)$ denote the set of functions
$F: \Bcal_j \to \Ncal$ which obey the field locality and symmetry
conditions of Definition~\ref{def:Kspace}.  Given $F:\Bcal_{j}
\rightarrow \Ncal$ we extend $F$ to $\Pcal_{j}$ by
\begin{equation}
    \label{e:B-extension}
    F (X) = F^{X} = \prod_{B\in \Bcal_{j} (X)}F (B)
    \quad\quad (X\in \Pcal_j)
    .
\end{equation}
The appearance of the set $X$ as an exponent introduces our convention
that such exponents signal functions that factorise over blocks.
Using
\eqref{e:B-extension}, an element $F \in \BKspace_{j}$ extends to an
element $F \in \Kspace_{j}$.
An important use of $\BKspace_{j}$ is the map
$I_j: \Qcalnabla\to \BKspace_j (\Lambda)$ defined in \refeq{IVB}.

The individual properties of Definition~\ref{def:Kspace} play
different roles in our analysis.  The property of field locality is of
fundamental importance and its preservation under iteration of the
renormalisation group map relies on the finite-range property of the
covariance decomposition via \refeq{Efaczz}, as illustrated in
Section~\ref{sec:example} above.  The symmetry properties are enjoyed
by any $V_j \in \Qcal^{(0)}$, and the symmetry assumption on $K$
ensures that the effect of $K_j$ on the construction of $V_{j+1}$ is
such that these symmetries are inherited from $V_j$ by $V_{j+1}$, and
in particular that $V_{j+1}$ does not contain additional terms not
present in $\Qcal^{(0)}$.  It is possible to relax the assumption of
supersymmetry by a suitable enlargement of $\Qcal$.  For example, in
the analysis of the $|\varphi|^4$ model in \cite{BBS-phi4-log} we
forego supersymmetry in Definition~\ref{def:Kspace} at the cost of
including an additional constant term in $\Qcal$; this is discussed in
Remark~\ref{rk:uphi4} below.

\subsection{Definition of norms}
\label{sec:norms}

We use specific norms as detailed in this section.  This particular
specification is made so that we can apply estimates on $I$ (e.g., in
Section~\ref{sec:ie}) and an important contraction property (namely
Proposition~\ref{prop:cl}); these results are proved in
\cite{BS-rg-IE}.  It also paves the way for applications of our
results in \cite{BBS-saw4-log,BBS-saw4,BBS-phi4-log}.  However,
accepting the results of \cite{BS-rg-IE}, the majority of this paper
can be read without knowing what the norms are, beyond the facts that
the norm of a product is less than the product of the norms, and the
norm of an expectation is less than the expectation of the norm.

\subsubsection{Parameters}
\label{sec:2np}

We use the norms and regulators for $\Ncal$ defined in
\cite[Section~\ref{IE-sec:reg}]{BS-rg-IE}, including the $\Phi$ norm
on test functions, the $\tilde\Phi$ norm on boson fields, the $T_\phi$
semi-norm on $\Ncal$.  The parameters $\h=\ell_j$ and $\h=h_j$ for
these norms are specified in \cite[Section~\ref{IE-sec:hex}]{BS-rg-IE}
and we repeat the definition of these parameters here.  They depend,
in particular, on two numbers $\ggen_j$ and $\ggen_{j+1}$, which we
assume can be taken to be as small as desired (uniformly in $j$, and
depending on $L$), and which obey
\begin{equation}
\label{e:gbarmono}
    \frac 12 \ggen_{j+1} \le
    \ggen_j \le 2 \ggen_{j+1}.
\end{equation}
This permits us to apply results from \cite{BS-rg-IE} which rely on
\refeq{gbarmono}.  The parameters $\h$ are given in terms of a (large)
$L$-dependent constant $\ell_0$ and a (small) universal constant $k_0$
by
\begin{equation}
\label{e:hl}
    \h_j =
    \begin{cases}
    \ell_0 L^{-j[\phi]}  & \h=\ell
    \\
     k_0 \ggen_j^{-1/4}L^{-jd/4} & \h=h,
     \end{cases}
\end{equation}
\begin{equation}
\lbeq{newhsig}
    \h_{\sigma,j}=
    \begin{cases}
    \ggen_j L^{(j\wedge j_{ab})[\phi]} 2^{(j-j_{ab})_+} & \h=\ell
    \\
    \ggen_j^{1/4} L^{(j\wedge j_{ab})[\phi]} 2^{(j-j_{ab})_+} & \h=h,
    \end{cases}
\end{equation}
where $[\phi]=\frac{d-2}{2}$, $x_+= \max \{x,0\}$, and where the
coalescence scale $j_{ab}$ is defined in \refeq{Phi-def-jc}.

\subsubsection{Norm for perturbative coordinate}

As a vector space, $\Qcal$ is isomorphic to $\C^{8}$ since a
polynomial in $\Qcal$ is determined by eight coupling constants.
Although all norms on $\C^{8}$ are equivalent, the coupling constants
$\nu,\lambdaa,\lambdab, \qa,\qb$ have natural scaling factors and we
use a norm that takes this into account.  We define a norm on $\Qcal$
by
\begin{equation}\label{e:Vnorm}
    \|V\|_{\Qcal_j}
    =
    \max \left\{
    |g|, \, |z|, \, |y|, \, L^{2j}|\nu|,\,
    \ell_j\ell_{\sigma,j}|\lambdaa|,\,
    \ell_j\ell_{\sigma,j}|\lambdab|,\,
     \ell_{\sigma,j}^2|\qa|,\, \ell_{\sigma,j}^2|\qb|
    \right\}
    .
\end{equation}
The scaling in \refeq{Vnorm} reflects the fact that the coupling
constants $g,z,y$ are associated to marginal field monomials (for
$d=4$), whereas the $L^{2j}$ reflects the fact that $\nu$ is
associated to the relevant monomial $\tau$.  The scaling of the
observable coupling constants includes factors of $\ell_j$ or
$\ell_{\sigma,j}$ for each boson or observable field, respectively, in
the corresponding monomials in $V$.

Two useful subspaces of $\Qcal$ are the subspace $\Qcal^{(0)} \simeq
\C^{5}$ consisting of elements of $\Qcal$ with $y=\qa=\qb=0$, and the
subspace $\Qcal^{(1)} \simeq \C^{7}$ consisting of elements with
$y=0$.

With \cite[Lemma~\ref{IE-lem:T0ep}]{BS-rg-IE} and its proof, it
follows that there is a $j$-independent constant $c>0$ such that
\begin{equation}
\label{e:T0Q}
    c^{-1} \max_{B \in \Bcal_j} \|V(B)\|_{T_{0,j}(\ell_j)}
    \le
    \|V\|_{\Qcal_j}
    \le
    c \max_{B \in \Bcal_j}
    \|V(B)\|_{T_{0,j}(\ell_j)}
    .
\end{equation}

\subsubsection{Norms for non-perturbative coordinate}
\label{sec:Knorms}

Recall from \cite[Section~\ref{IE-sec:reg}]{BS-rg-IE} the
definition of the $T_{\phi,j} (\h_{j})$ seminorm. Recall also from
\cite[Definition~\ref{IE-def:Gnorms}, \eqref{IE-e:GPhidef},
\eqref{IE-e:9Gdef}]{BS-rg-IE} the definition of the two norm pairs on
$\Ncal(X^\Box)$ given, for $F \in \Ncal(X^\Box)$, by
\begin{equation}
\label{e:np1}
    \|F\|_j
    = \sup_{\phi \in \C^\Lambda}
    \frac{\|F\|_{T_{\phi,j} (\ell_{j})}}{G_{j}(X,\phi)},   \quad
    \|F\|_{j+1} = \|F\|_{T_{0,j+1}(\ell_{j+1})},
\end{equation}
\begin{equation}
\label{e:np2}
    \|F\|_j
    =
    \sup_{\phi \in \C^\Lambda}
    \frac{\|F \|_{T_{\phi,j} (h_{j})}}{\tilde{G}_{j}(X,\phi)},
    \quad
    \|F\|_{j+1}
    =
    \sup_{\phi \in \C^\Lambda}
    \frac{\|F \|_{T_{\phi,j+1}(h_{j+1})}}{\tilde{G}^{\Gtilp}_{j+1}(X,\phi)},
\end{equation}
in terms of an arbitrary parameter $\Gtilp\in (0,1]$.  To handle these
norms simultaneously we will write them all as $\|F\|_{\Gcal_{k}}$
with $k=j,j+1$. For the first pair. we write $\Gcal_{j} = G_{j}
(\ell_{j})$ and $\Gcal_{j+1} = T_{0,j+1}(\ell_{j+1})$, and for the second
pair we write $\Gcal_{j} = \tilde{G}_{j} (h_{j})$ and $\Gcal_{j+1} =
\tilde{G}^{\Gtilp}_{j}(h_{j+1})$. Sometimes we omit parameters
such as $j$ and $\h_{j}$ when we think their values are clear from
context. Note that the notation is potentially misleading because the dependence
on the parameter $\h_{k}$ refers to the $T_{\phi}$ part of this norm,
not the regulators which are defined
in \cite[\eqref{IE-e:GPhidef}, \eqref{IE-e:9Gdef}]{BS-rg-IE}
always in terms of $\ell_k$.

In \eqref{e:np1} we actually only have a $T_0$ \emph{semi}-norm, not a
norm.  Let $\Ical (\volume)= \{F \in \Ncal(\volume) \mid
\|F\|_{T_{0}}=0 \}$.  The set $\Ical (\volume)$ is an ideal in the
algebra $\Ncal$, since the $T_{0}$ semi-norm has the product property.
Thus the $T_{0}$ semi-norm on $\Ncal$ defines a norm on the quotient
space $\Ncal/\Ical$.  We work in the quotient space, and thus regard
$T_0$ as a norm rather than a semi-norm.

The above norms are defined on $\Ncal (X)$, but to measure the
size of elements of $\Kcal$, which are maps $X \mapsto F (X)$ from
polymers $X$ into $\Ncal (X^{\Box})$, we include a weight for the size
of $X$ as well.  Thus we let $W :\Pcal_{j} \times \C^\volume \to
(0,\infty)$ be a fixed strictly positive weight function.  We say that
$F\in \Kcal_{j}$ \emph{vanishes at weighted infinity} if for each $X
\in \Pcal_j$,
\begin{equation}
\label{e:defnFcal}
    \lim_{\|\phi\|_{\Phi_{j}(X)} \rightarrow \infty}
    \|F (X)\|_{T_{\phi} (\h)}W^{-1} (X,\phi)
    =
    0
.
\end{equation}
Let
$\Fcal_{j} (W)$ be the vector subspace of $\Kcal_j$ consisting of
elements $F$ which vanish at weighted infinity. We define a norm on
$\Fcal_{j} (W)$ by
\begin{equation}
    \|F \|_{\Fcal_{j} (W)}
    =
    \sup_{X \in \Ccal, \; \phi \in \C^\volume}
    \|F (X)\|_{T_{\phi} (\h)}W^{-1} (X,\phi)
.
\end{equation}

Now we make choices of $W=W_j$ that connect these norms to the two
norm pairs \eqref{e:np1}--\eqref{e:np2}.  For $\amain > 0$ and $X \in
\Pcal_{j}$, let
\begin{equation}
    \label{e:f0def}
    f_{j} (\amain ,X)
    =
    \amain (|X|_j-2^d)_+ .
\end{equation}
Note that $f_j(a,X)=0$ for any small set $X$.  For $\Gcal_{j}$ a
regulator, and given $\rhogen_{j}\in(0,1)$, let
\begin{equation}
    \label{e:Wdef}
    W(X,\phi) = \rhogen_{j}^{f_{j} (\amain ,X)}\Gcal_{j} (X,\phi)
.
\end{equation}
The factor $\rhogen_j^{f_j(a,X)}$ replaces the \emph{constant}
$A^{-1}$ used in many other papers in a version of \eqref{e:KFcal},
e.g., in \cite[(6.10)]{Bryd09}.  Then for each of the four norms in
the two norm pairs we have a choice of $W$ and scale $k=j,j+1$ such
that
\begin{gather}
    \label{e:KFcal}
    \|F \|_{\Fcal_{k} (W)}
    =
    \sup_{X \in \Ccal_k}
    \rhogen_{k}^{-f_{k} (a,X)}
    \|F(X)\|_k
,
\end{gather}
with norms on the right-hand side as in \eqref{e:np1} and
\eqref{e:np2}.
We denote the four normed spaces determined by \eqref{e:KFcal} by
$\Fcal_{j}(G)$, $\Fcal_{j+1} (T_{0})$ and $\Fcal_{j} (\tilde{G})$,
$\Fcal_{j+1} (\tilde{G}^{\Gtilp})$.  The space $\Fcal_{j+1} (T_{0})$
is special, in that it has no dependence on $\phi$, and we have simply
\begin{equation}
    \label{e:FT0}
    \|F \|_{\Fcal_{j+1} (T_{0})}
    =
    \sup_{X \in \Ccal_{j+1}}
    \rhogen_{j+1}^{-f_{j+1} (a,X)}
    \|F (X)\|_{T_{0,j+1} (\h)}
.
\end{equation}
The space $\Fcal_{j+1}(T_0)$ is the set of elements of $\Kcal_{j+1}$
for which the above norm is finite.
We do have a norm here, rather than a semi-norm, because we have taken
the quotient space that factors out elements of semi-norm zero, as
discussed in Section~\ref{sec:norms}.

Fix $\Omega>1$ (a good choice is $\Omega=2$) and recall from
\cite[\eqref{IE-e:mass-scale}--\eqref{IE-e:chidef}]{BS-rg-IE} the
$\Omega$-scale $\jm$ and the sequence $\chi_j = \Omega^{-(j-\jm)_+}$.
We make two choices of $\rho$, namely
\begin{equation}
    \label{e:rhoFcaldef}
    \rhogen_{j} = \epdV_{j}(\h_j)
    =
    \begin{cases}
    \chicCov_{j}^{1/2} \ggen_{j} & \h_{j}=\ell_{j}
    \\
    \chicCov_{j}^{1/2} \ggen_{j}^{1/4}   & \h_{j}= h_{j},
    \end{cases}
\end{equation}
consistent with the definition of $\epdV_{j}$ in
\cite[\eqref{IE-e:epdVdef}]{BS-rg-IE}.  The $\h=\ell$ choice of
$\rhogen$ is used for $\Fcal(G)$ and $\Fcal(T_0)$, whereas the $\h=h$
choice is used for $\Fcal(\tilde G)$ and $\Fcal(\tilde G^{\Gtilp})$.
We set
\begin{equation}
    \label{e:ratiodef}
    \ratio_{j}
    = \frac{\epdV_{j}(\ell)}{\epdV_{j}(h)}
    = \ggen_{j}^{3/4},
\end{equation}
and define another norm on $\Kcal_j$ by
\begin{equation}
\label{e:9Kcalnorm}
    \|K\|_{\Wcal_{j}}
    =
    \max
    \Big\{
    \|K \|_{\Fcal_j(G)},\,
    \ratio_{j}^{3}
    \|K \|_{\Fcal_j(\tilde{G})}
    \Big\}
.
\end{equation}
By definition,
\begin{equation}
\lbeq{T0dom}
    \|K(X)\|_{T_{\phi,j}(\ell_j)}
    \le
    \|K (X)\|_{G_j(\ell_j)}G_j(X,\phi)
    \le
    \|K\|_{\Wcal_{j}}G_j(X,\phi)
    \quad
    \text{for any $X \in \Scal_j$}.
\end{equation}
On the right-hand side of \refeq{9Kcalnorm}, we choose $a\in (0,\frac 14 2^{-d})$ as the
value of $a$ in the exponent $f_j$ in the weight $\rhoFcal_j$ appearing in
the definitions of $\Fcal_j(G)$ (we make the same choice for $\Fcal_j(T_0)$),
whereas we choose $\tilde
a = 4a \in (0, 2^{-d})$ in the definition of $\Fcal_j(\tilde{G})$.
This particular choice produces the same power of $\ggen$ for each of
$\rhoFcal(\ell)^{f_j(a,X)}$ and $\rhoFcal(h)^{f_j(\tilde a, X)}$, and this plays a role
in the proof of Lemma~\ref{lem:KKK} below.
Let $\Wcal_{j} = \Wcal_{j} (\volume)$ denote the vector space
$\Fcal_{j} (G)\cap \Fcal_j(\tilde G)$ on $\volume$ with norm $\|\cdot\|_{\Wcal_{j}(\volume)}$.

Each the four norms \refeq{np1}--\refeq{np2} obeys the product property
\cite[\eqref{IE-e:norm-fac}]{BS-rg-IE},
and our analysis relies heavily on this.
The product property is spoiled in an unequally weighted maximum of two of these norms,
due to the weight.  For this reason, we do not have a version of
the $\Wcal$
norm obeying the product property, and consequently we often work directly with $\Fcal$ norms
instead.
The following proposition is proved in Proposition~\ref{prop:closed-subspace}.

\begin{prop}\label{prop:Wcal-completeness}
For either of the two choices $\volume = \Zd$ or $\volume = \Lambda$,
each of the spaces $\Fcal(G)$, $\Fcal(\tilde G)$, $\Fcal(T_0)$ and
$\Wcal$ is a Banach space.
\end{prop}

The $\Wcal$ norm depends on the parameter $\ggen_j$ appearing in
\refeq{rhoFcaldef}, and also through the parameter $h_j =k_0L^{-jd/4}\ggen_j$
appearing in the norm of $\Fcal_j(\tilde G)$.  In addition, it depends
on $m^2$ since $\chi$ of \refeq{rhoFcaldef} depends on $m^2$.  The
following lemma measures the effect on the norm under variation of
these two parameters.  The lemma is not used in the present paper but
it is recorded here for use in \cite{BBS-saw4-log}.

\begin{lemma}
\label{wishlist-lem:Wcalnormequiv} The norms $\Wcal_j(m^2,\ggen_j)$
and $\Wcal_j(0,\ggen_j)$ are identical when $j \le \jm$.  In addition,
if $\ggen_j' \leq \ggen_j< 1$ and $m'^2 \ge m^2>0$, then in the limit
of small $\ggen_j-\ggen_j'$, for all $K \in \Wcal_j(m'^2,\ggen')$,
\begin{equation} \label{new-e:normequiv}
    \|K\|_{\Wcal_j(m^2,\ggen_j)}
    \leq
    (1+O(\ggen_j-\ggen'_j)) \|K\|_{\Wcal_j(m'^2,\ggen_j')}
    .
\end{equation}
\end{lemma}

\begin{proof}
The first statement holds because $\chi_j(m^2)=\chi_j(0)=1$ when $j
\le \jm$, by definition.

For the second statement, we first consider the dependence on $m^2$.
It follows from the definition of $\chi_j$ in
\cite[\eqref{IE-e:mass-scale}--\eqref{IE-e:chidef}]{BS-rg-IE} that
$\chi_j$ is monotone non-increasing in $m^2$, and hence $1/\chi_j$ is
monotone non-decreasing in $m^2$.  Consequently, increasing $m^2$
causes $1/\rho_j$ to increase, consistent with
\eqref{new-e:normequiv}.

Next, we consider the $\ggen$ dependence.  The norm
$\|K\|_{\Fcal_j(G)}$ is monotone decreasing in $\ggen_j$, by
definition.  By definition, $h_j =k_0L^{-jd/4} \ggen_j^{-1/4}$ is
also monotone decreasing in $\ggen_j$, so the norm
$\|K\|_{\Fcal_j(\tilde G)}$ is also monotone decreasing in $\ggen_j$
by definition.  The factor $\ratio_j^3$ in the $\Wcal_j(\ggen)$ norm
is however monotone increasing, but since it is continuous, the claim
follows.
\end{proof}

\subsubsection{Norm for scale \texorpdfstring{$N$}{N}}
\label{sec:normN}

Special attention is required for the norm at scale $N$, but there is also
increased flexibility.  Our need to have both the $G$ and $\tilde G$ norms
is explained in \cite[Section~\ref{IE-sec:lfp}]{BS-rg-IE}, and it is connected
with the need to propagate estimates from one scale to another.  Once scale $N$
has been reached, there is no further propagation.
In particular, it is not a problem if there is degradation
of the $G$ regulator at the final scale $N$.  We employ the $T_0$ and $\tilde G$ norms
precisely to prevent such degradation from accumulating over an unbounded number
of scales, but for a single scale it is permissible.

At scale $N$, the torus $\Lambda$ is the only polymer, and it is a single block.
With the above in mind, for scale $N$ we define the $\Wcal_N$ norm of $F:\Pcal_N \to \Ncal$ by
\begin{equation}
    \|F\|_{\Wcal_N}
    = \sup_{\phi \in \C^\Lambda}
    \frac{\|F(\Lambda)\|_{T_{\phi,N}}}{G_N(\Lambda,\phi)^{10}}.
\end{equation}
The power ``10'' in the denominator reflects the regulator degradation mentioned above,
and any fixed larger value could be used instead.  (Cf.~\cite[Remark~\ref{IE-rk:scaleNnorm}]{BS-rg-IE}.)

\subsection{Main results}
\label{sec:mr}

In this section, we present our main results.  Throughout we typically
omit the subscript $j$ and abreviate the subscript $j+1$ to $+$.  Thus
we write $(V,K)$ rather than $(V_j,K_j)$, and write $(V_+,K_+)$ rather
than $(V_{j+1},K_{j+1})$.  We first state results for the finite
volume renormalisation group map on a torus, and then describe the
explicit construction of the map $(V,K) \mapsto V_+$.  Following this,
we extend the definition of the renormalisation group map to infinite
volume, and state results for the infinite volume map.  The infinite
volume map is important in \cite{BBS-saw4-log}, to define a dynamical
system that is not limited to flow through only a finite number of
scales.

\subsubsection{Main result in finite volume}

To simplify the notation, we write $V=V_j$, $I=I_j(V)$, $K=K_j$, and
we wish to construct $\delta q_+=\delta q_{j+1}$, $V_+=V_{j+1}$,
$I_{+} =I_{j+1}(V_{+})$, $K_{+}=K_{j+1}$ such that the action of
$\Ex_{+}\theta = \Ex_{j+1}\theta$ is as stated in \refeq{rgmapdef0},
i.e.,
\begin{equation}
    \label{e:rgmapdef-bis}
    \Ex_{+}\theta \big(I (V)\circ K\big)(\Lambda)
    =
    e^{\dq_+ \sigma\sigmab}\big(I_{+} (V_+)\circ K_{+}\big)(\Lambda)
.
\end{equation}
At \refeq{rgmapdef0}, we defined the renormalisation group map
$(V,K) \mapsto (\delta q_+,V_+,K_+)$, with $V,V_+\in \Qcal^{(0)}\simeq
\C^{5}$ and $\delta q_+ \in \C$.

We define a mapping $V \mapsto V^{(0)}$ from $\Qcalnabla \simeq \C^8$
to $\Qcalnabla^{(0)}\simeq \C^6$, by replacing
$z\tau_{\Delta}+y\tau_{\nabla\nabla} + q_{\pp\qq} \sigma\bar\sigma$ in
$V\in \Qcalnabla$ by $(z+y)\tau_{\Delta}$ in $V^{(0)}\in \Qcal^{(0)}$.
Similarly, we define $V \mapsto V^{(1)}$ from $\Qcalnabla \simeq \C^8$
to $\Qcalnabla^{(1)} \simeq \C^7$ by replacing
$z\tau_{\Delta}+y\tau_{\nabla\nabla}$ in $V\in \Qcalnabla$ by
$(z+y)\tau_{\Delta}$ in $V^{(1)}\in \Qcal^{(1)}$.  Recall the map $V
\mapsto \Vpt (V)$ from $\Qcal$ to $\Qcal$ defined in
\cite[\eqref{pt-e:Vptdef}]{BBS-rg-pt}.  Given $V,V_+\in \Qcal^{(0)}$
and $\delta q_+^a,\delta q_+^b \in \C$, we define $R_{+} \in
\Qcal^{(1)}$ and $\delta q_+ \in \C$ by
\begin{equation}
\lbeq{Rplusdef}
    (V_+,\delta q_+^a , \delta q_+^b) = \Vpt^{(1)}(V) + R_{+},
    \qquad
    \delta q_+ = \frac 12 (\delta q_+^a + \delta q_+^b).
\end{equation}
Conversely, given $V, R_+$, \refeq{Rplusdef} determines $(V_+,\delta
q_+^a , \delta q_+^b)$, and we state our results about the map $(V,K)
\mapsto (R_{+},K_+)$.  This then uniquely specifies a map $(V,K)
\mapsto (\delta q_+,V_+,K_+)$.  The construction of $R_{+}$ is
explicit and relatively simple, and its formula is written in
Section~\ref{sec:Rconstruction} below.

To state our estimates on $R_{+}$, we recall the definition of $\Scal$
from Definition~\ref{def:blocks2}, write $B_{\Qcal^{(0)}}(r) = \{V\in
\Qcal^{(0)} : \|V\|_\Qcal < r\}$, and define
\begin{equation}
\lbeq{BT0r}
    B_{T_0}(r)
    =
    \{ K \in \Kcal :
    \sup_{Y \in \Scal}
    \|K(Y)\|_{T_0(\ell)} < r \}.
\end{equation}
Also, for $j<N$, the covariances $C_j$ are identified with those in the decomposition
of the infinite volume covariance $(-\Delta_{\Zd}+m^2)^{-1}$, and these are
defined and obey the required estimates when
$m^2\in [0,\delta]$ for small $\delta$.
For $C_{N,N}$, we restrict to
$m^2 \in [\delta L^{-2(N-1)},\delta]$ as discussed in Section~\ref{sec:cd}.
Thus we define the intervals
\begin{equation}
\lbeq{massint}
    \Iint_j = \begin{cases}
    [0,\delta] & j<N
    \\
    [\delta L^{-2(N-1)},\delta] & j=N.
    \end{cases}
\end{equation}
We can now state our estimates on $R_{+}$.  The analyticity statement
concerns an analytic map from one complex Banach space to another.  By
definition, such a map is analytic on an open domain if it is
continuously Fr\'echet differentiable on that domain (see, e.g.,
\cite[Appendix~A]{PT87} or \cite{Chae85} for the elements of Banach
space analyticity).  In the derivative estimates, the $L^{p,q}$ norm
is the norm of a multi-linear operator from $\Qcal^p \times \Kcal_j^q$
to $\Qcal_{j+1}$.  The continuity in $m^2$ is in the interval $[0,\delta]$
for \emph{all} scales $j$; the restriction for $j=N$ occurs later.

The proof of Theorem~\ref{thm:mr-R} is given in
Section~\ref{sec:Iconstruction}.

\begin{theorem}
\label{thm:mr-R} Let $\volume = \Lambda$ and $j < N(\Lambda)$.  There
exists $r_Q>0$ (small) such that the map $R_{+} : B_{\Qcal^{(0)}}(r_Q)
\times \Kcal \times \Iint_+ \rightarrow \Qcal_{+}$
is analytic in $V$, quadratic in $K$,
continuous in $m^2 \in [0,\delta]$, and independent of $N$.  There
exists $M$ (large, dependent on $p,q \in \N_0$, independent of
$r_0,r_Q$) such that for $r_0\in (0,r_Q)$ and $(V,K,m^2) \in
B_{\Qcal^{(0)}}(r_Q) \times B_{T_0}(r_0) \times [0,\delta]$,
\begin{align}
\label{e:Rmain}
    \|D_V^p D_K^qR_{+}\|_{L^{p,q}}
    & \le
    M
    \begin{cases}
    r_0 r_Q^{-p} & p\ge 0,\, q=0\\
     r_Q^{1-p-q} &p\ge 0,\, q = 1,2\\
    0  &p\ge 0,\, q \ge  3.
    \end{cases}
\end{align}
Each Fr\'echet derivative $D_V^p D_K^qR_{+}$, when applied as a
multilinear map to directions $\dot{V}$ in $\big(\Qcal^{0}\big)^{p}$
and $\dot{K}$ in $\Kcal^{q}$, is jointly continuous in all arguments,
$m^{2}, V,K, \dot{V}, \dot{K}$.  In particular, it is jointly
continuous on the boundary $m^{2}=0$.
\end{theorem}

Next, we specify domains for the $K_+$ part of the RG map.  Let
$j<N(\Lambda)$.  We fix $\ggen_j,\ggen_{j+1}$ obeying
\refeq{gbarmono}.  As in
\cite[\eqref{IE-e:h-coupling-def-1-bis}]{BS-rg-IE}, we fix a universal
constant $C_\DV$ and for $x=\nu,z,\lambdaa,\lambdab$ define $r_{x,j}$
by
\begin{gather}
\label{e:h-coupling-def-1-bis}
    L^{2j}r_{\nu,j} = r_{z,j} = C_{\DV} \ggen_j,
    \quad\quad
    r_{\lambda,j} =
      C_{\DV}.
\end{gather}
We then define
\begin{align}
    \DV_j = \{(g,\nu,z,\lambda)\in \C^4 : 
    &
    C_{\DV}^{-1} \ggen_j < {\rm Re}\, g <
    C_{\DV} \ggen_j,
    \; |{\rm Im} \,g| < \textstyle{\frac {1}{10}} {\rm Re} \,g,
    \nnb & |x| \le r_x
    \; \text{for $x=\nu,z,\lambdaa,\lambdab$}\}
    \label{e:DV1}
,
\end{align}
which is the important stability domain defined in
\cite[\eqref{IE-e:DV1-bis}]{BS-rg-IE} restricted to $y=q_{\pp\qq}=0$.
The mass $m^2$ determines the sequence $\chicCov_j$ defined above
\refeq{rhoFcaldef} (in particular, $\chi_j=1$ for all $j$ when
$m^2=0$).  For $j<N$ and $\domr>0$ (large), we define domains
$\domRG_{j}=\domRG_{j}(\volume) \subset \Qcal \times
\Kspace_j(\volume)$ by
\begin{equation} \label{e:domRG}
  \domRG_j(\volume)
  =
  \DV_j
  \times
  B_{\Wcal_j(\volume)}(\domr\chi_j^{3/2}\ggen_j^3)
  .
\end{equation}
The radius $\domr\chi_j^{3/2}\ggen_j^3$ of the ball in \refeq{domRG}
depends on $m^2$ via $\chicCov_j$, and increases as $m^2$ decreases.
By definition,
\begin{equation}
    B_{\Wcal_j}(\domr\chi_j^{3/2}\ggen_j^3)
    \subset
    B_{\Fcal_j(G)}(\domr\chi_j^{3/2}\ggen_j^3)
    \subset
    B_{T_0}(\domr\chi_j^{3/2}\ggen_j^3),
\end{equation}
so with the choices $r_Q=C_\DV\ggen_j$ and
$r_0=\domr\chi_j^{3/2}\ggen_j^3$, the domain of Theorem~\ref{thm:mr-R}
is larger than $\domRG_j$:
\begin{equation}
\lbeq{Dbigger}
    \domRG_j \subset B_{\Qcal^{(0)}}(C_\DV \ggen_j) \times B_{T_0}(\domr\chi_j^{3/2}\ggen_j^3).
\end{equation}

The following theorem, which constructs the $K_+$ part of the
renormalisation group map, is our main result.  The construction of
$K_+$ is explicit, but it is not simple.  The theorem is a local
existence theorem for the dynamical system that $\RG$ generates: it
says in particular that the map $(V,K) \mapsto K_{+}$ is defined and
contractive when $(V,K)$ is in the domain $\domRG_j$ (which in
particular requires that $K$ be in a small ball).  The contractivity
appears in \refeq{DVKbd}, due to $\kappa <1$.  It is also evidenced by
the fact that we can choose $\domr$ to be large without affecting the
value of $M$, so in particular if we choose $\domr=2M$ then we see
from the $p=q=0$ case of \refeq{DVKbd} that the radius of the ball for
$K_+$ is half that of the ball for $K$ in the domain $\domRG_j$.  In
the derivative estimates, the $L^{p,q}$ norm is the norm of a
multi-linear operator from $\Qcal^p \times \Kcal_j^q$ to
$\Kcal_{j+1}$.

\begin{theorem} \label{thm:mr}
Let $\volume = \Lambda$ and $j < N(\Lambda)$.
  Fix any $a \in (0, 2^{-d})$,
  $\domr>0$,
  $C_\DV$ (both as large as desired),
  and
  let $L$ be sufficiently large (depending on $\domr$).  Let $p,q\in\N_0$.
  There exist $\delta$  (depending on $\domr,L$),  $M>0$ (depending on $p,q,L$ but not $\domr$)
  and $\kappa = O(L^{-1})$ such that for all $\ggen \in (0,\delta)$
  and $m^2 \in \Iint_{j+1}$, there exists a map
\begin{equation}
    K_+:\domRG_{j}(\Lambda)   \to \Wcal_{j+1}(\volume)
\end{equation}
  such that \refeq{rgmapdef-bis} holds.
  The map $K_+$ is analytic
  in $(V,K)$, and, pointwise in $(V,K)$, satisfies the estimates
\begin{alignat}{3}
\lbeq{DVKbd}
    \|D_{V}^pD_{K}^{q}K_+\|_{L^{p,q}}
    &\le
    \begin{cases}
    \kappa & p=0,\, q=1
    \\
    M  \chicCov_+^{3/2} \ggen_+^{3-p}
    &
    p
    \ge 0
    \\
    M  \ggen_+^{-p}
    \left(
    \chicCov_+^{1/2}
    \ggen_+^{10/4}
    \right)^{1-q}
    &
    p \ge 0,\, q \ge 1
    .
    \end{cases}
\end{alignat}
\end{theorem}

By Theorem~\ref{thm:mr-R} and \refeq{Dbigger}, under the hypotheses of
Theorem~\ref{thm:mr}, we also have
\begin{align}
\label{e:Rmain-g}
    \|D_V^p D_K^qR_{+}\|_{L^{p,q}}
    & \le
    \begin{cases}
    M
    \chicCov^{3/2}\ggen^{3-p} & p\ge 0,\, q=0\\
    M
     \ggen^{1-p-q} &p\ge 0,\, q = 1,2\\
    0  &p\ge 0,\, q \ge  3.
    \end{cases}
\end{align}
Furthermore, by \eqref{e:gbarmono}, we can replace $\chicCov$
and $\ggen$ in \eqref{e:Rmain-g} by $\chicCov_{+}$ and $\ggen_{+}$ at
the cost of increasing $M$ by a bounded multiple depending only on
$\Omega$.

Our construction of $K_{+}$ gives it a local dependence on
$K$, as formulated in the next proposition.

\begin{prop}
\label{prop:KplusU} For $U\in \Pcal_{j+1}(\Lambda)$, the value of
$K_{+}(U)$ depends on $K$ only via the restriction $K|_{U^\Box}$ of
$K$ to polymers in $\Pcal_j(U^\Box)$.
\end{prop}

To gain some insight into the meaning of the norm estimates, suppose
that the $p=q=0$ estimate of \refeq{DVKbd} holds at the final scale
$j+1=N$, i.e., $\|K_N\|_{\Wcal_N} \le M\chi_N^{3/2}\ggen_N^{3/2}$.  In
\cite{BBS-saw4-log}, we use the $\theta$ which appears in
\refeq{rgmapdef-bis} at all scales, but in \cite{BBS-saw4} the simpler
case in which $\theta$ is omitted at the final scale is sufficient.
We consider here the simpler case, in which in \refeq{rgmapdef-bis}
the final integration leaves no dependence on the fields.  There is
only one non-empty polymer at the final scale, namely $\Lambda$
itself. We denote the effect of setting the boson and fermion fields
to zero by a superscript $0$.  Then $K_N^0(\Lambda)$ is a
complex scalar, and we write its direct sum decomposition, as in
\refeq{Ncaldecomp}, as $K_N^0(\Lambda) =
K_{N;\varnothing}^0 + \sigma K_{N;\sigma}^0 + \sigmab
K_{N;\sigmab}^0 + \sigma \sigmab K_{N;\sigma\sigmab}^0$.
By \refeq{T0dom} and the definition of the norm in
\cite[\eqref{loc-e:Fnormsum}]{BS-rg-loc},
\begin{equation}
\lbeq{KWbd}
    \|K_N\|_{\Wcal_N} \ge \|K_N(\Lambda)\|_{T_{0,N}(\ell_N)}
    =
    |K_{N;\varnothing}^0| + \ell_{\sigma,N} |K_{N;\sigma}^0|
    + \ell_{\sigma,N} |K_{N;\sigmab}^0| + \ell_{\sigma,N}^2 |K_{N;\sigma\sigmab}^0|,
\end{equation}
where
\begin{equation}
\label{e:ellsigdef}
    \ell_{\sigma ,N}
=
    2^{N-j_{ab}}L^{ j_{\pp\qq}[\phi]}\ggen_{N}
,
\end{equation}
by \refeq{newhsig}.  We always assume that $N$ is larger than the
coalescence scale $j_{ab}$, so that $a,b$ can be identified with
points on the torus.  Also, it follows from \refeq{Phi-def-jc} that
$L^{j_{ab}}$ is bounded above and below by multiples of $|a-b|$ (in
particular, $|a-b| \ge \frac 12 L^{j_{ab}}$).  Thus we conclude that
\begin{equation}
\lbeq{Kg1}
    |K_{N;\varnothing}^0|
    \le M\chi_N^{3/2} \ggen_N^3
    \quad \text{and} \quad
    \left|
    K_{N;\sigma\bar\sigma}^0
    \right|
    \le
    \frac{M'}{4^{N-j_{ab}}}\frac{1}{|a-b|^{2[\phi]}}
    \chi_N^{3/2} \ggen_N,
\end{equation}
for some $M'$.  This is used in
\cite{BBS-saw4}.

We also consider the continuity of $K_+$ in the mass parameter $m^2
\in I_{j+1}$.  This issue is complicated by the fact that the radius
of the ball in $\Kcal_j$ in the domain $\domRG_j$ of \refeq{domRG}
depends on $\chi_j$, which itself depends on $m^2$.  Similarly, the
space $\Wcal_j$ depends on $\rho_j$, which also depends on $\chi_j$
and hence on $m^2$.  To disentangle the domain from the mass parameter
we wish to vary, we fix $\mgen^2 \in I_{j+1}$ and define
$\chigen_j=\chi_j(\mgen^2)$, and use this to define the domain and
space.  Thus we define the spaces $\tilde\Wcal_j$ by replacing
$\chi_j$ by $\chigen_j$ in \refeq{rhoFcaldef}, and we define the
domains
\begin{equation}
\label{e:domRGgen}
  \tilde\domRG_j(\volume)
  =
  \DV_j
  \times
  B_{\tilde\Wcal_j(\volume)}(r\chigen_j^{3/2}\ggen_j^3)
  .
\end{equation}
By definition, $\chigen_j$ increases as $\mgen^2$ decreases.  Consequently
the domain $\tilde\domRG_j$ increases as $\mgen^2$ decreases, and hence if
$(V,K) \in \tilde\domRG_j(\mgen^2)$ for a fixed value of $\mgen^2$, then
$(V,K) \in \tilde\domRG_j((\mgen')^2)$ for all $\mgen' \le \mgen$.
We also define
the intervals
\begin{equation}
\lbeq{Itilint}
    \Igen_j = \Igen_j(\mgen^2) =
    \begin{cases}
    [\frac 12 \mgen^2, 2 \mgen^2] \cap \Iint_j & (\mgen^2 \neq 0)
    \\
    [0,L^{-2(j-1)}] \cap \Iint_j & (\mgen^2 =0).
    \end{cases}
\end{equation}

\begin{theorem}
\label{thm:Kmcont} Let $\volume = \Lambda$ and $j < N(\Lambda)$.  Let
$a,\domr,C_\DV,L,\delta,M,\kappa$ be as in Theorem~\ref{thm:mr}.  Let
$\mgen^2 \in \Iint_{j+1}$.  The map $K_+$ of Theorem~\ref{thm:mr}
extends to a map
\begin{equation}
    K_+:\tilde\domRG_{j}(\Lambda)  \times  \Igen_{j+1}(\mgen^2)
    \to
    \tilde\Wcal_{j+1}(\Lambda),
\end{equation}
which is analytic in $(V,K)$, and obeys the estimates
\refeq{DVKbd}.  For $j+1<N$, every Fr\'echet derivative $D_V^p
D_K^qR_{+}$, when applied as a multilinear map to directions $\dot{V}$
in $\big(\Qcal^{0}\big)^{p}$ and $\dot{K}$ in $\Wcal^{q}$, is jointly
continuous in all arguments $m^{2}, V,K, \dot{V}, \dot{K}$. The domain
of joint continuity includes the boundary $m^{2}=0$, provided $(V,K)$
is in the domain $\tilde\domRG_{j}(\Lambda) \times
\Igen_{j+1}(\mgen^2)$ defined with $\mgen^2=0$.
\end{theorem}

Our main results all include the presence of observables,
corresponding to the observable fields $\sigma,\bar\sigma$.  However,
our construction is \emph{triangular}, in the sense that the bulk part
of $(V_+,K_+)$, obtained by setting $\sigma=\bar\sigma =0$, is the
same as if no observables were present in the original $(V,K)$, i.e.,
\begin{equation}
\label{e:piVKplus}
    \pi_\varnothing V_+(V,K) = V_+(\pi_\varnothing V, \pi_\varnothing K),
    \quad\quad
    \pi_\varnothing K_+(V,K) = K_+(\pi_\varnothing V, \pi_\varnothing K).
\end{equation}
The map $\pi_{\varnothing} : \Ncal \rightarrow \Ncal_{\varnothing}$ is
linear and bounded in $T_{0}$ norm, and therefore it is continuous in
the topology of this norm.  Furthermore $\pi_{\varnothing} : \Ncal
\rightarrow \Ncal_{\varnothing}$ is a homomorphism of algebras,
because it is evaluation at $\sigma =\bar{\sigma} =0$.  Therefore, for
any polynomial $F (V,K)$ in $V$ and $K$, we have $\pi_{\varnothing}F
(V,K) = F(\pi_{\varnothing}V,\pi_{\varnothing}K)$ and the same is true
for $T_{0}$ limits of polynomials.  The first equation in
\eqref{e:piVKplus} then follows from the analyticity statement in
Theorem~\ref{thm:mr-R}, which implies that $R_{+}$ is the limit in
$T_{0}$ norm of truncations of its power series in $V,K$.  To obtain
the second equation in \eqref{e:piVKplus}, we similarly use
Theorem~\ref{thm:mr} to approximate $K_{+} (V,K)$ in $T_{0}$ norm by a
polynomial in $V$ and $K$.

In the presence of observables, \refeq{piVKplus} is supplemented by
the statement that, for $x=a$ or $x=b$,
\begin{equation}
\lbeq{plusindep}
\begin{aligned}
    &\text{if $\pi_x V=0$ and $\pi_x K(X)=0$ for all $X \in \Pcal$ then}
    \\
    &\text{$\pi_x R_+=\pi_{ab} R_+=0$ and $\pi_x K_+(U)
    =\pi_{ab} K_+(U)=0$ for all $U \in \Pcal_+$.}
\end{aligned}
\end{equation}
In addition, $\lambda^a_{+}$ is independent of each of $\lambda^b$,
$\pi_b K$, and $\pi_{ab}K$, and the same is true with $a,b$
interchanged.  The statement in \refeq{plusindep} concerning $K_+$ is
proved in Theorem~\ref{thm:1}(v), and the statements about $R_+$ and
$\lambda_+$ are proved in Proposition~\ref{prop:Rcc}.

\subsubsection{Flow of coupling constants in finite volume}
\label{sec:Rconstruction}

In this section, we explicitly define the map $R_{+}$ of
Theorem~\ref{thm:mr-R}.  The proof that this map obeys the
estimates of Theorem~\ref{thm:mr-R} is deferred to
Section~\ref{sec:Iconstruction}.

We define $R_{+}$ in such a way that the relevant and marginal parts
of $K$ become incorporated into $V_+$.  The operator $\LT$ defined in
\cite{BS-rg-loc} is designed expressly for this purpose.  More
precisely, given $Y \subset X \subset \Lambda$, the operator
$\LT_{X,Y} : \Ncal_X \to \Qcalnabla(Y)$ is defined in
\cite[Definition~\ref{loc-def:LTXYsym}]{BS-rg-loc}, and we employ here
the field dimensions specified in
\cite[Section~\ref{pt-sec:loc-specs}]{BBS-rg-pt}.  The specific
details of the definition of $\LT$ do not play a role in the present
paper, but properties of $\LT$ are important.

The following three steps define $q_+\in \C$ and $V_{+}\in
\Qcalnabla^{(0)}$ as explicit functions of $V,K$.

\begin{enumerate}
\item For $\volume =\Lambda$, given $(V,K)$ and $B \in \Bcal
(\Lambda)$, we define
\begin{equation}
    \label{e:Q-def}
    Q (B)
    =
    \sum_{Y \in \Scal (\Lambda) : Y \supset B}
    \LT_{Y,B} I^{-Y} K (Y),
\end{equation}
where $I=I(V)$ and the negative exponent denotes the reciprocal,
namely $I^{-Y}=\frac{1}{I(V,Y)}=\prod_{B \in
\Bcal(Y)}\frac{1}{I(V,B)}$.  The fact that \refeq{Q-def} defines an
element $Q \in \Qcal$ is proved in Lemma~\ref{lem:Qapp}.  This defines
a map
\begin{equation}
    \label{e:Vhat}
    (V,K) \mapsto
    \Vhat
    =
    V
    -
    Q
     \in \Qcalnabla
.
\end{equation}
\item
We compose the map \refeq{Vhat} with the quadratic function
$V \mapsto \Vpt (V)$
(defined in \cite[\eqref{pt-e:Vptdef}]{BBS-rg-pt}) to obtain the map
\begin{equation}
    (V,K) \mapsto
    \Vpt (\Vhat)
    =
    \Vpt(V-Q)
.
\end{equation}
The map $\Vpt=V_{\pt,j+1}$ is \emph{independent} of $N$; see
\cite[Proposition~\ref{pt-prop:Vptg},
Definition~\ref{pt-def:VptZd}]{BBS-rg-pt}.
\item
Finally, we set
\begin{equation}
    \label{e:Vjplus1}
    V_{+}
    =
    \Vpt^{(0)} (\Vhat)
    ,
    \quad\quad
    \qcpl_{+}
    =
    \qcpl + {\textstyle{\frac{1}{\sigma\bar\sigma}}} \pi_{ab}\Vpt(\Vhat)
,
\end{equation}
with the superscript $(0)$ denoting the operation described under
\refeq{rgmapdef-bis} (replacement of $z\tau_\Delta +
y\tau_{\nabla\nabla} + q\sigma\bar\sigma$ by $(z+y)\tau_\Delta$).
\end{enumerate}

We then define $I_{+} \in \Bcal \Kcal_{j+1} (\Lambda)$ by
\begin{equation}
    I_{+} = I_{j+1} (V_{+})
.
\end{equation}
The above definition of $(V_+,q_+)$ determines $R_{+}: \Qcal^{(0)}
\times \Kcal_{j} (\Lambda) \to \Qcal^{(1)}$ by
\begin{equation}
\label{e:rhoflow}
    R_{+} (V,K)  =
    \Vpt^{(1)}(\Vhat) - \Vpt^{(1)}(V)
    .
\end{equation}
By definition, $R_{+}$ is a quadratic function of $K$; its dependence
on $V$ is nontrivial due to the dependence in $Q$ of $I$ on $V$.

We now interpret more explicitly the meaning of the estimate
\refeq{Rmain} for the flow of coupling constants determined by
Theorem~\ref{thm:mr-R}.  By \refeq{Rplusdef} and \refeq{rhoflow},
$V_+,\dq_+$ are determined by $(V,K)\in\domRG_j$ by
\begin{align}
\lbeq{Vflow}
    V_+(V,K) &=
    \Vpt^{(0)}(V) + ( \Vpt^{(0)}(\Vhat)-\Vpt^{(0)}(V)),
    \\
\lbeq{qflow}
    (\textstyle{\frac 12}\delta q_+^a,\textstyle{\frac 12}\delta q_+^b)
    &=  {\textstyle{\frac{1}{\sigma\bar\sigma}}} \pi_{ab}\Vpt(V)
    +  {\textstyle{\frac{1}{\sigma\bar\sigma}}} \pi_{ab}( \Vpt(\Vhat)-\Vpt(V))
    ,
    \\
\lbeq{qqq}
    \dq_+(V,K) &
    =
    \textstyle{\frac 12}(\delta q_+^a +\textstyle{\frac 12}\delta q_+^b).
\end{align}
The first terms on the
right-hand sides of \refeq{Vflow}--\refeq{qflow} are independent of
$K$ and constitute the pertubative flow discussed at length in
\cite{BBS-rg-flow}.  The last terms on the right-hand sides of
\refeq{Vflow}--\refeq{qflow} do depend on $K$ and constitute the
non-perturbative correction to the perturbative flow.  We write these
non-perturbative corrections to the coupling constants
$(g_+,z_+,\nu_+,\lambda_+^a,\lambda_+^b,q_+^a,q_+^b)$ as $v_{x,j}$,
with
$x=g,z,\nu,\lambda^a,\lambda^b,q^a,q^b$.
The following
proposition gives estimates for these correction terms.

\begin{prop}
\label{prop:Rcc} Let $j<N$, $(V,K)\in \tilde\domRG_j$, and $m^2 \in
\Iint_{j+1}$.  The bounds
\begin{align}
    v_{g,j} &= O(\chi_j^{3/2} \ggen_j^3), \quad v_{z,j} = O(\chi_j^{3/2} \ggen_j^3), \quad
    v_{\nu,j} = O(\chi_j^{3/2} L^{-2j} \ggen_j^3), \nnb
    v_{\lambda,j} & =O(\chi_j^{3/2}  \ggen_{j}^{2}) \1_{j< j_{\pp \qq}}
    \quad
    v_{q,j} = \frac{1}{|a-b|^{2[\phi]}} O(\chi_j^{3/2} 4^{-(j-j_{ab})} \ggen_{j})\1_{j\ge j_{\pp \qq}}
\label{e:vjbds}
\end{align}
hold with $L$-dependent constants, where $\lambda$ represents either
of $\lambda^a,\lambda^b$ and similarly for $q$.  For $x=a$ or $x=b$,
if $\pi_x V=0$ and $\pi_x K(X)=0$ for all $X \in \Pcal$ then $\pi_x
R_+=\pi_{ab} R_+=0$.  In addition, $\lambda^a_{+}$ is independent of
each of $\lambda^b$, $\pi_b K$, and $\pi_{ab}K$, and the same is true
with $a,b$ interchanged.  Finally, each $v_{j}$ is continuous in $m^2
\in [0,\delta]$.
\end{prop}

\begin{proof}
Recall the definition of the $\Qcal$ norm in \refeq{Vnorm} and the
definition of $\ell,\ell_\sigma$ from \eqref{e:hl}--\eqref{e:newhsig}.
With these, \refeq{Rmain} gives the estimates \eqref{e:vjbds}, where
the indicator functions for $v_{\lambda,j},v_{q,j}$ arise as follows.

The last term on the right-hand side of \refeq{qflow} determines
$v_{q,j}$.  To justify the indicator function in \eqref{e:vjbds} we
have to show that $v_{q,j}$ is zero for $j < j_{ab}$.  By the
definition of $j_{\pp \qq}$ the distance between $\pp$ and $\qq$ is at
least $\frac{1}{2}L^{j_{\pp \qq}}$. A small set of scale $j$ has
diameter at most $cL^j$ for some $c$ depending only on $d=4$. For
$j<j_{\pp \qq}$, since $L$ is large no small set at scale $j$ can
contain both points $a$ and $b$, so $\pi_{\pp \qq}Q=0$ and hence
$\pi_{\pp \qq}\Vhat=\pi_{\pp \qq}V$.  Since $\Vpt(\Vhat) -\Vpt(V)$ is
quadratic in $\Vhat$ we must also consider $\sigma \bar{\sigma}$ cross
terms. Cross terms between $\sigma \bar{\phi}_{\pp}$ and $\bar{\sigma}
\phi_{\qq}$ are zero because $\Ex_{j+1} \bar{\phi}_{\pp}\phi_{\qq} =
C_{j+1;\pp \qq} = 0$ when $j < j_{\pp \qq}$ (see
\cite[Lemma~\ref{pt-lem:Pobs}]{BBS-rg-pt}).  Thus $v_{q,j}$ is zero
for $j < j_{ab}$.

Let $j \ge j_{\pp \qq}$. We have to prove that $v_{\lambda
,j}=0$. This holds if $\pi_{\pp}\Vhat=\pi_{\pp}V$ for the $\Vhat$ and
$V$ in the second term of \eqref{e:Vflow}. By \eqref{e:Q-def} this
holds if $\sigma \phib$ and $\sigmab\phi$ are not in the range of
$\LT$ at scale $j$.  This is discussed in
\cite[Section~\ref{pt-sec:loc-specs}]{BBS-rg-pt}, where it is
explained that the parameters in $\LT$ are indeed selected so that for
$j \ge j_{\pp \qq}$, $\sigma \phib$ and $\sigmab\phi$ are not in the
range of $\LT$.

Suppose now that $\pi_a V=0$ and $\pi_a K(X)=0$ for all $X \in \Pcal$.
Then $\pi_a\Vpt(V)=0$ by the formula for $\lambdapt$ in
\cite[\eqref{pt-e:lambdapt2}]{BBS-rg-pt}, and $\pi_a \Vhat = 0$ by
\refeq{Q-def}--\refeq{Vhat}.  From this it follows that as required,
$\pi_x R_+=\pi_{ab} R_+=0$.  A similar argument applies when $a$ is
replaced by $b$.

To see that $\pi_a V_{+}$ is independent of each of $\pi_b V$, $\pi_b
K$, $\pi_{ab}K$, we argue as follows.  Since the flow of $\lambda^a$
stops at the coalescence scale, we may assume that $j<j_{ab}$.  Let
$X\in \Scal_j$ be a small set that contains $a$.  Then $X$ cannot also
contain $b$, so by the field locality assumption in
Definition~\ref{def:Kspace}, $\pi_b K(X)=\pi_{ab}K(X)=0$, and hence
$\Vhat$ does not depend on $\pi_b K$ or $\pi_{ab}K$.  We appeal again
to the formula for $\lambdapt$ in
\cite[\eqref{pt-e:lambdapt2}]{BBS-rg-pt} to conclude that
$\pi_a\Vpt(\Vhat)$ does not depend on $\lambda^b$ either.  A similar
argument applies when $a$ is replaced by $b$.

The continuity in $m^{2}$ of $v_{x,j}$ holds because the coefficients
of $\Vpt$ (given explicitly in
\cite[\eqref{pt-e:gpt2a}--\eqref{pt-e:qpt2}]{BBS-rg-pt}) are
continuous in $m^2\in [0,\delta]$ by
\cite[Proposition~\ref{pt-prop:rg-pt-flow}]{BBS-rg-pt}.
\end{proof}

Finally, for use in \cite{BBS-saw4}, we make the
following additional observation.  Let $\nu^+=\nu + 2g C_{0,0}$.  We
claim that
\begin{equation}
\lbeq{factnu}
    |\nu_+-\nu^+| =O( \chi_j^{3/2} L^{-2j}\ggen_j^2).
\end{equation}
To see this, we apply \eqref{e:Vflow} and \eqref{e:vjbds} to
obtain $|\nu_+-\nu_{\pt}| =O( \chi_j^{3/2} L^{-2j} \ggen_j^3)$, so it
suffices to show that $|\nu_{\pt}-\nu^+| =O( \chi_j^{3/2} L^{-2j}
\ggen_j^2)$.  For the latter, we see from
\cite[\eqref{pt-e:nupta}]{BBS-rg-pt} that $\nu_{\pt}-\nu^+$ is a sum
of terms that are each quadratic in the bulk coupling constants, and
the claim then follows using $\ggen$ bounds on the coupling constants
and \cite[Lemma~\ref{pt-lem:wlims}]{BBS-rg-pt}.

\subsubsection{Main result in infinite volume}
\label{sec:rg-iv}

Theorem~\ref{thm:Kmcont} concerns the renormalisation group map on a
torus $\Lambda$.  We now develop a framework which permits an
extension of the map to the infinite volume $\Zd$, and state results
concerning this extension.  The main result is Theorem~\ref{thm:VKZd}.

To begin, we fix a scale $j<\infty$, and now regard
Theorem~\ref{thm:Kmcont} as simultaneously a statement about every
torus $\Lambda$ with $N(\Lambda)>j$.  We write the $\Lambda$-dependent
input to Theorem~\ref{thm:Kmcont} as $K_\Lambda$, so we have a family
$(K_\Lambda)$ for all $\Lambda$ with $N(\Lambda)>j$, with each
$K_\Lambda \in \Kcal_j(\Lambda)$.  The output of
Theorem~\ref{thm:Kmcont} includes a family $(K_{+,\Lambda})$, with
each $K_{+,\Lambda} \in \Kcal_{j+1}(\Lambda)$.   We associate to
an embedding of a torus into a larger torus a compatibility condition
on the family $(K_\Lambda)$ that is preserved by the renormalisation
group map, and use this compatibility to construct the renormalisation
group map in infinite volume.

For a nonempty polymer $X\in\Pcal_{*} (\volume)$ (with $\volume$
either $\Lambda$ or $\Zd$) and a torus $\Lambda'$, we say that $\iota$
is a \emph{coordinate map} from $X$ to $\Lambda'$ if (i) $\iota:X
\rightarrow \Lambda'$ is an injective map that maps nearest neighbour
points in $X$ to nearest-neighbour points in the image set $\iota X$,
(ii) nearest-neighbour points in $\iota X$ are mapped by $\iota^{-1}$
to nearest-neighbour points in $X$, (iii) if $X$ contains a
point $x$ where there is an observable then $\iota x$ is the location
of the observable in $\Lambda '$.  When we write $\iota^{-1}$, we
always understand it to be the inverse defined on the image $\iota X$.

Next, we define the maps on $\Ncal$ induced by $\iota$.
Let $X$ be a polymer in $\Pcal_{k} (\volume)$ for some scale $k$ and
let $\iota$ be a coordinate map from $X$ to $\Lambda'$. For $\phi$ in
$\C^{\Lambda'}$ we define $\phi_{\iota }$ in $\C^{X}$ by $(\phi_{\iota
})_{x} = \phi_{\iota x}$, and similarly for the Grassmann generators,
$(\psi_{\iota })_{x} = \psi_{\iota x}$.  To define the action of
$\iota$ on $\Ncal$, it suffices to define the action of $\iota$
separately on the summands $\Ncal^\varnothing$, $\Ncal^\pp$,
$\Ncal^\qq$, $\Ncal^{\pp\qq}$ in \eqref{e:Ncaldecomp}.  We define an
algebra isomorphism $\iota: \Ncal^\varnothing (X) \to
\Ncal^\varnothing (\iota X)$ (the same name $\iota$ is used also for
this map), as follows.  An element
\begin{equation}
    \label{e:Fform}
    F = \sum_{y \in \vec\Lambdabold^*}
    \frac{1}{y!}F_y \psi^y
\end{equation}
of $\Ncal^\varnothing(X)$ is defined in terms of coefficients $F_{y}$
which are functions of fields in $X$, i.e., $F_y : \C^X \to \C$.  We
define $\iota (F_{y}):\C^{\iota X}\to \C$ by $\iota
(F_{y})(\phi)=F_y(\phi_{\iota})$ and then set
\begin{align}
\label{e:iotaNcal}
    \iota F
    &=
    \sum_{y \in \vec\Lambdabold^*}
    \frac{1}{y!}\iota ( F_{y}) \psi_{\iota }^{y}
.
\end{align}
According to the definition of $\psi_{\iota}$ the product
$\psi_{\iota}^{y}$ is a product of generators attached to points in
$\iota X$, as it should be. The correspondence between $\iota$ as a
coordinate map and $\iota$ as an algebra isomorphism is functorial: if
$j = \iota \circ\iota '$ as coordinate maps then $j = \iota \circ
\iota '$ as maps on $\Ncal$.  To define the action of $\iota$ on
$\Ncal^{\pp} (X)$, recall that the elements of $\Ncal^{\pp} (X)$ have
the form $\sigma F$ with $F\in \Ncal^{\varnothing} (X)$. Then we
set $\iota \sigma F = \sigma \iota F$. Thus $\iota$ does nothing to
the observable fields $\sigma$ and $\bar{\sigma}$, which makes it clear how
the action of $\iota$ on $\Ncal^\qq$, $\Ncal^{\pp\qq}$ is defined.

Any polymer $X$ on a torus $\Lambda$ whose diameter is less than that
of $\Lambda$ will have a coordinate map to any larger torus $\Lambda'$
(meaning $N(\Lambda') \ge N(\Lambda)$), and we say that $X$ is a
\emph{coordinate patch} on $\Lambda$ if $\diam{X} \le \frac 12
\diam{\Lambda}$.  In particular, coordinate patches cannot wrap around
the torus. We always assume that $L>2^{d}$, so that for scales $j<N
(\Lambda)$ small sets are coordinate patches.  The next
definition introduces the compatibility condition mentioned above.  It
is called Property~$(\Zd)$ and it relates $K_{\Lambda}$ to $K_{\Lambda
'}$. Notice that the definition allows $\Lambda '=\Lambda$.  In this
case Property~$(\Zd)$ is equivalent to the Euclidean invariance statement in
Definition~\ref{def:Kspace}.

\begin{defn}\label{defn:KZd}
We say that a family $(K_\Lambda)$ with each $K_\Lambda \in
\Kcal_j(\Lambda)$ has Property~$(\Zd)$ if
\begin{align}
\label{e:KZd}
    \iota K_{\Lambda} (X)
    =
    K_{\Lambda '} (\iota X)
    &\;\;
    \text{for all
    coordinate patches $X \in \Pcal_{j} (\Lambda)$,
    all $\Lambda'$ larger than $\Lambda$,}
    \nnb
    & \;\;\text{and all coordinate
    maps $\iota : X^\Box \to \Lambda'$}
.
\end{align}
Given a family $(K_\Lambda)$ that has Property~$(\Zd)$ we define
$K_{\Zd} \in \Kcal_{j} (\Zd)$ by
\begin{equation}
    K_{\Zd} (X)
    =
    \iota^{-1}\big(K_{\Lambda} ( \iota X)
    \big)
    \quad\quad
    (X \in \Pcal_j(\Zd))
,
\end{equation}
for some choice of $\Lambda$ with $\diam{\Lambda}\ge 2\diam{X}$, and
some choice of a coordinate map
$\iota : X \to \Lambda$.
\end{defn}

We claim that if $K$ has Property~$(\Zd)$ then $K_{\Zd} (X)$ does not
depend on $\iota$ or $\Lambda$.  To see this, suppose we have two
coordinate maps $\iota_{1} ,\iota_{2}$ from $X$ to tori $\Lambda$ and
$\Lambda '$, with $\Lambda '$ the larger torus. Then there exists a
coordinate map $\iota_{3}$ from $(\iota_{1} X)^{\Box}$ to $\Lambda '$
such that $\iota_{2} = \iota_{3}\circ\iota_{1} $ on
$X^{\Box}$. Property~$(\Zd)$ implies that
\begin{equation}
    \iota_{3} K_{\Lambda} ( \iota_{1} X)
    =
    K_{\Lambda '}( \iota_{3}\circ\iota_{1} X
    )
    =
    K_{\Lambda '}( \iota_{2}X)
,
\end{equation}
and the claim then follows by applying $\iota_{2}^{-1} =
\iota_{1}^{-1}\circ\iota_{3}^{-1}$ to both sides.

For a function $F$ defined on polymers in $\Pcal_{j} (\volume)$
and a polymer $Y$ in $\Pcal_{j} (\volume)$, let $F|_{Y}$ denote the
restriction of $F$ to $\Pcal_j(Y)$, i.e. to scale-$j$ polymers $X
\subset Y$.  According to Proposition~\ref{prop:KplusU},
$K_{+,\Lambda}(U)$ depends on $K_\Lambda$ only via
$K_\Lambda|_{U^\Box}$, and for fixed $V$ we can therefore regard the
map $K_\Lambda \to K_{+,\Lambda}$ defined by Theorem~\ref{thm:Kmcont}
as a family of maps $g_\Lambda : K_\Lambda|_{U^\Box} \mapsto
K_{+,\Lambda}(U)$ indexed by $\Lambda$.

We will prove the following proposition (see
Theorem~\ref{thm:1}(iii)).

\begin{prop}
\label{prop:gLam} Let $U \in \Pcal_{j+1}(\Lambda)$ be a coordinate
patch and let $\iota : U^\Box \to \Lambda'$ with $\Lambda'$ larger
than $\Lambda$.  Then $\iota g_\Lambda (K_\Lambda|_{U^\Box})
= g_{\Lambda'}(\iota K_\Lambda |_{U^\Box})$.
\end{prop}

The following proposition shows that Property~$(\Zd)$ is preserved by
the renormalisation group map.

\begin{prop}
\label{prop:KplusZd} If the collection $(K_{\Lambda})$ has
Property~$(\Zd)$ then $(K_{+,\Lambda})$ produced by
Theorem~\ref{thm:Kmcont} also has Property~$(\Zd)$.
\end{prop}

\begin{proof}
Let $U \in \Pcal_{j+1}(\Lambda)$ be a coordinate patch and let $\iota
: U^\Box \to \Lambda'$ with $\Lambda'$ larger than $\Lambda$.  Then
\begin{align}
    \iota K_{+,\Lambda}(U)
    & =
    \iota g_\Lambda(K_\Lambda |_{U^\Box})
    =
    g_{\Lambda'}(\iota (K_\Lambda |_{U^\Box}))
    =
    g_{\Lambda'}(K_{\Lambda'} |_{(\iota U)^\Box})
    =
    K_{+,\Lambda'}(\iota U),
\end{align}
by Proposition~\ref{prop:gLam} for the second equality, and by
Property~$(\Zd)$ of $(K_\Lambda)$ for the third.
\end{proof}

Now we define the infinite volume map $(V, K_{\Zd})\mapsto K_{+,\Zd}$.
We fix $V$ and drop it from the notation.  Let $K_{\Zd}\in \Kcal
(\Zd)$ and $U\in\Pcal_{j+1}(\Zd)$. We choose a torus $\Lambda$ with $N
(\Lambda)>j+1$ and a coordinate map $\iota : U^\Box \to \Lambda$.  We
first aim to apply Lemma~\ref{lem:extendEuclideanK} to define
$K_\Lambda \in \Kcal_j(\Lambda)$ appropriately associated to
$K_{\Zd}$.  For this, let $\Xcal = \iota \Ccal_{j+1}( U^\Box)$, which
is a class of subsets of $\Lambda$.  Define $F:\Xcal \rightarrow
\Ncal$ by $F = \iota \circ K_{\Zd} \circ \iota^{-1}$. For a Euclidean
automorphism $E$ of $\Lambda$, and for $X \in \Xcal$ such that
$EX\in\Xcal$, there is an automorphism $E'$ of $\Zd$ such that $E'
(\iota^{-1}X) = \iota^{-1} (EX)$.  It follows from the Euclidean
covariance of $K_{\Zd}$ that $K_{\Zd}\circ E' = E' \circ K_{\Zd}$, and
it is then straightforward to check that $F (EX) = E (F (X))$, which
is the main hypothesis for Lemma~\ref{lem:extendEuclideanK}. The
hypothesis involving $W$ can be vacuously satisfied by choosing
$W=\infty$, and the other hypotheses hold because $K_{\Zd} \in
\Kspace_{j} (\Zd)$. Therefore, by Lemma~\ref{lem:extendEuclideanK},
there exists an extension $\hat F \in \Kspace_{j} (\Lambda)$ of $F$
such that $K_{\Lambda}$ defined by $K_{\Lambda}= \hat F$ satisfies
\begin{equation}
\lbeq{Kextension}
    K_{\Lambda}|_{\iota U^\Box} = \iota \circ K_{\Zd} \circ
    \iota^{-1}|_{\iota U^\Box}
.
\end{equation}
We then define $K_{+,\Zd}(U)\in\Ncal (U^{\Box})$ by
\begin{equation}
\label{e:KZdmap}
    K_{+,\Zd} (U)
    =
    \iota^{-1} \circ g_\Lambda (K_{\Lambda }|_{\iota U^\Box}),
\end{equation}
and we must prove that this definition assigns
the same value regardless of how we choose $\Lambda$ and $\iota$.

Let $\iota '$ be another coordinate map from $U^{\Box}$ into
another torus $\Lambda '$ with $N (\Lambda ')>j+1$, and let
\begin{equation}
    K_{\Lambda'}|_{\iota' U^\Box} = \iota' \circ K_{\Zd} \circ
    \iota'^{-1}|_{\iota' U^\Box}
.
\end{equation}
Let $j=\iota
'\circ \iota^{-1}$. Then $j$ is a coordinate map from $\iota U^{\Box}
\subset \Lambda$ into $\Lambda '$. By
\eqref{e:KZdmap} and Proposition~\ref{prop:gLam},
\begin{align}
    K_{+,\Zd} (U)
    &=
    (\iota^{-1} \circ g_\Lambda ( K_{\Lambda}|_{\iota U^\Box})
    =
    \iota'^{-1} \circ j \circ g_\Lambda ( K_{\Lambda}|_{\iota U^\Box})
    \nnb &
    =
    \iota'^{-1} \circ g_\Lambda' ( jK_{\Lambda}|_{\iota U^\Box})
    =
    \iota'^{-1} \circ g_\Lambda' ( K_{\Lambda'}|_{\iota' U^\Box}).
\end{align}
Therefore the definition of $K_{+,\Zd} (U)$ does not depend on the
choices in the definition.  Furthermore, this defines a map $K_{\Zd}
\mapsto K_{+,\Zd}$.  Because the finite volume map preserves the
symmetries of Definition~\ref{def:Kspace} by Theorem~\ref{thm:mr}, the
infinite volume map also preserves these symmetries.  The infinite
volume map $K_{\Zd} \mapsto K_{+,\Zd}$ is the unlabelled arrow in the
commutative diagram:

\setlength{\unitlength}{.8mm}
\begin{center}
\begin{picture}(120,40)
\put(38,30){$K_\Lambda|_{(\iota U)^\Box}$}
\put(77,32){\vector(-1,0){21}}
\put(68,36){$\iota$}
\put(80,30){$K_{\Zd}|_{U^\Box}$}
\put(42,28){\vector(0,-1){21}}
\put(36,15){$g_\Lambda$}
\put(82,28){\vector(0,-1){21}}
\put(38,0){$K_{+,\Lambda}(\iota U)$}
\put(59,2){\vector(1,0){19}}
\put(80,0){$K_{+,\Zd}(U)$}
\put(69,5){$\iota^{-1}$}
\end{picture}
\end{center}

The map $R_+$ of Section~\ref{sec:Rconstruction} depends on
$\Lambda$ because it is a function of $K\in\Kcal (\Lambda)$. We now
make this dependence explicit and write $R_{+,\Lambda}$ and
$V_{+,\Lambda}$ in place of $R_{+}$ and $V_{+}$.
To complete the definition of the renormalisation group map in
infinite volume, we define the infinite volume map $(V,K_{\Zd})\mapsto
R_{+,\Zd}$.  This is similar to the construction of the map
$K_{+,\Zd}$, except $R_{+,\Lambda}$ has
values in $\Qcal$ as opposed to values in $\Kcal$.

To distinguish between scale-$(j+1)$ blocks in $\Zd$ and blocks in a
torus, we write $B$ for the former and $C$ for the latter.  In
particular, $R_{+,\Lambda} (C)$ (as in \eqref{e:VXdef}) is an element
of $\Ncal (C)$.  By \refeq{rhoflow}, $R_{+,\Lambda}$ is defined in
terms of $\Vpt$ and in terms of $Q$ of \refeq{Q-def}.  By definition,
$\Vpt$ evaluated on a block $C$ depends only on fields and their
derivatives on $C$, and hence depends on the values of fields in a
cube obtained by extending $C$ by a few vertices in each direction.
The same is true for $Q$ on a scale-$j$ block.  Together, these facts
much more than imply that $R_{+,\Lambda} (C)$ depends only on the
restriction of $K_{\Lambda}$ to polymers in $\Pcal (C^{\Box})$.

Let $K_{\Zd}\in\Kcal (\Zd)$ and $B\in \Bcal_{+} (\Zd)$. We choose a
torus $\Lambda$ with $N (\Lambda)>j+1$ and a coordinate map
$\iota:B^{\Box}\to \Lambda$.  As in the definition of $K_{\Zd}$,
choose $K_{\Lambda}\in\Kcal(\Lambda)$ such that
\begin{equation}
    \label{e:KrB}
    K_{\Lambda}|_{\iota B^\Box}
    =
    \iota \circ K_{\Zd} \circ \iota^{-1} |_{\iota B^\Box}
.
\end{equation}
We define $R_{+,\Zd} (B)\in\Ncal (B)$ by
\begin{equation}
    \label{e:RZd-def}
    R_{+,\Zd} (B)
    =
    \iota^{-1}\Big(R_{+,\Lambda} (V,K_{\Lambda}) (\iota B)\Big).
\end{equation}
The values of $(R_{+,\Zd} (B), B \in \Bcal_{+} (\Zd))$ determine a
unique element $R_{+,\Zd} \in\Qcal$ and therefore we have a map
$(V,K_{\Zd}) \mapsto \Qcal$.  The following proposition shows that
this map does not depend on the choices of $\Lambda$ and $\iota$ made
in its definition.

\begin{prop}
\label{prop:VZd} (i) Let $(K_{\Lambda})$ be a family that has property
$(\Zd)$. Then for any tori $\Lambda$ and $\Lambda '$ with $N
(\Lambda),N (\Lambda ') \ge j+1$,
\begin{equation}
    R_{+,\Lambda}(V,K_{\Lambda})
=
    R_{+,\Lambda'} (V,K_{\Lambda'})
.
\end{equation}
(ii) The definition of $R_{+,\Zd} (B)$ in \eqref{e:RZd-def} does not
depend on the choice of torus $\Lambda$ or coordinate map
$\iota $.
\end{prop}

\begin{proof}
The proofs of (i) and (ii) require the following preparation.  We fix
$V$ and drop it from the notation, and let $\Lambda,\Lambda'$ be as in
part (i) of the proposition.  Let $C\in\Bcal (\Lambda)$ and let
$j:C^\Box\to \Lambda'$ be a coordinate map.  We use subscripts
$\Lambda$ and $\Lambda '$ to indicate membership in $\Ncal (\Lambda)$
or $\Ncal (\Lambda ')$.  Let $K_{\Lambda}\in \Kcal (\Lambda)$ and
$K_{\Lambda'}\in\Kcal (\Lambda ')$ be any elements that satisfy
\begin{equation}
    \label{e:Zd-restricted}
    j K_{\Lambda} (X) = K_{\Lambda '} (jX)
    \quad \quad
    \text{for $X \in \Pcal (C^{\Box})$}
.
\end{equation}
Recall the definition \refeq{Q-def} of $Q$, which here we write as
$Q_\Lambda$ since it depends on $K_\Lambda$.  We claim that
\begin{gather}
    j\big(Q_{\Lambda} (C)\big)
    =
    Q_{\Lambda '} (jC)
.
\end{gather}
By definition,
\begin{gather}
    j\big(Q_{\Lambda} (C)\big)
    =
    \sum_{Y \in \Scal (\Lambda) : Y \supset C}
    j\left(
    \LT_{Y,C} I_{\Lambda }^{-Y} K_{\Lambda } (Y)
    \right)
,
\end{gather}
and by a small variation of
\cite[Proposition~\ref{loc-prop:9LTdef}]{BS-rg-loc} followed by
\eqref{e:Zd-restricted},
\begin{gather}
    j\Big(\LT_{Y,C} I_{\Lambda }^{-Y} K_{\Lambda } (Y)\Big)
    =
    \LT_{jY,jC} j\big(I_{\Lambda }^{-Y}\big) j\big (K_{\Lambda } (Y)\big)
    =
    \LT_{jY,jC} I_{\Lambda'}^{-jY} K_{\Lambda'} (jY)
    \label{e:jLT}
.
\end{gather}
Therefore
\begin{align}
    \label{e:QLambda}
    j\big(Q_{\Lambda} (C)\big)
    &=
    \sum_{Y' \in \Scal (\Lambda') : Y' \supset jC}
    \LT_{Y',jC} I_{\Lambda'}^{-Y'} K_{\Lambda'} (Y')
    =
    Q_{\Lambda '} (jC)
.
\end{align}

\smallskip\noindent (i) Since $(K_{\Lambda})$ has the $(\Zd)$ property,
\eqref{e:QLambda} holds for all blocks $C\in
\Bcal(\Lambda)$. Therefore, $Q_{\Lambda'} = Q_{\Lambda }$ as elements of $\Qcal$.  By the
$\Lambda$-independence of the map $V \mapsto \Vpt$ constructed in
\cite[Section~\ref{pt-sec:fcc}]{BBS-rg-pt} and the definition
of $R_{+}$ in \refeq{rhoflow}, it follows that
$R_{+,\Lambda}(V,K_{\Lambda}) = R_{+,\Lambda'} (V,K_{\Lambda'})$. This
concludes the proof of (i).

\smallskip\noindent (ii) Given $K_{\Zd}\in \Kcal(\Zd)$, choose
$\Lambda,\iota,K_\Lambda$ and $\Lambda',\iota',K_{\Lambda'}$
so that \refeq{KrB} holds for both choices.  In this case, recall that
$N (\Lambda)$ and $N (\Lambda ')$ are greater than $j+1$, and $\iota
,\iota '$ are defined on $B$ in $\Bcal_{+} (\Zd)$.
Then
\eqref{e:Zd-restricted} holds with $j=\iota '\circ \iota^{-1}$ and
$C=\iota B$. Therefore, by part (i),
\begin{gather}
    \iota^{-1}\Big(R_{+,\Lambda} (K_{\Lambda}) (\iota B)\Big)
    =
    \iota'^{-1}j\Big(R_{+,\Lambda} (K_{\Lambda}) (\iota B)\Big)
    =
    \iota'^{-1}\Big(R_{+,\Lambda} (K_{\Lambda}) (j\iota B)\Big)
    \nnb
    =
    \iota'^{-1}\Big(R_{+,\Lambda'} (K_{\Lambda'}) (j\iota B)\Big)
=
    \iota'^{-1}\Big(R_{+,\Lambda'} (K_{\Lambda'}) (\iota' B)\Big).
\end{gather}
This shows that the definition of $R_{+,\Zd} (B)$ in \eqref{e:RZd-def}
does not depend on the choice of $\Lambda$ or $\iota$, and completes
the proof.
\end{proof}

By combining \refeq{KZdmap} and Proposition~\ref{prop:VZd}, we obtain
the \emph{infinite volume map}
\begin{equation}
\label{e:ivmap}
    (V,K_{\Zd})\mapsto (R_{+,\Zd},K_{+,\Zd} )
\end{equation}
for all scales $j<\infty$.  In contrast to the finite volume case, it
is not a defining feature of this map that it preserves a circle
product under expectation.  Indeed we do not have an interpretation of
the expectation nor of the circle product in infinite volume.  What
the infinite volume map does achieve is a simultaneous encoding of the
restriction of the finite volume map to coordinate patches for
\emph{all} volumes (this effectively ignores the part of the finite
volume map that concerns polymers that wrap around the torus).  As
such, we regard the infinite volume map as an inductive limit of the
finite volume maps, which given a polymer $U$ captures the behaviour
of $K_{+,\Lambda}(U)$ for all volumes $\Lambda$ whose diameter is at
least twice that of $U$.  The following theorem is an analogue of
Theorems~\ref{thm:mr-R} and \ref{thm:Kmcont} (the former specialised
as in \refeq{Rmain-g}) that summarises the properties of the infinite
volume map.  It follows directly from Theorems~\ref{thm:mr-R} and
\ref{thm:Kmcont} and the definition of the infinite volume map.

\begin{theorem}
\label{thm:VKZd} Let $\volume = \Zd$ and $j < \infty$.  In
\eqref{e:np2}, set $\Gtilp =1$. Let $a,\domr,C_\DV,L,\delta,M,\kappa$
be as in Theorem~\ref{thm:mr}.  Let $\mgen^2 \in I_{j+1}$.  Then
\refeq{ivmap} defines maps
\begin{equation}
    R_{+,\Zd}:\tilde\domRG_{j}(\Zd)  \times  \Igen_{j+1}(\mgen^2)
    \to
    \Qcal^{(1)}_{j+1},
    \quad
    K_{+,\Zd}:\tilde\domRG_{j}(\Zd)  \times  \Igen_{j+1}(\mgen^2)
    \to \tilde\Wcal_{j+1}(\Zd),
\end{equation}
which are analytic in $(V,K)$, and obey the estimates
\refeq{DVKbd}--\refeq{Rmain-g} with all the norms on $\Zd$ (rather
than on $\Lambda$). Every Fr\'echet derivative $D_V^p
D_K^qR_{+,\Zd}$, when applied as a multilinear map to directions
$\dot{V}$ in $\big(\Qcal^{0}\big)^{p}$ and $\dot{K}$ in $\Wcal^{q}$,
is jointly continuous in all arguments $m^{2}, V,K, \dot{V},
\dot{K}$. The domain of joint continuity includes the boundary
$m^{2}=0$, provided $(V,K)$ is in the domain $\tilde\domRG_{j}(\Zd)
\times \Igen_{j+1}(\mgen^2)$ defined with $\mgen^2=0$.
\end{theorem}

\subsubsection{Generalisations of main results}
\label{sec:gmr}

We have formulated our results in the specific setting of the supersymmetric
representation of the 4-dimensional weakly self-avoiding walk model, defined by
the polynomial $V$ given in terms of bulk and observable terms in \refeq{Vx}.
However, the results are flexible, and can be extended with little effort in
(at least) the following two directions.

\smallskip \noindent \emph{Other observables.}
The first extension is to consider a different choice of
observables.  The observable terms in \refeq{Vx}
are suitable for the analysis of the two-point function in \cite{BBS-saw4}.
The choice of $\h_\sigma$ made in \refeq{newhsig} for the observable field $\sigma$
is designed to be as large as possible so that the observable terms in $V$
remain benign for
the stability estimates of \cite{BS-rg-IE}, and in particular for
\cite[Proposition~\ref{IE-prop:monobd}]{BS-rg-IE}.  Different choices of observables
could be made with corresponding different choices of $\h_\sigma$; what needs to
be checked is that: (i)~the stability estimates continue to hold with the new
observables, which requires that $\h_\sigma$ be not too large, and (ii)~the analogue
of the second estimate of \refeq{Kg1} applies in the new setting, which requires that
$\h_\sigma$ be not too small.  A specific example where both of these objectives
can be met for other observables
is given in \cite{ST-phi4}, where
watermelon networks for the 4-dimensional
weakly self-avoiding walk are analysed at the critical point.  These are networks of
$p$ weakly mutually- and self-avoiding walks from the origin to a distant point $x$,
and we study the asymptotic behaviour as $|x|\to\infty$, for $p>1$.

\smallskip \noindent \emph{The $|\varphi|^4$ model.}
The second extension concerns the 4-dimensional $n$-component $|\varphi|^4$ spin model,
at weak coupling.
In \cite{BBS-phi4-log}, we apply the results of the present paper to analyse the
critical
behaviour, in particular of the susceptibility.
A simplification is that the $|\varphi|^4$ model is purely bosonic---there is no fermion
field.  A small complication is that the model is $O(n)$-symmetric rather than supersymmetric.
Consequently, $V_+$ acquires a constant term $\delta u$ from $\Vpt$, in a similar manner
to the occurrence of
$\delta q \sigma\bar\sigma$ in \refeq{rgmapdef-bis}.  The constant term is a bulk rather
than an observable term, and consequently it occurs in $V(X)$ as $\delta u|X|_0$, where
$|X|_0$ is the cardinality of $X$.  In \cite{BBS-phi4-log}, we extract the constant
term from the circle product, just as we do for $\dq\sigma\bar\sigma$.  This requires
a small adaptation to the proof of Lemma~\ref{lem:K7a}, discussed
in Remark~\ref{rk:uphi4}.  The inclusion of observables for the $|\varphi|^4$ model is
studied in \cite{ST-phi4}.

\section{Reduction to a key theorem}
\label{sec:mr-red}

In this section, we reduce the proofs of Theorems~\ref{thm:mr} and \ref{thm:Kmcont},
as well as Propositions~\ref{prop:KplusU} and \ref{prop:gLam}, to the key result
Theorem~\ref{thm:1}.  We also prove Theorem~\ref{thm:mr-R} concerning the map $R_{+}$.
Finally, we prove
Theorem~\ref{thm:VKZd}, by transferring the finite volume statements of
Theorem~\ref{thm:mr} to infinite volume.
The proof of Theorem~\ref{thm:1}
is substantial and is carried out in the remainder of the paper.

\subsection{Flow of coupling constants}
\label{sec:Iconstruction}

We now prove Theorem~\ref{thm:mr-R}, which concerns the map $R_{+}$.
The proof uses the fact, proved in
\cite[Proposition~\ref{loc-prop:opLTdefXY}]{BS-rg-loc}, that if $Y
\subset X \in \Scal_j$ and $F \in \Ncal (X^\Box)$ then
\begin{equation}
\label{e:LTXY5-00}
    \|\LT_{X,Y}F\|_{T_0} \le C \|F\|_{T_0}.
\end{equation}
It also uses the fact that for a block $B \in \Bcal$ and small
$r_Q>0$, the reciprocal $I(V,B)^{-1}$ of $I(V,B)$ is an analytic
function of $V\in B_{\Qcal^{(0)}}(r_Q)$ taking values in $\Ncal,
\|\cdot\|_{T_0}$.  This and related facts are further discussed in
Section~\ref{sec:ie}, for the study of $K_+$.

A basic tool we use is the Cauchy estimate for analytic functions on
Banach spaces, to infer estimates on the derivatives of an analytic
function from estimates on the function itself.  Cauchy's formula can
be found in \cite[p.~134]{PT87}.  For complex Banach spaces $X,Y$, $f$
an analytic map from a domain in $X$ to $Y$, a positive integer $p$
and directions $\dot{x}= (\dot{x}_{1},\dots ,\dot{x}_{p})$ in $X^{p}$,
it follows from Cauchy's formula that the Fr\'echet derivative of
order $p$ of $f$ is given by
\begin{equation}
    \label{e:cauchy-derivs}
    D^{p}f (x)\dot{x}
=
    \oint  \dots \oint
    f \big(x + \sum t_{j}\dot{x}_{j}\big)
    \frac{dt_{1}}{2\pi i \,t_{1}^{2}} \dots \frac{dt_{p}}{2\pi i \,t_{p}^{2}}
    ,
\end{equation}
where the contours of integration are circles in the complex plane, whose
radius $r$ is such that the polydisc
\begin{equation}
    \{x + \sum_{j}t_{j}\dot{x}_{j}:|t_{j}|<r, j=1,\dots ,p\}
\end{equation}
is contained in the domain of $f$.
The Cauchy estimate follows from Cauchy's
formula in the same way that it does for analytic
functions of a single variable, and can be found explicitly in
\cite[Theorem~9.16]{Chae85}.

The following proposition is used in our proofs of continuity statements
in our main results, e.g., in Theorems~\ref{thm:mr-R}--\ref{thm:mr}.

\begin{prop}
\label{prop:joint-continuity}
Let $X$ and $Y$ be Banach spaces and let $U$ be an open subset of $X$.
Let $E$ be a compact topological space.  Let $f:(s,x) \mapsto f_{s}
(x)$ be a uniformly bounded map from $E\times X$ to $Y$ such that
$x \mapsto f_{s} (x)$ is analytic and $s \mapsto f_{s} (x)$ is
continuous. Then for $p\in \N_0$, the map $(s,x,\dot{x})\mapsto
D^{p}f_{s} (x)\dot{x}$ from $E\times U \times X^{p}$ to $Y$ is jointly
continuous.
\end{prop}

\begin{proof}
Let $x\in U$ and $p\in\N_0$.  By the uniform bound on $f$, and
by \eqref{e:cauchy-derivs}, for directions
$\dot{x}$ of unit norm, the multilinear map $D^{p}f_{s} (y)$ is
bounded in norm uniformly both in $s$ and in $y$ in some
neighbourhood $U_{x}$ of $x$.  Also, by the Cauchy integral formula and
dominated convergence, the map $s \mapsto D^{p}f_{s} (x)\dot{x}$ is
continuous in $s$. Since $D^{p}f_{s} (y)$ is itself differentiable,
and the Fr\'echet derivative is also bounded uniformly, the map $y \mapsto
D^{p}f_{s} (y)$ is norm continuous on $U_{x}$ uniformly in $s$.

We demonstrate the case $p=1$, and omit the proof for $p=0$ as well as the
inductive proof for $p>1$.
Let $p=1$, $s,t \in E$, $x,y \in U$ and $\dot{x},\dot{y} \in X$. We must show that
$D f_{t} (y)\dot{y}$ converges to $Df_{s} (x)\dot{x}$ as
$(t,y,\dot{y})$ tends to $(s,x,\dot{x})$, and we start with
\begin{equation}
    D f_{t} (y)\dot{y}
    -
    Df_{s} (x)\dot{x}
    =
    Df_{t} (y) \big(\dot{y} - \dot{x}\big)
    +
    \big(
    Df_{t} (y)\dot{x}
    -
    Df_{t} (x)\dot{x}
    \big)
    +
    \big(
    Df_{t} (x)\dot{x}
    -
    Df_{s} (x)\dot{x}
    \big)
.
\end{equation}
The first term tends to zero because $Df_{t} (y)$ is bounded in norm
uniformly in $y$ near $x$.
The second term tends to zero because $y \mapsto
Df_{t} (y)$ is norm continuous at $y=x$ uniformly in $t$. The third
term tends to zero because $t \mapsto Df_{t}
(x)\dot{x}$ is continuous at $t=s$. This concludes the proof for $p=1$.
\end{proof}

\begin{proof}[Proof of Theorem~\ref{thm:mr-R}]
By \refeq{rhoflow},
\begin{equation}
    R_{+}(V,K) = \Vpt^{(1)}(V-Q) -\Vpt^{(1)}(V),
\end{equation}
with
\begin{equation}
    Q(B) = \sum_{Y \in \Scal : Y \supset B} \LT_{Y,B}I^{-Y}K(Y).
\end{equation}
The map $\Vpt$ is quadratic in $V$ by definition, and hence entire
analytic in $V$.  The map $\LT_{Y,B}$ is a bounded linear map (with
respect to the $\|\cdot\|_{T_0}$ norm), by \refeq{LTXY5-00}, and, as
noted above, $I^{-Y}$ is analytic in $V \in B_{\Qcal^{(0)}}(r_Q)$.
Therefore $Q$ is analytic in $V \in B_{\Qcal^{(0)}}(r_Q)$ and linear
in $K$, and hence $R_{+}$ is analytic in $V \in B_{\Qcal^{(0)}}(r_Q)$
and quadratic in $K$.  It is also continuous in $m^2 \in [0,\delta]$,
since the coefficients of $\Vpt$ (given explicitly in
\cite[\eqref{pt-e:gpt2a}--\eqref{pt-e:qpt2}]{BBS-rg-pt}) are
continuous in $m^2\in [0,\delta]$ by
\cite[Proposition~\ref{pt-prop:rg-pt-flow}]{BBS-rg-pt}.

Next we prove the estimates of \refeq{Rmain}, which we repeat here as
\begin{align}
\label{e:Qalpha}
    \|D_V^p D_K^qR_{+}\|_{L^{p,q}}
    & \le
    M
    \begin{cases}
    r_0 r_Q^{-p} & p\ge 0,\, q=0\\
    r_Q^{1-p-q} &p\ge 0,\, q = 1,2\\
    0  &p\ge 0,\, q \ge  3.
    \end{cases}
\end{align}
The $q\ge 3$ case of \refeq{Qalpha} holds since $R_{+}$ is quadratic
in $K$.  For the remaining cases of \refeq{Qalpha}, we use the Cauchy
estimate for analytic functions.  Recall from \refeq{T0Q}
that the
$\Qcal$ norm is equivalent to the $T_0$ norm on a block.  We choose
$r_Q$ small enough that $Q$ is analytic in $V \in
B_{\Qcal^{(0)}}(2r_Q)$ and $\|I(V)^{-Y}\|_{T_0} \le 2$ for $Y \in
\Scal$.  Then $\|Q(B)\|_{T_0(\ell)} \le O(r_0)$ for $K \in
B_{T_0}(2r_Q)$.  It follows from a small extension of
\cite[Proposition~\ref{IE-prop:monobd}]{BS-rg-IE} that
$\|\Vpt^{(1)}(V-Q,B)\|_{T_0(\ell)} \le O( r_Q)$.  Let $\dot K$ have
unit $T_0$ norm, and let $Q_t=Q(K+t\dot{K})$ and $f(t)=\Vpt^{(1)}(V-Q_t(B))$.
Then $\|f(t)\|_{T_0} \le O(r_Q)$, and $f(t)$ is
analytic in $t \in \C$, as long as $K+t\dot{K}\in B_{T_0}(2r_Q)$.  We
seek estimates for $K$ in the smaller ball of radius $r_Q$, so $f (t)$
is analytic in $t$ for $|t|< 2r_Q-r_Q=r_Q$.  By the Cauchy estimate,
for $K$ in the smaller ball we have
\begin{align}
    \|D_K \Vpt^{(1)}(V-Q,B) \dot{K}\|_{T_{0,+}}
    &=
    \|f'(0) \|_{T_{0,+}}
    \label{e:cauchyR}
    \le
    \frac{O(r_Q)}{2r_Q-r_Q}
    = O(1)
.
\end{align}
By taking the supremum over $\dot{K}$ we obtain the $(p,q)=(0,1)$ case
of \eqref{e:Qalpha}.

For the $(p,q)=(0,0)$ case of \refeq{Qalpha}, we define
$g(s)=\Vpt^{(1)}(V-Q(sK),B)$ with $K \in B_{T_0}(r_0)$, so that
$R_{+}(B)=\int_0^1 g'(s)ds$.  Application of the previous Cauchy
argument to bound $g'(s)$ leads to $\|R_{+}(B)\|_{T_0} \le O(r_0)$ on
$B_{\Qcal^{(0)}}(2r_Q) \times B_{T_0}(r_0)$ (with $r_0\le r_Q$).  For
$(p,q)=(0,2)$, as in \refeq{cauchyR} where we lost a factor $r_Q$ for
the $K$-derivative in the Cauchy estimate, in a second application of the
Cauchy estimate we lose another $r_Q$ and obtain
$\|D_K^2\Vpt^{(1)}(V-Q,B)\|_{T_0} \le O(r_Q^{-1})$.  This completes
the proof of the $p=0$ case of \refeq{Qalpha}.

We bound the $V$ derivatives similarly, using the fact that a distance
$r_Q$ separates the ball $B_{\Qcal^{(0)}}(r_Q)$ from the boundary of
the larger ball $B_{\Qcal^{(0)}}(2r_Q)$.  For each $V$ derivative, the
Cauchy estimate causes one power of $r_Q$ to be lost.  This is the
origin of the $r_Q^{-p}$ in \refeq{Qalpha}, and this completes the
proof of \eqref{e:Qalpha}.

Finally, in view of the analyticity and continuity established in the first paragraph
of the proof, the joint continuity of $D_V^p D_K^qR_{+}$ follows from the
uniform bound on $R_{+}$ and Proposition~\ref{prop:joint-continuity}.
\end{proof}

\subsection{Flow of non-perturbative coordinate}
\label{sec:Kflow}

In this section, we state Theorem~\ref{thm:1} and show that it implies the
statements concerning $K_+$ in
Theorem~\ref{thm:Kmcont} and hence also Theorem~\ref{thm:mr}.
We include the statements of Propositions~\ref{prop:KplusU} and \ref{prop:gLam}
as part of Theorem~\ref{thm:1}, so as to combine what must be
proved about the finite volume $K_+$ in one place.
Theorem~\ref{thm:1} is stated in terms of $\Fcal$ norms, and subsequently we
show that estimates in terms of $\Fcal$ can be combined to produce estimates
in terms of $\Wcal$
as in Theorems~\ref{thm:mr} and \ref{thm:Kmcont}.
The structure of the proof of Theorem~\ref{thm:1} is discussed in
Section~\ref{sec:thm1structure} below;
the proof is carried out in the
remainder of the paper.

The following theorem holds for either of the norm pairs
$\Fcal=\Fcal_j(G)$, $\Fcal_{+} = \Fcal_{j+1}(T_0)$, or
$\Fcal=\Fcal_j(\tilde G)$, $\Fcal_{+} = \Fcal_{j+1}(\tilde
G^{\Gtilp})$.  These spaces depend on parameters $\ggen,m^2$.  The map
$K_+$ asserted to exist in the theorem is an explicit function of
$(V,K)$ which is the \emph{same} for each of the norm pairs (on the
intersection of the domains).  An important element of
Theorem~\ref{thm:1} is the fact that $\kappa <1$, in fact $\kappa$ can
be made as small as desired by taking $L$ large.  This contractive
property of the map $K_{+}$ is an essential feature in our
applications in \cite{BBS-phi4-log,BBS-saw4-log}. Recall that
$\rhoFcal $ is given in \eqref{e:rhoFcaldef}.

\begin{theorem}
\label{thm:1}
Let $\volume = \Lambda$ and $j < N(\Lambda)$.
\\
(i) Fix any $a \in (0, 2^{-d})$, $C_\DV$ (as large as desired), and
let $L$ be sufficiently large.  There exist $\rD$ (small, independent
of $L$), $\delta$ (small, dependent on $\rD,L$), $M_0>0$ (large,
dependent on $L$), $\cgam^*>0$ (large, independent of $L$), such that
for all $\ggen \in (0,\delta)$
and $m^2 \in \Iint_{j+1}$, and with $\kappa = \cgam^* L^{-1}$, there
exists a map
\begin{equation}
    \label{e:Kballs}
    K_{+} : \DV \times B_{\Fcal} (\rD\rhoFcal)
    \to
    B_{\Fcal_{+}} (\kappa \rD \rhoFcal_+),
\end{equation}
such that the expectation preserves the circle produce in the sense
that \refeq{rgmapdef-bis} holds.  The map $K_+$ is analytic in
$(V,K)$, and
\begin{align}
\label{e:KTay0}
    &
    \|K_{+}(V,0)\|_{\Fcal_{+}}
    \le
    M_0 \epdV_+^3
    .
\end{align}
Moreover, there exist $a_{+}>a$ and $\h_{++}>\h_{+}$ such that
\eqref{e:Kballs}--\eqref{e:KTay0} hold with $a$ replaced by $a_+$
and $\h_{+}$ replaced by $\h_{++}$ in the $k=j+1$ definitions
\eqref{e:KFcal} and \eqref{e:np1}--\eqref{e:np2}.
\\
(ii)
For $U\in \Pcal_{+}(\Lambda)$, the value of
$K_{+}(U)$ depends on $K$ only via $K(X)$ for $X \in \Pcal(U^\Box)$.
\\
(iii)
Let $U \in
\Pcal_{+}(\Lambda)$ be a coordinate patch and let $\iota : U^\Box \to
\Lambda'$ with $\Lambda'$ larger than $\Lambda$.  Let $g_\Lambda$ be
the map defined above Proposition~\ref{prop:gLam}
(given by \refeq{Kballs}).  Then $\iota
g_\Lambda( K_\Lambda |_{U^\Box}) = g_{\Lambda'}(\iota
( K_\Lambda |_{U^\Box}))$.
\\
(iv)
Let $\mgen^2 \in \Iint_{j+1}$.
The map $K_+$ extends to a map
  \begin{equation}
    K_+:
    \DV \times B_{\tilde\Fcal} (\rD\rhoFcal)
    \times  \Igen_{+}(\mgen^2) \to B_{\tilde\Fcal_{+}}(\kappa\rD\rhoFcal_{+}),
  \end{equation}
  which is
  analytic in $(V,K)$,
  continuous in $m^2\in\Igen_{j+1}(\mgen^2)$, and obeys \refeq{KTay0}.
  Here $\tilde \Fcal$ is defined in
  terms of $\chigen = \chi(\mgen^2)$, whereas $m^2$ is the mass in the original
  covariance $(-\Delta+m^2)^{-1}$.
\\
(v)
For $x =\pp$ or $x=\qq$, if $\pi_x V=0$ and $\pi_x K(X)=0$ for all
$X \in \Pcal$ then $\pi_x K_+(U)$ for all $U \in \Pcal_+$.
\end{theorem}

\begin{defn}
\label{def:Kplusprops} For later convenience, we refer to the
analyticity statement of part~(i), and to the statements of parts
(ii,iii,iv), simply as $(V,K)$-\emph{analyticity}, the
\emph{restriction} property, the \emph{isometry} property, and
\emph{mass continuity}, respectively.  There is also a \emph{vanishing
at weighted infinity} property of $K_+$ inherent in the definition of
$\Fcal_+$ (see \refeq{defnFcal}), and the \emph{field locality},
\emph{symmetry} and \emph{component factorisation} properties of $K_+$
inherent in the definition of $\Kcal_+$ in
Definition~\ref{def:Kspace}.  We use these terms when verifying these
eight properties of $K_+$ in later sections.
\end{defn}

Our next goal is to conclude our main results for the finite volume $K_+$,
from Theorem~\ref{thm:1}.  The statement of Theorem~\ref{thm:1} includes
the statements of Propositions~\ref{prop:KplusU} and \ref{prop:gLam},
and we show now that Theorems~\ref{thm:mr} and \ref{thm:Kmcont} follow
from Theorem~\ref{thm:1}.
This requires the conversion of $\Fcal$ estimates to $\Wcal$ estimates.

Let $\ratio = \ggen^{3/4}$, as in \refeq{ratiodef}.  We begin with the
following lemma, which uses the $\Ycal_+$ norm defined by
\begin{equation}
\label{e:Ycaldef}
    \|K\|_{\Ycal_+}
    =
    \max\{ \|K\|_{\Fcal_{+} (T_0)}, \ratio_{+}^{3}\|K\|_{\Fcal_{+} (\tilde{G})} \}.
\end{equation}

\begin{lemma}
\label{lem:KKK} There is a
constant $c_9>0$ such that for any $K \in
\Kcal_+$,
\begin{equation}
\lbeq{YWcal}
    \|K\|_{\Wcal_+}
    \le c_9
    \|K \|_{\Ycal_+}
.
\end{equation}
\end{lemma}

\begin{proof}
It follows from \cite[Proposition~\ref{norm-prop:KKK}]{BS-rg-norm},
for $X \in \Pcal_+$, $K(X) \in \Ncal(X)$, and for any positive integer
$A<p_\Ncal$, that there is a constant $c_A$ such that
\begin{equation}
\label{e:KKK1norm}
    \|K (X)\|_{G_+ ,\ell_+}
\le
    c_A
    \max
    \left\{
    \|K(X) \|_{T_{0,+}(\ell_+)} ,
    \left( \frac{\ell_+}{h_+} \right)^{A+1}
    \|K(X) \|_{\tilde{G}_+ ,h_+ }
    \right\}.
\end{equation}
We apply \eqref{e:KKK1norm} with $A=9$; it is for this reason that we
require $p_\Ncal \ge 10$.  To account for observables, the ratio
$\ell_+/h_+$ here is understood as the maximum of the two ratios
$\ell_{j+1}/h_{j+1}$ and $\ell_{\sigma,j+1}/h_{\sigma,j+1}$.  By
\refeq{hl}--\refeq{newhsig}, both ratios are bounded above by an
$L$-dependent multiple of $\ggen_+^{1/4}$.  This gives
\begin{equation}
\label{e:KKK1}
    \|K (X) \|_{G_{+} ,\ell_{+}}
\le
    c_9\max
    \left\{
    \|K (X) \|_{T_{0,+},\ell_{+}},
    c (L)\ggen_{+}^{10/4}
    \|K(X) \|_{\tilde G_{+},h_{+} }
    \right\}.
\end{equation}
As discussed below \refeq{T0dom}, we make the choice $\tilde a =4a$,
and this choice gives
\begin{gather}
\label{e:rhoequality}
    \rhoFcal_+(\ell)^{f_+(a,X)}
    =
    (\chi_+^{1/2} \ggen_+)^{a(|X|-2^d)_+}
    \nnb
    \hspace{2cm}
    \ge
    (\chi_+^{1/2} \ggen_+^{1/4})^{4a(|X|-2^d)_+}
    = \rhoFcal_+(h)^{f_+(\tilde a, X)}.
\end{gather}
With \eqref{e:KFcal}, this implies that
\begin{equation}
\label{e:Gnormbd}
    \|K \|_{\Fcal_{+}(G)}
    \le
    c_9
    \max
    \left(
    \|K   \|_{\Fcal_{+}(T_{0})},
    c (L)\ggen_{+}^{10/4}
    \|K  \|_{\Fcal_{+}(\tilde G )}
    \right)
,
\end{equation}
and since $c (L)\ggen_{+}^{10/4} \le \ggen_{+}^{9/4} =
\omega_{+}^{3}$ for $\ggen_{+}$ sufficiently small depending on $L$,
this implies that
\begin{equation}
\lbeq{YWcal-bis}
    \|K\|_{\Wcal_+}
    \le c_9
    \|K \|_{\Ycal_+}
.
\end{equation}
This completes the proof.
\end{proof}

We now show that
Theorems~\ref{thm:mr} and \ref{thm:Kmcont} follow from Theorem~\ref{thm:1}.
We begin with Theorem~\ref{thm:mr}, and afterwards consider the
mass continuity statement of Theorem~\ref{thm:Kmcont}.

\begin{proof}[Proof of Theorem~\ref{thm:mr}]
Fix $\domr >0$ (large).  The proof uses $\Wcal$ balls of radii $a\ll
A$, defined in terms of small $r$ and large $R$ by
\begin{equation}
    a = R\chi^{3/2} \ggen^3, \quad\quad A = r \chi^{1/2} \ggen^{10/4}.
\end{equation}
We use analyticity in the ball of larger radius to prove estimates in
the ball of smaller radius, using Cauchy estimates. The radius $a$
appears in the definition \eqref{e:domRG} of $\domRG_{j}(\Lambda)$ and
$A$ is chosen so that $\omega^{-3} A = \rD \rhoFcal (h)$ where
$\omega$ is defined by \eqref{e:ratiodef}.

For $K$ in the larger ball $B_\Wcal(A)$ of radius $A$, the
definition \eqref{e:9Kcalnorm} of the $\Wcal$ norm translates into the
$\Fcal$ estimates
\begin{equation}
    \|K\|_{\Fcal(G)} \le A
    \le r\rhoFcal (\ell),
    \quad \quad
    \|K\|_{\Fcal(\tilde G)} \le \omega^{-3} A
    = \rD \rhoFcal (h)
\end{equation}
and the inclusions
\begin{equation}
    \label{e:W2Fcal}
    B_\Wcal(A) \subset B_{\Fcal(G)}(A),
    \quad \quad
    B_\Wcal(A) \subset B_{\Fcal(\tilde G)}(\omega^{-3}A).
\end{equation}
Recall that Theorem~\ref{thm:1} asserts that $(K,V) \mapsto K_{+}$ is
analytic (continuously differentiable) as a map from $\DV \times
\Fcal$ to $\Fcal_{+}$ for two choices of the pair $(\Fcal,
\Fcal_{+})$, namely $(\Fcal (G), \Fcal_{+} (T_{0}))$ and $(\Fcal
(\tilde{G}), \Fcal_{+} (\tilde{G}))$. Since the inclusions
\eqref{e:W2Fcal} are bounded linear maps they are analytic. Therefore,
the composition of these inclusions with $(K,V) \mapsto K_{+}$ is
analytic.  It follows that $(K,V) \mapsto K_{+}$ is an analytic map from
$\DV \times B_{\Wcal} (A)$ into the intersection of the two choices of
$\Fcal_{+}$, which is the space $\Ycal$ defined by \eqref{e:Ycaldef}.
According to Lemma~\ref{lem:KKK}, $\Ycal$ is continuously embedded into
$\Wcal_{+}$, so with a further composition with this embedding we find
that $(K,V) \mapsto K_{+}$ is an analytic map from $\DV \times
B_{\Wcal} (A)$ to $\Wcal_{+}$. Since $\DV \times B_{\Wcal} (A)$
contains $\domRG$ we have proved that $(K,V) \mapsto K_{+}$ is
analytic on $\domRG$ as claimed in Theorem~\ref{thm:mr}.

Our first task is to prove case $(p,q)=(0,1)$ of the estimates
claimed in Theorem~\ref{thm:mr} for $(V,K) \in \domRG$, namely
\begin{align}
    \label{e:DVKbd-bis}
    \|D_{V}^pD_{K}^{q}K_+(V,K)\|_{L^{p,q}}
    &\le
    \begin{cases}
    \kappa'' & p=0,\, q=1
    \\
    M '' \chicCov_+^{3/2} \ggen_+^{3-p}
    &
    p
    \ge 0
    \\
    M''  \ggen_+^{-p}
    \left(
    \chicCov_+^{1/2}
    \ggen_+^{10/4}
    \right)^{1-q}
    &
    p \ge 0,\, q \ge 1
    ,
    \end{cases}
\end{align}
We will prove that case $(p,q)=(0,1)$
holds on the larger domain $\DV \times B_{\Wcal} (A/2)$ and this
stronger statement is used in the proof of the other cases.

Let $(V,K)\in\DV \times B_{\Wcal} (A/2)$. Let
$\dot{K} \in \Fcal$ and set $T = D_K K_{+}(V,K)$. We first prove that
\begin{equation}
    \label{e:K-derivative}
    \|T \dot{K}\|_{\Fcal_{+}}
    \le
    \kappa ' \|\dot{K}\|_{\Fcal}
,
\end{equation}
where $\kappa ' = O (L^{-1})$.  The argument is the same for both
norm pairs.  We start with the pair $\Fcal = \Fcal (\tilde{G})$ and
$\Fcal_{+} = \Fcal_{+} (\tilde{G})$.  Let
$f(t)=K_{+}(V,K+t\dot{K})$. By Theorem~\ref{thm:1} $f$ is analytic at
$t \in \C$ such that $K+t\dot{K}$ is in the ball $B_{\Fcal} (\rD
\rhoFcal (h))$, and $f (t)$ has values in the ball $B_{\Fcal_{+}}
(\kappa \rD \rhoFcal_{+}(h))$. Since $K$ is in the smaller ball
$B_\Wcal(A/2) \subset B_{\Fcal} (\rD \rhoFcal (h)/2)$, $f (t)$ is
analytic in $t$ for $|t|< \rD \rhoFcal (h)/2$.  By the Cauchy
estimate,
\begin{equation}
    \label{e:cauchy}
    \|T \dot{K}\|_{\Fcal_{+}}
    =
    \|f'(0) \|_{\Fcal_{+}}
    \le
    \frac{\sup_{t} \|f (t)\|_{\Fcal_{+}}}{\rD \rhoFcal (h) - \rD \rhoFcal (h)/2}
    \le
    \frac{\kappa \rD \rhoFcal_{+} (h)}{\rD \rhoFcal (h)/2}
    =
    O (\kappa)
,
\end{equation}
where the last equality follows from \eqref{e:gbarmono}.
By Theorem~\ref{thm:1}, $\kappa = O (L^{-1})$.
Since $T$ is a linear operator, the above bound on
unit norm $\dot{K}$ implies \eqref{e:K-derivative} for this norm pair.

Now we consider the same argument for the other norm pair, $\Fcal =
\Fcal (G)$ and $\Fcal_{+} = \Fcal_{+} (T_{0})$.  This time $f (t)$ is
analytic in $t$ for $|t|< \rD \rhoFcal (\ell) - A/2$ and the
right-hand side of the Cauchy estimate is $\kappa \rD \rhoFcal_{+} (\ell)/ (\rD
\rhoFcal (\ell) - A/2)$.  Since $A = O (\chi^{1/2}\ggen^{10/4})$, it
is negligible relative to $\rD \rhoFcal (\ell)$, which is given by
\eqref{e:rhoFcaldef}, and we have proved \eqref{e:K-derivative} for
both norm pairs.

By \eqref{e:K-derivative}
\begin{gather}
    \|T\dot{K}\|_{\Fcal_{+} (T_{0})}
    \le
    \kappa' \|\dot{K}\|_{\Fcal (G)}
    \le
    \kappa' \|\dot{K}\|_{\Wcal},
\\
    \omega_{+}^{3}\|T\dot{K}\|_{\Fcal_{+} (\tilde{G})}
    \le
    \omega_{+}^{3}\kappa' \|\dot{K}\|_{\Fcal (\tilde{G})}
    \le
    2\kappa'
    \|\dot{K}\|_{\Wcal}
,
\end{gather}
where we used $\omega_{+} \le 2 \omega$. We combine this with
Lemma~\ref{lem:KKK} and the
definition \eqref{e:Ycaldef} of $\Ycal$, to
obtain
\begin{gather}
    \label{e:K-derivative2}
    \|T\dot{K}\|_{\Wcal_{+}}
    \le
    c_{9}\|T\dot{K}\|_{\Ycal_{+}}
    \le
    \kappa'' \|\dot{K}\|_{\Wcal}
,
\end{gather}
where $\kappa '' = 2 c_{9}\kappa '$.
We have proved \eqref{e:K-derivative2} for $(V,K) \in \DV \times B_{\Wcal}
(A/2)$, which is a larger domain than $\domRG$.
Since $T=D_{K}K_{+} (V,K)$, this proves case
$(p,q)=(0,1)$ of \eqref{e:DVKbd-bis}.

Next we prove case $(p,q)=(0,0)$ of \eqref{e:DVKbd-bis}. By
integrating the $t$ derivative of $K_{+}(V,tK)$ with respect to $t$
and estimating the integrand with \eqref{e:K-derivative2}, we have, for
$K$ in $B_{\Wcal} (A/2)$,
\begin{equation}
    \label{e:Kplus-lip}
    \|K_+(V,K) - K_+(V,0)\|_{\Wcal_+}
    \le
    \kappa'' \|K\|_{\Wcal}
.
\end{equation}
Furthermore, it follows from \refeq{KTay0} and Lemma~\ref{lem:KKK}
that
\begin{align}
    \label{e:Kplus0}
    \|K_+(V,0)\|_{\Wcal_+}
    &
    \le c_9 \|K_+(V,0)\|_{\Ycal_+}
    \le
    c_9
    \max \{ M_0 \chi_+^{3/2} \ggen_+^3, \ggen_+^{9/4} M_0 \chi_+^{3/2} \ggen_+^{3/4}\}
    \nnb
    & = c_9M_0\chi_+^{3/2} \ggen_+^3.
\end{align}
Now let $K$ be in the small ball $B_{a} (\Wcal)$ required for case
$(p,q)=(0,0)$.  By combining \eqref{e:Kplus-lip}--\eqref{e:Kplus0}, we obtain
\begin{equation}
\lbeq{Kbis}
    \|K_{+}(V,K)  \|_{\Wcal_{+}}
    \le
    c_9M_0\chi_+^{3/2} \ggen_+^3 + \kappa'' \|K\|_{\Wcal}
    \le
    c_9(M_0   + \kappa R' ) \chi_+^{3/2} \ggen_+^3
    =M_1 \chi_+^{3/2} \ggen_+^3,
\end{equation}
where $R'$ is a multiple of $R$,
we take  $L$ large enough that $\kappa R' \le 1$, and
$M_1=c_9(M_0+1)$.
This proves case $(p,q)=(0,0)$ of \eqref{e:DVKbd-bis}.  For later use,
note that \eqref{e:Kplus-lip}--\eqref{e:Kplus0} imply that
$K_{+}$ maps $\DV \times B_{\Wcal} (A/2)$ into $B_{\Wcal_{+}} (A)$.

To obtain \eqref{e:DVKbd-bis} for the case $q=0$, $p>0$, we fix $K \in
B_{\Wcal}(a)$, for which we have just established that $K_+\in
B_{\Wcal_+} (M_1 \chi^{3/2} \ggen^3)$.  This bound is understood to
hold for $V \in \DV(2C_\DV)$; this is the domain $\DV$ with $C_\DV$
doubled, and since $C_\DV$ is arbitrary in Theorem~\ref{thm:1} we can
use its conclusions with the doubled value.  Let $\|\dot{V}_i\|_\Qcal
=1$, and let $f(s) = K_{+}\big(V+\sum_i s_{i}\dot{V}_{i},K\big)$.  We
choose $\epsilon>0$ so that $V+\sum_i s_i\dot{V}_i \in \DV(2C_\DV)$
for $|s_i|\le \epsilon\ggen$.  Then we apply the Cauchy estimate to
$f$ as an analytic function of $s_{1},\dots ,s_{p}$ in the domain
$|s_{i}| < \epsilon\ggen$.  The denominator in the analogue of
\eqref{e:cauchy} is the distance from $V\in\DV$ to the boundary of the
domain $\DV(2C_\DV)$, which is at least $\epsilon\ggen$. For $p$
derivatives there is one such denominator for each derivative.  This
gives a factor proportional to $\ggen_j^{-p}$ so the Cauchy estimate
bounds the derivative by $O (\chi^{3/2} \ggen^3 \ggen_j^{-p})$ as
stated in the second estimate of \refeq{DVKbd-bis}.

To obtain \eqref{e:DVKbd-bis} for the case $p\ge 1,\, q\ge 1$, we use
the Cauchy estimate on $f (s,t) = K_{+}\big(V+\sum
s_{i}\dot{V}_{i},K+\sum t_{i}\dot{K}_{i}\big)$ as an analytic function
of $s_{1},\dots ,s_{p}$ and $t_{1},\dots ,t_{q}$ in the domain
$|s_{i}|< \epsilon\ggen$ and $|t_{i}|< \frac 12 A$.  The denominator
in the analogue of \eqref{e:cauchy} is the distance from $(V,K)$ to
the boundary of the domain, and for $p+q$ derivatives there is one
such denominator for each derivative, which gives a factor
proportional to $\ggen^{-p} A^{-q}$. As we have proved above the image
of this domain lies in $B_{\Wcal_{+}} (A)$, so the Cauchy estimate bounds
the derivative by $O ( \ggen^{-p} A^{1-q})$ as desired.
\end{proof}

\begin{proof}[Proof of Theorem~\ref{thm:Kmcont}] By
Theorem~\ref{thm:1}(iv), $K_+$ is a continuous function of $m^2\in
\Igen(\mgen^2)$ as a map into $\tilde\Fcal_+$.  This is the case for
each of the norm pairs, so $\tilde\Fcal_+$ can be either
$\tilde\Fcal_+(T_0)$ or $\tilde\Fcal_+(\tilde G)$.  Therefore $K_+$ is
continuous as a map into the space $\tilde\Ycal_+$ defined as in
\refeq{Ycaldef}, and so by Lemma~\ref{lem:KKK} it is continuous also
as a map into $\tilde\Wcal_+$.  The joint continuity of $D_V^p
D_K^qK_{+}$ follows from the uniform bound on $K_{+}$ and
Proposition~\ref{prop:joint-continuity}.
\end{proof}

\subsection{Proof of main result for infinite volume}
\label{sec:reduction}

We now deduce our main result Theorem~\ref{thm:VKZd} for the infinite
volume map, from the finite volume result Theorem~\ref{thm:1}.  The
proof for $R_{+,\Zd}$ is similar to but simpler than the proof for
$K_{+,\Zd}$, and we only present the details for $K_{+,\Zd}$.  In
Section~\ref{sec:Kflow}, it is shown that the statements of
Theorem~\ref{thm:mr} for $K_+$ in finite volume are a consequence of
Theorem~\ref{thm:1}(i,iv).  The sufficiency of
Theorem~\ref{thm:1}(i,iv) was established via Cauchy estimates based
on analyticity, together with an argument to conclude estimates in
$\Wcal$ norm from those in $\Fcal$ norm.  Joint continuity of
the Fr\'echet derivatives was a consequence of
Proposition~\ref{prop:joint-continuity}. These items apply in
the same way to an infinite volume version of
Theorem~\ref{thm:1}(i,iv), so it suffices to prove such an infinite
volume version.  This is the content of Theorem~\ref{thm:1Zd} below.

The infinite volume version of Theorem~\ref{thm:1}(iii) is omitted
because it is meaningless in the infinite volume context.  We do not
need the infinite volume version of Theorem~\ref{thm:1}(ii), but we
note that it does hold by the definition of $K_{+,\Zd}$.  Namely, for
$U\in \Pcal_{+}(\Zd)$, the value of $K_{+,\Zd}(U)$ depends on $K$ only
via $K(X)$ for $X \in \Pcal(U^\Box)$.  In addition, the field locality
and symmetry properties for $K_{+,\Zd}$, required for membership in
$\Fcal_+(\Zd)$, follow from the corresponding finite volume properties
by definition \refeq{KZdmap} of $K_{+,\Zd}$.

\begin{theorem}
\label{thm:1Zd}
Let $j=0,1,\dots$ be any scale.
In \eqref{e:np2} set $\Gtilp =1$.
\\
(i) Fix any $a \in (0, 2^{-d})$, $C_\DV$ (as large as desired), and
let $L$ be sufficiently large.  There exist $\rD$ (small, independent
of $L$), $\delta$ (small, dependent on $\rD,L$), $M_0>0$ (large,
dependent on $L$), $\cgam^*>0$ (large, independent of $L$), such that
for all $\ggen_j \in (0,\delta)$ and $m^2 \in \Iint_{j+1}$, and with
$\kappa = \cgam^* L^{-1}$,
\begin{equation}
    K_{+,\Zd} : \DV \times B_{\Fcal (\Zd)} (\rD\rhoFcal) \to
    B_{\Fcal_{+} (\Zd)} (\kappa \rD \rhoFcal_+)
.
\end{equation}
The map $K_{+,\Zd}$ is analytic in $(V,K)$, and
\begin{align}
\label{e:KTay00}
    &
    \|K_{+,\Zd}(V,0)\|_{\Fcal_{+}}
    \le
    M_0 \epdV_+^3
    .
\end{align}
\\
(iv) Let $\mgen^2 \in \Iint_{j+1}$.  The map $K_{+,\Zd}$ extends to a
map
\begin{equation}
    K_{+,\Zd}:
    \DV \times B_{\tilde\Fcal} (\rD\rhoFcal)
    \times  \Igen_{+}(\mgen^2) \to B_{\tilde\Fcal_{+}}(\kappa\rD\rhoFcal_+),
\end{equation}
which is analytic in $(V,K)$, continuous in
$m^2\in\Igen_{j+1}(\mgen^2)$, and obeys \refeq{KTay00}.  Here $\tilde
\Fcal$ is defined in terms of $\chigen = \chi(\mgen^2)$, whereas $m^2$
is the mass in the original covariance $(-\Delta+m^2)^{-1}$.
\end{theorem}

The proof relies on the facts about coordinate maps $\iota$ and
extensions by symmetry given in
Lemmas~\ref{lem:extendK}--\ref{lem:extendEuclideanK}.

\begin{proof}
(i) Let $\mathfrak{r} = \rD\rhoFcal$, $\mathfrak{r}_{+} = \kappa
\rD\rhoFcal_{+}$, $W (X) = \rhoFcal^{f (a,X)}$, $W_{+} (X) =
\rhoFcal_{+}^{f_{+} (a,X)}$.
Let $(V,K)\in\DV \times B_{\Fcal (\Zd)} (\mathfrak{r})$, so that
(recall the definitions \eqref{e:KFcal} and \eqref{e:np1}--\eqref{e:np2})
\begin{equation}
    \label{e:KZd-in}
    \|K (X)\|_{\Gcal}
\le
    \mathfrak{r} W (X)
    \quad\quad
    (X \in \Ccal (\Zd)),
\end{equation}
with $\Gcal$ equal to $G_{j}$ or $\tilde{G}_{j}$.  We will prove that
$K_{+,\Zd}\in B_{\Fcal_{+} (\Zd)} (\kappa \rD
\rhoFcal_{+})$. Since we set
$\Gtilp=1$, it is equivalent to prove that
\begin{align}
&
    \|K_{+,\Zd}(U)\|_{T_{0,+}}
    \le
    \mathfrak{r}_{+}
    W_{+} (U),
\quad
    \label{e:fKeta2}
    \|K_{+,\Zd}(U)\|_{\tilde{G}_{+}}
    \le
    \mathfrak{r}_{+}
    W_{+} (U)
\end{align}
hold for all connected polymers $U$ in $\Zd$.  We must also show that
$K_{+,\Zd}$ vanishes at weighted infinity (see \refeq{defnFcal}), since this is part of the
definition of $\Fcal_{+}$. In our present context, we must show that
\begin{equation}
    \label{e:Kinfinity}
    \lim_{\|\phi\|_{\Phi(U)}\rightarrow \infty}
    \|K_{+,\Zd} (U)\|_{T_{\phi ,+}} \tilde{G}_{+}^{-1}(U,\phi)
    =
    0
.
\end{equation}
Let $U$ be a connected polymer in $\Zd$. We will construct
$K_{+,\Zd}(U)$ as the image of $K$ under a composition
$\iota_{\Lambda}^{-1} \circ C_{U} \circ B \circ A$ of four maps.

Let $U'=U^{\Box}$, and let $\iota_{\Lambda}: U' \to \Lambda$ be a
coordinate map to a torus. By \eqref{e:KZd-in} and
Lemma~\ref{lem:extendK},
\begin{equation}
    \|\iota_{\Lambda} K (\iota_{\Lambda}^{-1}X)\|_{\Gcal} 
\le
    \mathfrak{r} W (X)
    \quad\quad
    (X \in \Ccal(\iota_{\Lambda} U'))
.
\end{equation}
Let $\Xcal = \Ccal (\iota_{\Lambda} U')$.  By
Lemma~\ref{lem:extendEuclideanK}, with Euclidean symmetry hypothesis
verified as above \refeq{Kextension}, the map $X\mapsto
\iota_{\Lambda} K (\iota_{\Lambda}^{-1}X)$ extends from $\Xcal$ to an
element $\hat{K} \in B_{\Fcal
(\Lambda)}(\mathfrak{r})$. Lemma~\ref{lem:extendEuclideanK} implies
that the map $A_{U}:K \mapsto \hat{K}$ preserves the vanishing at
weighted infinity property and is a linear contraction from $\Fcal
(\Zd)$ to $\Fcal (\Lambda)$.  In particular, the evaluation map is analytic.  We
next apply Theorem~\ref{thm:1Zd}, with $a$ and $\h_{+}$ replaced by the values
$a_{+}$ and $\h_{++}$ provided by Theorem~\ref{thm:1Zd}. To remind us
that we have these stronger values, we write
$\Fcal_{++}$ in place of $\Fcal_{+}$ and $\Ncal_{++}$ in place of
$\Ncal_{+}$. By Theorem~\ref{thm:1}, the map $B:(V,\hat{K}) \mapsto
K_{+,\Lambda} (V,\hat{K})$ is analytic as a map from the
$\mathfrak{r}$ ball in $\Fcal(\Lambda)$ to the $\mathfrak{r}_{+}$ ball
in $\Fcal_{++} (\Lambda)$.
Now consider evaluation on the polymer $\iota_{\Lambda} U$ as a map
$C_{U}:K_{+,\Lambda} \mapsto K_{+,\Lambda} (\iota_{\Lambda} U)$.  By
definition of the $\Fcal_{++} (\Lambda)$ norm, this is a bounded linear
map into the space $\Ncal_{++} (\iota_{\Lambda} U')$, and we have the
$++$ analogue of \eqref{e:Kinfinity}.  In particular it is
analytic. Furthermore, by \eqref{e:KZdmap}, the composition
$C_{U}\circ B\circ A_{U}$ is the map $(V,K) \mapsto \iota_{\Lambda}
K_{+,\Zd} (U)$ because $K_{+,\Lambda} (\iota_{\Lambda} U)$ does not
depend on the choice $\hat{K}$ of extension of $\iota_{\Lambda} K
(\iota_{\Lambda}^{-1}X)$ off $\Xcal$.  In summary, we have proved that
$(V,K) \mapsto \iota_{\Lambda} K_{+,\Zd} (U)$ is analytic as a map
from the $\mathfrak{r}$ ball in $\Fcal (\Zd)$ to $\Ncal_{++}
(\iota_{\Lambda} U')$, and putting the estimates together we have
\begin{equation}
    \label{e:iotaK}
    \|\iota_{\Lambda} K_{+,\Zd}(U)\|_{\Gcal_{++}}
    \le
    \mathfrak{r}_{+}
    W_{++} (U)
.
\end{equation}

We now pass to an estimate on $K_{+,\Zd}(U)$, exploiting the fact that
\eqref{e:iotaK} holds for all $\Lambda$. According to
Lemma~\ref{lem:extendK2} with $\h$ replaced by $\h_{++}$, when
$\Lambda$ is sufficiently large, the inverse $\iota_{\Lambda}^{-1}$ is
a bounded linear operator from $\Ncal_{++} (U^{\Box})$ to $\Ncal_{+}
(U^{\Box})$.  This is the step where we use the parameter $\Gtilp <1$
for the regulator $\tilde{G}^{\Gtilp}$, and
where we use $\h_{++} > \h_{+}$. We obtain
\begin{align}
&
    \|K_{+,\Zd}(U)\|_{T_{0,+}}
    \le
    \mathfrak{r}_{+}
    W_{++} (U),
\\
&
    \label{e:fKeta}
    \|K_{+,\Zd}(U)\|_{\tilde{G}_{+}}
    \le
    \mathfrak{r}_{+}
    W_{++} (U)
.
\end{align}
Therefore \eqref{e:fKeta2} holds and, since the bounded linear
operator $\iota_{\Lambda}^{-1}$ is an analytic map from $\Ncal_{++}
(U^{\Box})$ to $\Ncal_{+} (U^{\Box})$, we can compose with the
previous maps and conclude that $(V,K) \mapsto \iota_{\Lambda}
K_{+,\Zd} (U)$ is analytic as a map from the $\mathfrak{r}$ ball in
$\Fcal (\Zd)$ to $\Ncal_{+} (U')$. Furthermore,
\eqref{e:Kinfinity} holds. Therefore $K_{+,\Zd } \in B_{\Fcal_{+}
(\Zd)} (\mathfrak{r}_{+})$, as desired.

If we set $K=0$ so that $\hat{K}=0$ and use \eqref{e:KTay0}, namely
$\|K_{+}(V,0)\|_{\Fcal_{+}} \le M_0 \epdV^3$, then an argument
analogous to the one above gives \eqref{e:KTay00}.

We now strengthen the above analyticity, which is pointwise in $U$, to
the desired analyticity statement.  It is here that we take advantage
of the fact that $a_+>a$.  For a positive integer $M$, let
$\Ucal_{0}=\{U \in \Pcal_+ : |U|_{j+1} \le M, \, U\ni 0\}$. Let
$\Ucal'_{0}=\{U' \in \Pcal_+ : U\in \Ucal_{0}\}$.  Let $g_{M,0}: \Dcal
\times B_{\Fcal(\Zd)}(\mathfrak{r}) \to \Ncal_{+}^{\Ucal'_{0}}$ be the
map that takes $(V,K)$ into $\big(\pi_{\varnothing}K_{+,\Zd} (U),U \in
\Ucal_{0}\big)$.  The latter is a Banach space with the norm
\begin{equation}
    \|F\|
=
    \sup_{U \in \Ucal_{0}} \|F (U)\|_{\Ncal_{+} (U')}
    W_{++}^{-1}(U)
.
\end{equation}
Since $\Ucal_{0}$ is finite, it follows from
Lemma~\ref{lem:prod-analyticity} and the analyticity pointwise in $U$
that $g_{M,0}$ is analytic.  Let $g_{M} (V,K) = \pi_{\varnothing
}K_{+,\Zd}\1_{M}$, where $\1_{M} (U)=1$ if $|U|_{j+1} \le M$ and
otherwise is zero.  For every polymer $U$ such that $|U|_{j+1}\le M$,
there exists a translation $E_{U}$ of $\Zd$ such that $E_{U}U$
contains the origin.  By the Euclidean invariance of
$\pi_{\varnothing}K_{+,\Zd}$,
\begin{equation}
    \label{e:eu-invariance}
    \pi_{\varnothing}K_{+,\Zd} (U)
    =
    E_{U}^{-1}\pi_{\varnothing}K_{+,\Zd} (E_{U}U)
.
\end{equation}
These relations imply that $g_{M}$ equals the composition of $g_{M,0}$
with a bounded linear extension map into the space $\Kcal_{+} (\Zd)$
with the norm
\begin{equation}
    \|F\|_{\Fcal_{+} (W_{++})}
=
    \sup_{U \in \Ccal_{+} (\Zd)} \|F (U)\|_{\Ncal_{+} (U')}
    W_{++}^{-1}(U)
.
\end{equation}
Therefore, for each $M$ the map $g_{M}$ is analytic as a map with
values in the space $\Kcal_{+} (\Zd)$ with this norm.  When $a_{+}$ is
replaced by $a$, the weight $W_{++}$ becomes $W_{+}$, and this norm
becomes the $\Fcal_{+} (\Zd)$ norm.  Furthermore, since $a_{+}>a$, the
space $\Fcal_{+} (\Zd)$ is a larger space, and uniformly in $V,K$ we
have
\begin{equation}
    \label{e:uniformX}
    \lim_{M\rightarrow \infty}
    \|\pi_{\varnothing}K_{+,\Zd} (1 - \1_{M}) \|_{\Fcal_{+} (\Zd)}
=
    0
.
\end{equation}
Therefore the sequence $g_{M}$ of analytic functions converges to
$\pi_{\varnothing}K_{+,\Zd}$ in $\Fcal_{+} (\Zd)$ uniformly in $K$ as
$M\rightarrow \infty$.  According to \cite[Theorem 2, p.~137]{PT87},
as a uniform limit, $\pi_{\varnothing}K_{+,\Zd}$ is analytic as a map
into $\Fcal_{+} (\Zd)$.  The observable component $\pi_{*}K_{+,\Zd}$
is also analytic as a map into $\Fcal_{+} (\Zd)$ using the same
argument with $\Ucal_{0}$ replaced by $\Ucal_{*}=\{U \in \Pcal_+ :
|U|_{j+1} \le M, \, U\cap \{\pp ,\qq \} \not = \varnothing\}$ and
omitting \eqref{e:eu-invariance} and the line below about composition
with an extension.  Having proved that the $\varnothing$ and $*$
components of $K_{+,\Zd}$ are analytic it follows from
Lemma~\ref{lem:prod-analyticity} that $K_{+,\Zd}$ is analytic as a map
into $\Fcal_{+} (\Zd)$.  This concludes the proof of part~(i).

\smallskip \noindent (iv) Let $\mgen^2 \in \Iint_{j+1}$.  We return to
the map $B$ defined in part~(i). By Theorem~\ref{thm:1}(iv), $B$
extends to a map that depends on $m^2$, and by fixing $(V,\hat{K})$
this extension becomes a continuous map $m^2 \mapsto K_{+,\Lambda}$
into $\tilde{\Fcal}_{+} (\Lambda)$. The other maps $A_{U}, C_{U},
\iota_{\Lambda}^{-1}$ do not depend on $m^{2}$. It follows that the
composition $m^2 \mapsto K_{+,\Zd} (U)$ is continuous as a map into
$\Ncal (U')$ with norms $\|\cdot\|_{T_{0,+}}$ and
$\|\cdot\|_{\tilde{G}_{+}}$.  Finally, if we replace $\Fcal_{+} (\Zd)$
by the $m^2$ independent space $\tilde{\Fcal}_{+} (\Zd)$, then the
convergence in \eqref{e:uniformX} is also uniform in $m^2$ and since
the uniform limit of continuous functions is continuous, we also
achieve mass continuity as claimed. This completes the proof.
\end{proof}

\section{Preliminaries to proof of Theorem~\ref{thm:1}}
\label{sec:pre}

In Section~\ref{sec:thm1structure}, we describe the basic structure of
the proof of Theorem~\ref{thm:1},
and in Section~\ref{sec:thm1parameters} we specify several parameters
that occur in the proof.  In Section~\ref{sec:ie}, we recall several
useful results from \cite{BS-rg-IE}.

We assume throughout the paper
that $\ggen$ is sufficiently small to
carry out each steps that is encountered.
According to the definition of $\epdV$ in \refeq{rhoFcaldef}, taking $\ggen$ small
is equivalent to taking $\epdV$ small, and we often phrase smallness conditions in terms
of $\epdV$ instead of $\ggen$.
The assumption that $\ggen$ is small
is used so frequently that we often apply it without explicit mention.

Throughout the paper, we
use the following notation for non-negative real sequences $A=A_{j}$,
$B=B_{j}$:
\begin{align}
    \label{e:precdef}
    A \prec B
    & \quad
    \text{ if $A_j \le c B_j$ for all $j$, with $c$ independent of $L$},
\\
    A \prec_{L} B
    & \quad
    \text{ if $A_j \le c B_j$ for all $j$, with $c = c (L)$},
\\
    A \asymp B
    & \quad
    \text{ if $A \prec B$ and $B \prec A$}
.
\end{align}

\subsection{Structure of proof of Theorem~\ref{thm:1}}
\label{sec:thm1structure}

We fix $V$ and regard the map $(V,K) \mapsto K_{+}$ asserted
to exist in Theorem~\ref{thm:1} as a map $K \mapsto K_{+}$.
We
construct $K_{+}$ as a composition of six maps:
\begin{gather}
    ({\rm Map}~i) : K^{(i-1)}
    \mapsto
    K^{(i)}
    \quad
    i = 1,\dots ,6,
\quad
    \text{with $K^{(0)} = K$, \; $K^{(6)} = K_{+}$}
.
\end{gather}
The six maps are described in detail in
Sections~\ref{sec:Map1-2}--\ref{sec:Map5-6}, and are described briefly here.

Maps~1 and 2 are defined in such a way that
\begin{align}
    (I \circ K)(\Lambda)
    & =  (I \circ K^{(1)})(\Lambda)
    =  (\Ihat \circ K^{(2)})(\Lambda),
\label{e:KImaps12}
\end{align}
where $\Ihat \in \BKspace_j$ is defined by $\Ihat(V)=I(\Vhat)$,
with $\Vhat=V-Q$ defined by \eqref{e:Vhat} in terms of $Q=Q(V,K)$ given
by \eqref{e:Q-def}.
The combined effect of these two maps is to transfer the relevant and marginal
parts of $K(X)$ for $X \in \Scal$ into $V$, with the result that $V$ is replaced
by $\Vhat$.  The decay in $|X|$ that is encoded in our norms for large sets $X$
by \refeq{KFcal} allows us to forego any transfer of $K(X)$ for large sets $X$.
Map~1 takes advantage of the non-uniqueness of the circle product to replace $K$
by $K^{(1)}$, which results from the transfer of the relevant and marginal
parts of $K$ on small
sets other than blocks, so that they become concentrated
in $K^{(1)}$ on blocks instead.
This transfer is achieved using the important change of variables formula
given by Proposition~\ref{prop:change-of-variable-1}.
Then Map~2 transfers the relevant and marginal parts of $K^{(1)}$ that are
concentrated on single blocks into $V$ so as to form $\Vhat$.  Thus $\Ihat$ appears
on the right-hand side of \refeq{KImaps12}.  All three circle products in
\refeq{KImaps12} are on scale~$j$.

Map~3 is our implementation of the formal power series statement of
\refeq{EIapprox} that $\Ex_{j+1} \theta I_j (V,\Lambda) \approx
I_{j+1}(\Vpt,\Lambda)$, but now no longer merely as a statement about
formal power series.  The renormalised polynomial $\Vpt$ therefore
appears, but since Map~2 has replaced $V$ by $\Vhat$, we write
$\Vpt=\Vpt(\Vhat)=\Vpt(V-Q)$, We define $\tilde{I}_{j+1} \in
\BKspace_{j}$ (as in \cite[\eqref{IE-e:Itildef}]{BS-rg-IE}) by
\begin{equation}
\label{e:Itildef}
    \tilde I_{j+1}(V,b) =
    e^{-V (b)}(1+W_{j+1}(V,b))
    \quad\quad (b \in \Bcal_j)
    ,
\end{equation}
and define $\tilde{I}_\pt \in \BKspace_j$ by
\begin{equation}
    \tilde{I}_\pt = \Itilde_{j+1}(\Vpt).
\end{equation}
The expectation is performed in Map~3, and $K^{(3)}$ is constructed such that
\begin{align}
   \Ex_{+}\theta (\Ihat \circ K^{(2)})(\Lambda)
   &
     =  (\Ipttil \circ K^{(3)})(\Lambda).
\label{e:KImaps3}
\end{align}
The circle product on the left-hand side of \refeq{KImaps3} is at scale~$j$, whereas
on the right it is at scale~$j+1$.  This entails a slight abuse of notation,
in which we regard $\Ipttil$ in \refeq{KImaps3} as the element of $\BKspace_{j+1}$
defined for $B \in \Bcal_{j+1}$ by $\Ipttil(B) = \prod_{b \in \Bcal_j(B)}\Ipttil(b)$.
It is in Map~3 that we change scale in our estimates, with $K^{(2)}$ measured with
scale-$j$ norm but $K^{(3)}$ with scale-$(j+1)$ norm.
This change of scale is important in revealing the contraction encapsulated in the
small parameter $\kappa$ in Theorem~\ref{thm:1}.

The $K^{(3)}$ produced by Map~3 is larger than what is claimed for $K_+$ in
Theorem~\ref{thm:1}, due to the fact
that it includes perturbative contributions that arise because of the local manner
in which we implement the spirit of
the proof of \refeq{EIapprox} from \cite[Proposition~\ref{pt-prop:I-action}]{BBS-rg-pt}.
Map~4 reapportions these overly large parts of $K^{(3)}$ by a second application of
the change of variables of Proposition~\ref{prop:change-of-variable-1},
and thereby constructs a better $K^{(4)}$ such that
\begin{align}
   & (\Ipttil \circ K^{(3)})(\Lambda)
    =  (\Ipttil \circ K^{(4)})(\Lambda).
\label{e:KImaps4}
\end{align}

Maps~5 and 6 perform three final adjustments, all relatively minor.
One adjustment is to put $\Ipttil$ into the correct scale-$(j+1)$ form
of \refeq{IVB} rather than as a product over scale-$j$ blocks.
The other two deal with the fact that $\Vpt$ contains terms $y\tau_{\nabla\nabla}$
and $\dq \sigma\sigmab$ which are not present in $V_+$.
The term $y\tau_{\nabla\nabla}$ is converted to a term $z\tau_\Delta$ via
summation by parts, at the cost of an adjustment to $K$.  The term
$\dq \sigma\sigmab$ is pulled outside the circle product, at the cost of another
small adjustment to $K$.  This finally leads to
\begin{align}
   &
    (\Ipttil \circ K^{(4)})(\Lambda)
    =  (\Ipt^+ \circ K^{(5)})(\Lambda)
    = e^{\dq\sigma\sigmab} (I_{+} \circ K_{+})(\Lambda),
\label{e:KImaps56}
\end{align}
where the precise definition of $\Ipt^+$ is given in Map~5.
The circle products in \refeq{KImaps4}--\refeq{KImaps56} are all at scale~$j+1$.

The combination of \refeq{KImaps12} and \refeq{KImaps3}--\refeq{KImaps56} gives
\begin{equation}
    \Ex_{+}\theta (I \circ K)(\Lambda) =
    e^{\dq\sigma\sigmab} (I_{+} \circ K_{+})(\Lambda),
\end{equation}
which is \eqref{e:rgmapdef-bis}.
This shows that $(V_+,K_+)$ preserves the form of the circle product under expectation, as
required.  To complete the proof of Theorem~\ref{thm:1}(i), it is necessary also
to show that there exist $r,M>0$ and $\kappa =O(L^{-1})$ such that
$K_{+} : \DV \times B_{\Fcal} (\rD
\rhoFcal) \to B_{\Fcal_{+}} (\kappa \rD \rhoFcal)$ with $K_+$ an analytic map,
and such that $\|K_{+}(V,0)\|_{\Fcal_{+}} \le M \epdV^3$.
These facts are required for each of the norm pairs
$\Fcal = \Fcal_j(G)$, $\Fcal_+=\Fcal_{j+1}(T_0)$ and
$\Fcal = \Fcal_j(\tilde G)$, $\Fcal_+=\Fcal_{j+1}(\tilde G)$.
We carry out Maps~1--6 simultaneously for each of the two norm pairs, and
prove the estimates and analyticity map by map, culminating in Map~6 with
the desired statements for $K_+$.
Similarly, relevant observations concerning the statements of
Theorem~\ref{thm:1}(ii--iv) are made for each Map, and at Map~6 their proof for
$K_+$ is complete.  See Section~\ref{sec:pfthm1}, where the proof of Theorem~\ref{thm:1}
is summed up.

\subsection{Parameters for proof of Theorem~\ref{thm:1}}
\label{sec:thm1parameters}

For convenience, we gather here the specification of several parameters
that occur in the proof of Theorem~\ref{thm:1}.

For each Map~$i$ with $i=1,2,4,5,6$ ($i=3$ is excluded) there is an
associated $L$-independent constant $\mu_i \ge 1$; the values
(typically large) of these constants are determined in
Sections~\ref{sec:Map1-2}--\ref{sec:Map5-6}.  For Map~3, there is an
important constant $\kappa_3$ which is an $L$-independent multiple of
$L^{-1}$; $\kappa_3$ can be made as small as desired by taking $L$
sufficiently large.

For the constant $r$ that determines the size of balls appearing in
Theorem~\ref{thm:1}, we fix any value
\begin{equation}
\label{e:r0def}
    0 < r < \min \left\{ \frac{1}{\mu_1 \mu_2}, 
    1 \right\}.
\end{equation}
We set $r^{(0)}=r$, and define $r^{(i)}=\mu_i r^{(i-1)}$ for
$i=1,2,4,5,6$, whereas $r^{(3)}=\kappa_3 r^{(2)}$.  We define the
small parameter $\kappa$ by
\begin{equation}
    \kappa = \mu_1\mu_2 \kappa_3  \mu_4\mu_5\mu_6 < 1.
\end{equation}
Then $\kappa = \cgam^* L^{-1}$ for an $L$-independent constant
$\cgam^*$, $r^{(6)}=\kappa r^{(0)}$, and $r^{(i)} < 1$ for each $i$.

We fix $a\in (0,2^{-d})$ as in the statement of Theorem~\ref{thm:1}.
Let $\eta = \eta(d) >1$ be the geometrical constant of
Lemma~\ref{lem:small}.  For $i \in \{0,1,2,3,4,5,6\}$, we fix
$a^{(i)}$ such that
\begin{align}
\label{e:achoice}
    &
    0 < a^{(2)} < a^{(1)} < a^{(0)}=a <
    a^{(6)} < a^{(5)} < a^{(4)} < a^{(3)}
    < \eta a^{(2)} \le 2^{-d}
.
\end{align}
The parameters $a^{(i)}$ determine Banach spaces
$\Fcal_{k}^{(i)}=\Fcal_{k} (a^{(i)})$ which are defined by replacing
$a$ by $a^{(i)}$ in \eqref{e:f0def} and \eqref{e:KFcal}.

For $i=1,2$, Map~$i$ maps the ball of radius $r^{(i-1)}\rhoFcal$ in
$\Fcal_j(a^{(i-1)})$ into the ball of radius $r^{(i)}\rhoFcal$ in
$\Fcal_j(a^{(i)})$.  In Map~3, the scale increases from $j$ to $j+1$,
and the ball of radius $r^{(2)}\rhoFcal$ in $\Fcal_j(a^{(2)})$ is
mapped into the ball of radius $r^{(3)}\rhoFcal$ in
$\Fcal_{j+1}(a^{(3)})$.  Map~3 is the beneficial map, as $a^{(3)}$
improves by becoming larger, and also $r^{(3)}$ improves by becoming
smaller thanks to the factor $\kappa_3$.  These improvements undergo
degradations in Maps~4--6, in which the ball of radius
$r^{(i-1)}\rhoFcal$ in $\Fcal_{j+1}(a^{(i-1)})$ is mapped into the
ball of radius $r^{(i)}\rhoFcal$ in $\Fcal_{j+1}(a^{(i)})$.  However
the overall effect remains beneficial, with $a^{(6)}>a^{(0)}$, and
with $r^{(6)}=\kappa r^{(0)}$ for small $\kappa=\cgam^* L^{-1}$.  The
composition of the six maps is well defined and maps the ball
$B_{\Fcal_{j}} (\rD^{(0)} \rhoFcal)$ into a small ball $B_{\Fcal_{j}}
(\kappa\rD^{(0)} \rhoFcal)$.

\subsection{Interaction estimates}
\label{sec:ie}

The analysis of the six maps uses estimates on $I,\Ihat,\Ipttil$,
which we refer to generically as \emph{interaction estimates}.
These estimates, which include stability estimates, rely on the hypothesis that
\begin{equation}
\label{e:VKhyp}
    (V,K) \in \DV \times B_{\Fcal}(\rD \rhoFcal),
\end{equation}
with $\Fcal$ either $\Fcal(G)$ or $\Fcal(\tilde G)$, with corresponding
choice of $\rhoFcal$ in \refeq{rhoFcaldef}.
It is the purpose of \cite{BS-rg-IE} to provide the
interaction estimates, and we appeal frequently to results from
\cite{BS-rg-IE}, some of which we now recall.
Further results from \cite{BS-rg-IE} are recalled within the analysis
of Maps~3,5,6.

The analysis of \cite{BS-rg-IE} is missing an ingredient needed here,
which is that $K$ now plays a role in interaction estimates.
For example, $\Ihat \in \BKspace_j$,
is defined by $\Ihat(V)=I(\Vhat)$,
with $\Vhat=V-Q$ defined by \eqref{e:Vhat} in terms of $Q=Q(V,K)$ given
by \eqref{e:Q-def} as
\begin{equation}
    \label{e:Q-def-2}
    Q (B)
    =
    \sum_{Y \in \Scal (\Lambda) : Y \supset B}
    \LT_{Y,B} I^{-Y} K (Y),
    \quad\quad
    B \in \Bcal (\Lambda).
\end{equation}
Similarly, $\Ipttil = \Itilde(\Vpt(\Vhat))$ depends on $K$ as well as on $V$.
The following proposition combines results from \cite{BS-rg-IE} with new statements
concerning $K$-dependence.
For its proof, we  recall from
\refeq{LTXY5-00}
that $\LT_{X,Y}$ is a bounded operator, in the sense that for $Y\in \Pcal_j$ with
$Y \subset X \in \Scal_j$, and for $F' \in \Ncal (X^\Box)$,
\begin{equation}
\label{e:LTXY5-first}
    \|\LT_{X,Y}F'\|_{T_0} \prec \|F'\|_{T_0}.
\end{equation}
We also recall that the enlarged domain $\bar\DV \supset \DV$ is
defined in \cite[\eqref{IE-e:DVell}--\eqref{IE-e:DVh}]{BS-rg-IE}; its
precise definition does not play a direct role below.  Finally,
recall that we write $I^{-X} = 1/I(X)$.

\begin{prop}
\label{prop:Ianalytic1}
Let $I_*$ denote any one of $I,\Ihat,\Ipttil$, with $j_*$ respectively
equal to $j,j$, and to \emph{either} of $j$ or $j+1$.
Let $B\in \Bcal_j$.
Let $(V,K) \in \DV_j \times B_{\Fcal} (r^{(0)}\rhoFcal)$.
Then $I_*(B)$ is an analytic function of $(V,K)$ taking values in
$\Ncal(B^\Box), \|\cdot\|_{j_*}$.
In addition, for $F\in \Ncal(B^{\Box})$ a gauge-invariant
polynomial of bounded degree in the fields such that
$\pi_{ab}F=0$ when $j \le j_{\pp \qq}$,
\begin{align}
\label{e:IF}
    \|I_*(B) F\|_{j_*,\h}
    &
    \prec
    \|F\|_{T_{0,j} (\h)}
,
\\
\label{e:Iass1}
    \|I_*(B)\|_{j_*,\h}
    &
    \le 2
,
\\
\label{e:I-b}
    \|I_*^{-B}\|_{T_{0,j_*} (\h)}
    &
    \le 2
    .
\end{align}
In addition, $I^{-B}$ is an analytic function of $V \in \bar\DV_j$ taking values in
$\Ncal(B^\Box), \|\cdot\|_{T_{0,j}}$.
\end{prop}

\begin{proof}
For the case $K=0$, all the above statements are proved already in
\cite[Propositions~\ref{IE-prop:Istab}--\ref{IE-prop:Ianalytic1:5}]{BS-rg-IE}.
In particular, this gives the above statements concerning $I$, since there is
no $K$-dependence in $I$.
Our task here is to extend the statements of
\cite[Propositions~\ref{IE-prop:Istab}--\ref{IE-prop:Ianalytic1:5}]{BS-rg-IE}
to include the dependence of $\Vhat,\Vpt$ on $K \in B_{\Fcal} (r^{(0)}\rhoFcal)$,
as well as on $V$.
In fact, the statements of
\cite[Propositions~\ref{IE-prop:Istab}--\ref{IE-prop:Ianalytic1:5}]{BS-rg-IE} are all
proved
in the enlarged domain $\bar\DV \supset \DV$,
and this is important below.

We begin with a bound on $Q$, which repeats a step in the proof of
Theorem~\ref{thm:mr-R}.
Let $(V,K) \in \DV \times B_{\Fcal}(r^{(0)}\rhoFcal)$.
We claim that
\begin{equation}
\label{e:Qbound-2}
    \|Q(B)\|_{T_0 }
\prec
    r^{(0)}
    \rhoFcal
    .
\end{equation}
To prove \refeq{Qbound-2},
we apply \eqref{e:LTXY5-first}, together with the $I_*=I$ case of \refeq{IF}, \refeq{I-b},
and the product property of the norm, to conclude that
if $F\in \Ncal(B^{\Box})$  then
\begin{align}
    \label{e:JCK0I}
    \|I^{X}\LT_{X,Y}I^{-X} F\|_{j}
    &\prec
    \|\LT_{X,Y}I^{-X} F\|_{T_{0,j}}
    \prec
    \|F\|_{T_{0,j}}
    .
\end{align}
With \refeq{Q-def-2}, this gives
\begin{equation}
\label{e:Qboundpf}
    \|Q(B)\|_{T_0 }
\prec
    \sup_{X \in \Scal}\|K(X)\|_{T_{0}}
\le
    \|K\|_{\Fcal}
\le
    r^{(0)} \rhoFcal
    ,
\end{equation}
which proves \refeq{Qbound-2}.

Now that \refeq{Qbound-2} has been established, it follows immediately from
\cite[Proposition~\ref{IE-prop:JCK-app-1}]{BS-rg-IE} that $\Ihat$ obeys
the estimates \refeq{IF}--\refeq{I-b}, that $\Ihat$ is analytic in $V$
for fixed $Q$, and that $V-Q$ lies in the enlarged domain $\bar\DV'$
(the prime
denotes an unimportant change in constants defining $\bar\DV$) on which
stability and analyticity of $I$ is proved in
\cite[Proposition~\ref{IE-prop:Ianalytic1:5}]{BS-rg-IE}.
Moreover, it follows from the definition of $Q$ in \refeq{Q-def-2} and the results
already established for $I$ that $(V,K) \mapsto V-Q$ is analytic as a map from
$\DV \times B_{\Fcal}(r^{(0)}\rhoFcal)$ into $\bar\DV$.  Thus $\Ihat$ has
the claimed analyticity.  Similarly, since $V-Q \in \bar\DV$, it follows from
\cite[Proposition~\ref{IE-prop:monobd}]{BS-rg-IE} that $\Vpt(V-Q)$ also lies
in the analyticity domain of $I$ both for scale $j$ and scale $j+1$, and hence $\Ipttil$ is analytic
both as a map $\Vpt \mapsto \Ipttil$ defined on $\bar\DV$, and
as a function of $(V,K)$ as desired.
This completes the proof.
\end{proof}

The proof of
analyticity of $I^{-B}$ easily extends to a small
$\|V(B)\|_{T_0}$-ball, since there is no need for a positivity of the coupling
constant $g$ when employing the $T_0$ norm; this
small extension is used in Theorem~\ref{thm:mr-R}
above.
For future reference, we proved in \refeq{Qbound-2} that, for
$(V,K) \in \DV \times B_{\Fcal}(r^{(0)}\rhoFcal)$,
\begin{equation}
\label{e:Qbound}
    \|Q(B)\|_{T_0 }
\prec
    r^{(0)}
    \rhoFcal
    .
\end{equation}
Also, with the bounds \eqref{e:IF} and \eqref{e:I-b}, the proof of
\refeq{JCK0I} extends to show that
for any of the three choices of $I_*$ above,
if $Y\in \Pcal_j$, $Y \subset X \in
\Scal_j$ and $F \in\Ncal(X)$ then
\begin{align}
    \label{e:JCK0}
    \|I_*^{X}\LT_{X,Y}I_*^{-X} F\|_{j}
    &\prec
    \|\LT_{X,Y}I_*^{-X} F\|_{T_{0,j}}
    \prec
    \|F\|_{T_{0,j}}
    .
\end{align}
We use \eqref{e:JCK0} repeatedly.
Note that all norms in \eqref{e:JCK0} are at scale $j$, even when
$I_*=\Ipttil$.

The following lemma is also useful.  Its first two estimates can be
understood as a consequence of the fact that $\Ihat$ and $I$ differ by
the contribution of $Q$ to the interaction polynomial, with $Q$
obeying \eqref{e:Qbound}.  The third estimate in the lemma is a
reflection of the fact that $\rhoFcal$ provides a measure of the
difference between $V$ and $\Vpt$.

\begin{lemma}
\label{lem:JCK}
Let $(V,K) \in \DV_j \times  B_{\Fcal_j}(r^{(0)}\rhoFcal)$.
Let $Q$ be given by \eqref{e:Q-def}.
For $B \in \Bcal_j$,
$X \in \Scal_j$ and $F \in\Ncal(X^\Box)$,
\begin{align}
\label{e:JCK1}
    \|\Ihat(B)- I(B)\|_{j}
&    \prec
    \rhoFcal  ,
\\
\label{e:JCK2}
    \| \Ihat(B) - I(B)(1+ Q(B) ) \|_{T_{0} }
   & \prec_{L}
    \rhoFcal^{2} ,
\\
\label{e:JCK3}
    \big\|
    \LT_X \big( I^{-X}- \Itilde_{\pt}^{-X}\big) F
    \big\|_{T_{0} (\h)}
    &\prec_L \;
    \rhoFcal
    \|F\|_{T_0 (\h)}
    .
\end{align}
\end{lemma}

\begin{proof}
The bounds \refeq{JCK1}--\refeq{JCK2} are a consequence of
\cite[Proposition~\ref{IE-prop:JCK-app-1}]{BS-rg-IE},
with its hypothesis on $Q$ of provided by
\refeq{Qbound}.
For \refeq{JCK3}, we write
\begin{equation}
    I^{-X}- \Itilde_{\pt}^{-X}
    = \left(I^{-X}- \Ihat^{-X}\right)
    + \left(\Ihat^{-X} -\Itilde_{\pt}^{-X} \right),
\end{equation}
and apply the triangle inequality.
The second term is estimated in \cite[Lemma~\ref{IE-lem:JCK-app-2}]{BS-rg-IE}
(its hypothesis is satisfied since we have established in the proof of
Proposition~\ref{prop:Ianalytic1} that $V-Q\in \bar\DV$).
The first term can be estimated similarly, using \refeq{JCK1},
and we omit the details.
\end{proof}

The following proposition is used in the proof that $K_+$
vanishes at weighted infinity.

\begin{prop}
\label{prop:Ivan}
Let $(V,K) \in \DV_j \times  B_{\Fcal_j}(r^{(0)}\rhoFcal)$, let
$B\in\Bcal_j$,
and let $I_*,j_*$ denote any of the options in Proposition~\ref{prop:Ianalytic1}.
For $F \in \Ncal(B^\Box)$ a polynomial of degree at most $p_\Ncal$,
\begin{equation}
\lbeq{vai}
    \lim_{\|\phi\|_{\Phi_j(B^\Box,\h)} \to \infty}
    \|I_*(B)F\|_{T_{\phi,j_*}}\Gcal_{j_*}^{-1}(X,\phi) = 0,
\end{equation}
where $\Gcal$ represents
$G$ or $\tilde G$ according to whether $\h=\ell$ or $\h=h$.
\end{prop}

\begin{proof}
It is proved in \cite[\eqref{IE-e:vai}]{BS-rg-IE} that \refeq{vai} holds for
$I_*=I$ when $V\in \bar\DV$.
We have seen above that $\Vhat=V-Q \in \bar\DV'_j$ and that
$\Vpt=\Vpt(\Vhat) \in \bar\DV'_{j_*}$.  The primes represent an unimportant change
in constants defining the domain, so the result follows for the other options
for $I_*$.
\end{proof}

\subsection{Analyticity lemmas}
\label{sec:analyticity}

In establishing the analyticity of $K_+$, we apply
the useful and elementary facts about analytic functions
on complex Banach spaces presented in the next lemmas.
For a general introduction to analyticity in Banach spaces, see
\cite{Chae85,PT87}.

Let $X,Y$ be complex Banach spaces and let $D \subset X$ be an open
subset of $X$.  Let $L(X,Y)$ denote the space of bounded linear maps
from $X$ to $Y$.  A map $f:D\to Y$ is \emph{analytic} if it is
continuously Fr\'echet differentiable, i.e., if there exists a
continuous map $f'
: D \to L(X,Y)$ such that
\begin{equation}
    \big\|
    f (x+\dot{x}) - f (x) - f' (x) \dot{x}
    \big\|_{Y}
=
    o (\|\dot{x}\|_{Y} )
    \quad
    \text{as $\dot{x} \to 0$}.
\end{equation}
Let $\Acal$ be an index set.  For each $\alpha \in\Acal$, let
$Y_{\alpha}$ be a Banach space and let $w_\alpha \ge 1$ be a positive
weight.  Let $Y = \prod_{\alpha}Y_{\alpha}$ be the weighted product
Banach space: an element of $Y$ has the form $y = (y_{\alpha} \in
Y_{\alpha}\mid \alpha \in \Acal)$ with norm
\begin{equation}
    \|y\|_{Y}
=
    \sup_{\alpha \in \Acal} \|y_{\alpha}\|_{Y_{\alpha}}w_\alpha
.
\end{equation}
A collection of maps $f_\alpha : D \to Y_\alpha$, for $\alpha \in
\Acal$, naturally determines a map $f:D \to Y$.  In our applications,
the weights play a role in estimates but not in proving analyticity,
as the following lemma shows.

\begin{lemma}
\label{lem:prod-analyticity} Let $\Acal$ be a finite set and let $f =
(f_\alpha)$ as above.  Then $f : D \to Y$ is analytic if and only if
$f_\alpha : D \to Y_\alpha$ is analytic for each $\alpha \in \Acal$.
\end{lemma}

\begin{proof}
Suppose first that each $f_\alpha$ is analytic, and let $f_\alpha'$
denote its derivative.  In particular, $f_\alpha': D \to L(X,Y)$ is
continuous.  For $x\in D$, let $\hat f (x) = (f_{\alpha}' (x)\mid
\alpha \in \Acal)$. Since $\Acal$ is a finite set, $\hat f (x) \in L
(X,Y)$ and $x \mapsto \hat f (x)$ is a continuous map from $D$ to $L
(X,Y)$.  Also,
\begin{equation}
    \big\|
    f (x+\dot{x}) - f (x) - \hat f (x) \dot{x}
     \big\|_{Y}
=
    \sup_{\alpha\in \Acal}\big\|
    f_{\alpha} (x+\dot{x}) - f_{\alpha} (x) - f_{\alpha}' (x) \dot{x}
     \big\|_{Y_{\alpha}}
\le
    \sup_{\alpha\in \Acal}o\big(\|\dot{x}\|_{Y_\alpha}\big)
.
\end{equation}
Since $\Acal$ is finite, $\sup_{\alpha\in \Acal}o
(\|\dot{x}\|_{Y_{\alpha}} ) \le o (\|\dot{x}\|_{Y} )$, so $f:D \to Y$
is analytic and $\hat f$ is its derivative.

Conversely, suppose that $f$ is analytic.  Define $\pi_\alpha : Y \to
Y_\alpha$ by $\pi_{\alpha}y =y_{\alpha}$.  Since $w_{\alpha} \ge 1$
the map $\pi_\alpha$ is a bounded linear map from $Y$ to $Y_{\alpha}$,
and is therefore analytic.  Thus $f_{\alpha} = \pi_{\alpha}\circ f$ is
the composition of two analytic maps and hence is also analytic.
\end{proof}

Let $n$ be a positive integer. For $i=1,2,\dots ,n$, let $X_{i}$ be a
Banach space. Let $Y$ be a Banach space and let $M:X_{1}\times \dots
\times X_{n}\rightarrow Y$ be a multilinear map which is bounded. That
is, there is a constant $C_{M}$ such that  for any $n$-tuple $x=
(x_{1},\dots ,x_{n})$ in $X_{1}\times \dots \times X_{n}$,
\begin{equation}
    \|M (x)\|_{Y}
\le
    C_{M} \prod_{i=1}^{n} \|x_{i}\|_{X_{i}}
.
\end{equation}
For positive weights $w_{1},\dots ,w_{n}$ let $X = X_{1}\times \dots
\times X_{n}$ be the Banach space whose norm is
\begin{equation}
    \|x\|_{X}
=
    \sup_{i=1,\ldots,n} w_{i}\|x_{i}\|_{X_{i}}
.
\end{equation}

\begin{lemma}\label{lem:multilinear-analyticity}
The map from $X$ to $Y$ defined by $x \mapsto M (x)$ is analytic with
derivative $M' (x)\dot{x} = \sum_{i}M (x_{1},\dots
,x_{i-1},\dot{x_{i}},x_{i+1},\dots ,x_{n})$.
\end{lemma}

\begin{proof}
It is straightforward to verify that $\|M (x+\dot{x}) - M (x) - M'
(x)\dot{x}\|_{Y}$ is bounded by
\begin{equation}
    \left(
    C_{M}\prod_{i=1}^{n} w_{i}^{-1}
    \right)
    \sum_{p=2}^{n} \binom{n}{p}
    \|x\|_{X}^{n-p}\|\dot{x}\|_{X}^{p}
=
    o (\|\dot{x}\|_{X})
\end{equation}
as required.
\end{proof}

\section{Maps~1--2: Transfer of relevant parts of
\texorpdfstring{$K$}{K} to \texorpdfstring{$V$}{V}}
\label{sec:Map1-2}

We now begin the discussion of the six maps leading from $K$ to $K_{+}$,
with Maps~1--2.  In Maps~1--2, there is no change of scale, and all
objects are scale-$j$ objects.  To simplify the notation, we do not
indicate the scale explicitly.  The norms appearing in this section
are either all the $G$ norm, or all the $\tilde G$ norm, each together
with its corresponding $\Fcal$ space $\Fcal=\Fcal(G)$ or
$\Fcal=\Fcal(\tilde G)$.

Map~1 (and also Map~4) relies on an essential change of variables formula,
which provides a mechanism for rewriting a circle product $(I \circ
K)(\Lambda)$ as $(I \circ K')(\Lambda)$, with $K'$ ``better'' than
$K$.  The change of variables is given by
Proposition~\ref{prop:change-of-variable-1} below.
Map~1 uses the change of variables  to remove
the relevant and marginal parts of $K(X)$ when $X\in
\Scal$ is not a block, by transfering them into
$K(B)$ where $B$ is a block.  This creates $K^{(1)}$ and leaves $I$
unchanged.  Map~2 then removes the relevant and marginal parts of
$K(B)$ and transfers them to $V$, thereby creating $\Ihat$ and
$K^{(2)}$.

\subsection{Change of variables}
\label{sec:change-of-variable-section}

We work at a fixed scale and do not indicate the scale explicitly in
the notation.  This section applies for \emph{any} norm $\|\cdot\|$ on
$\Ncal$ which obeys the product property
\cite[\eqref{IE-e:norm-fac}]{BS-rg-IE}.

Let
\begin{equation}
\label{e:DcalJ}
    \Dcal (J)
    =
    \{(U,B) \in \Scal\times \Bcal: U \supset B\}
    .
\end{equation}
Suppose we have a mapping $J : \Ccal \times \Bcal \to \Ncal$ which
obeys, for $U \in \Ccal$ and $B \in \Bcal$, the condition $J (U,B)=0$
if $(U,B) \not \in \Dcal (J)$, as well as
\begin{gather}
    \label{e:kprimecondition}
    \sum_{U: (U,B) \in \Dcal (J)} J (U,B) = 0
    \quad\text{for fixed $B \in \Bcal$},
\\ \nonumber
    \text{$J(U,B)$ obeys the field locality (in its $B$ argument)
    }
    \\
    \text{and symmetry conditions of Definition~\ref{def:Kspace}}
\label{e:J-localisation}
.
\end{gather}
For example, field locality means that $J (U,B)\in \Ncal (B^{\Box})$
and Euclidean symmetry means that $E J(U,B) = J(EU,EB)$ for all
automorphisms $E$ of $\Lambda$.

Let $\IstabC \ge 1$, and suppose that
$\Iin \in \Bcal \Kspace$ is stable in the sense that
\begin{equation}
    \label{e:Iin-stability}
    \|\Iin (B)\| \le \IstabC
.
\end{equation}
For $\Kin \in \Kspace$ and $U \in \Ccal$, let
\begin{equation}
    \label{e:McalnowM-def}
    \McalnowM (U) = \Kin (U) - \Iin^{U} \sum_{B \in \Bcal (U)} J (U,B)
    .
\end{equation}
Given positive $\aout, \ain, \rhogen$ we define spaces
$\Fcal_{\mathrm{in}} = \Fcal (\ain,\rhogen)$ and
$\Fcal_{\mathrm{out}} = \Fcal (\aout,\rhogen)$,
as in \eqref{e:KFcal}.  The open balls of radius
$\rball$ in these spaces are denoted by $B_{\Fcal_{\rm in}} (\rball)$
and $B_{\Fcal_{\rm out}} (\rball)$.

\begin{prop}
\label{prop:change-of-variable-1} Let $\ain$ be small as specified in
Lemma~\ref{lem:Y0}.  Let $\aout<\ain$ and $z' > \frac 12 z$. Let
$\rhogen$ be sufficiently small depending on the difference
$\aout-\ain$.  Let $\epsilon \in (0,1)$.  Let $J,\Iin$ be as
specified by \eqref{e:DcalJ} -- \eqref{e:Iin-stability}.  Suppose
that $\Kin \in \Kspace$ and $J$ satisfy
\begin{align}
\label{e:Jbd}
    &
    \sup_{\Dcal (J)}
    \|\Iin^{U} J (U,B) \|
    \leq
    \epsilon \rhogen^{z'},
    \\
\label{e:Mcalbd}
    &
    \McalnowM \in B_{\Fcal_{\rm in}} (\epsilon \rhogen^{z})
.
\end{align}
Then there exists $\Kout \in \Kspace$ such that
\begin{align}
\label{e:Kout-IK}
    &
    (\Iin \circ \Kout) (\Lambda)
=
    (\Iin \circ \Kin) (\Lambda),
\\
\label{e:Kout-poly}
    &
    \text{$\Kout$ is polynomial in $\Iin,\bar J,\Kin$ (with $\bar J (U,B)=\Iin^UJ(U,B)$)},
\\
\label{e:Koutbd-KM}
    &
    \text{$\Kout = \McalnowM +E$
    with $E\in  B_{\Fcal_{\rm out}} (\epsilon \rhogen^{z+(\ain-\aout)/2})$}.
\end{align}
If $\Kin =0$ and
$J=0$, then $\Kout =0$.
\end{prop}

The essential conclusion of the
Proposition~\ref{prop:change-of-variable-1} is that $\Kout$
is both a suitable replacement for $\Kin$ by \eqref{e:Kout-IK}, and is
a small perturbation of $M$ by \eqref{e:Koutbd-KM}.  The latter
property will be useful in our applications in Maps~1 and 4, where $M$
will have desirable properties.  Note also that it follows immediately
from \eqref{e:Mcalbd} and \eqref{e:Koutbd-KM} that
\begin{align}
\label{e:Koutbd-K}
    &
    \Kout \in B_{\Fcal_{\rm out}} (2\epsilon \rhogen^{z}).
\end{align}
We defer the proof of Proposition~\ref{prop:change-of-variable-1} to
Appendix~\ref{sec:change-of-variable}.

\subsection{Map~1:  Transfer from small sets to block}

We set $I^{(1)}=I^{(0)}=I$ and $K^{(0)}=K$.  Map~1 determines
$K^{(1)}$.  The structure of the following lemma sets a pattern that
we follow throughout our analysis of the six maps.  Part~(i) is the
statement that the output pair $(I^{(1)}, K^{(1)})$ is an equivalent
representation of the input pair. Part~(ii) says that the range of the
map is contained in the domain of the next map. Part~(iii) finds the
image when $K=0$, needed in the proof of
\eqref{e:KTay0}. Part~(iv) identifies a property that the map
achieves, which did not hold for the input.

As mentioned previously, Map~1 transfers the relevant and marginal
parts of $K(X)$ for $X\in \Scal \setminus \Bcal$ into $K(B)$ where $B
\in \Bcal$.  Naively, to achieve this we would attempt to replace
$K(X)$ by $K(X) - \LT_X K(X)$.  However, to maintain stability of the
subtracted term, we replace it instead by $I^X\LT_{X}I^{-X}K(X)$,
which enjoys the decay properties of the factor $I^X$ and will still
provide the cancellation we seek.  (Recall that we write $I^{-X} =
1/I^X$.)  Thus, for $X \in \Scal \setminus B$ we wish to replace
$K(X)$ by $K(X) - I^X\LT_{X}I^{-X}K(X)$ via a corresponding adjustment
to $K(B)$.  This is what Proposition~\ref{prop:change-of-variable-1}
permits us to do.  To apply
Proposition~\ref{prop:change-of-variable-1}, we define $J(X,B)=0$ if
$(X,B) \not\in \Dcal (J)$ and
\begin{equation}
\label{e:Map1JXB}
    J(X,B) = \LT_{X,B}I^{-X}K(X) \quad \text{for
    $X \in \Scal$ with $X \supsetneqq B$},
\end{equation}
and to achieve the cancellation condition imposed by
\eqref{e:kprimecondition}, we are forced to define
\begin{equation}
\label{e:Map1JBB}
    J (B,B)
    =
    -
    \sum_{Y \in \Scal : Y\supsetneqq B} \LT_{Y,B}I^{-Y}K(Y).
\end{equation}
With these definitions, and with $\Iin = I$, it then follows from
\eqref{e:McalnowM-def} and $\sum_{B\subset X}\LT_{X,B} = \LT_{X}$ (the
latter due to \cite[\eqref{loc-e:LTsymXXi}]{BS-rg-loc}) that
\begin{equation}
\label{e:Mstep1def}
    M(X)
    =
    \begin{cases}
    K (X) & X \in \Ccal\setminus\Scal \\
    K (X) - I^X  \LT_{X} I^{-X}K (X)
    &  X \in \Scal \setminus \Bcal
    \\
    K (B) + I^B
    \sum_{Y \in \Scal, Y\supsetneqq B} \LT_{Y,B} I^{-Y}K  (Y)
    &X = B \in \Bcal.
    \end{cases}
\end{equation}

The important achievement of \eqref{e:Mstep1def} is the cancellation
of the local part of $K(X)$ when $X\in \Scal \setminus B$.  The cost
for this cancellation is that these local parts have been transferred
to $K(B)$.
Recall that $\LT_{X} = \LT_{X,X}$ by
\cite[\eqref{loc-e:LTXY6}]{BS-rg-loc}.
It then follows from \eqref{e:Mstep1def}
and the definition of $ Q(B)$ in \eqref{e:Q-def},
that
\begin{align}
\label{e:MQB}
    \LT_{B} I^{-B}\McalnowM (B)
    &=
    \sum_{Y \supset B} \LT_{Y,B} I^{-Y} K  (Y) =  Q (B)
    ,
\end{align}
and this will be cancelled in Map~2 by the $Q$ term in the definition
of $\Vhat$ in \eqref{e:Vhat}.
We now expand on the above by presenting the details for Map~1.

\begin{lemma}\label{lem:K1}
For $u>0$ sufficiently small and
$\rhoFcal$ sufficiently small,
there
exist $\mu_1 \ge 1$ and $K^{(1)}$ such that,
for $(V,K)$ obeying \eqref{e:VKhyp},
\begin{align}
    (i)&\quad
    (I \circ K )(\Lambda)
    =
    (I \circ K^{(1)})(\Lambda),
\nonumber
\\
    (ii)&\quad
    K^{(1)} \in B_{\Fcal^{(1)}} (r^{(1)} \rhoFcal)
    ,
\nonumber
\\
    (iii)&\quad
    \text{$K^{(1)} = 0$ when $K =0$},
\nonumber
\\
    (iv)&\quad
    \|\LT_{X}I^{-X}K^{(1)} (X)\|_{T_0}
    \le
    r^{(1)} \rhoFcal^{1+u}
    \quad\quad
    X \in \Scal \setminus \Bcal
.
\nonumber
\end{align}
\end{lemma}

\begin{proof}
(i)
By the hypothesis of Theorem~\ref{thm:1} that $K \in B_{\Fcal^{(0)}}
(r^{(0)} \rhoFcal)$, and by \eqref{e:JCK0}, we have
\begin{equation}
    \|I^{X} J (X,B) \|
    \prec
    r^{(0)} \rhoFcal
    .
\end{equation}
We choose $\mu_1$, depending on the constant in the above
inequality, so that with $r^{(1)}=\mu_1 r^{(0)}$ we can
conclude  that
\begin{equation}
\label{e:Jbd1}
    \|I^{X} J (X,B) \|
    \le
    \frac{1}{4} r^{(1)} \rhoFcal^{z'},
\end{equation}
and also, using \eqref{e:Mstep1def}, that
\begin{equation}
\label{e:Mstep1bd}
    \McalnowM \in B_{\Fcal^{(0)}} (\textstyle{\frac{1}{2}}r^{(1)} \rhoFcal)
    .
\end{equation}
We apply Proposition~\ref{prop:change-of-variable-1} with $J$
given by \eqref{e:Map1JXB}--\eqref{e:Map1JBB} and with
\begin{gather}
    \Iin = I, \quad
    \Kin = K, \quad
    \rhogen = \rhoFcal, \quad
    \ain = a^{(0)}, \quad
    \aout = a^{(1)},\quad
    z=z'=1  ,\quad
    \epsilon = \frac 12 r^{(1)}
    ,
\end{gather}
and we define the map $K \mapsto K^{(1)}$ by setting $K^{(1)} =
\Kout$.  Then (i) is an immediate consequence of
Proposition~\ref{prop:change-of-variable-1} as soon as we verify that
the hypotheses hold. The hypothesis \eqref{e:kprimecondition} holds by
construction.  The fact that \eqref{e:J-localisation} holds can be
seen from the fact that $\LT$ preserves the relevant symmetries and
cannot generate a non-zero constant part (see Lemma~\ref{lem:LTcp} and
the proof of Lemma~\ref{lem:Qapp}).
Hypothesis~\eqref{e:Iin-stability} follows from
Proposition~\ref{prop:Ianalytic1}, \eqref{e:Jbd} follows from
\eqref{e:Jbd1}, and \eqref{e:Mcalbd} follows from \eqref{e:Mstep1bd}.
\\
(ii)
The estimate follows from \eqref{e:Mstep1bd}.  For the vanishing at
weighted infinity
property inherent in the definition of $\Fcal_+$ (see \refeq{defnFcal}),
the vanishing as $\|\phi\|_\Phi
\to \infty$ is a consequence of Proposition~\ref{prop:Ivan}, the definition of $J$
in \refeq{Map1JXB}--\refeq{Map1JBB}, and the fact that $K^{(1)}$ is a polynomial
in $I,\bar J,K$ by Proposition~\ref{prop:change-of-variable-1}.
The field locality, symmetry and
component factorisation properties follow
from the fact that $K^{(1)}=\Kout \in \Kcal$ by
Proposition~\ref{prop:change-of-variable-1}.
\\
(iii) This is an immediate consequence of
Proposition~\ref{prop:change-of-variable-1}.
\\
(iv)
Let $X \in \Scal\setminus \Bcal$.  By \eqref{e:Mstep1def}
and the fact that $\LT_X \circ \LT_X = \LT_X$ by
\cite[\eqref{loc-e:oLT4}]{BS-rg-loc},
\begin{equation}
\label{e:MnoLoc}
    \LT_X I^{-X}\McalnowM(X)= 0,
    \quad \quad X \in \Scal \setminus \Bcal.
\end{equation}
Therefore (iv) is equivalent to
\begin{equation}
    \label{e:Kout-bound2M}
    \| \LT_{X}I^{-X}(K^{(1)} (X)-\McalnowM(X))\|_{T_0}
    \le
    r^{(1)} \rhoFcal^{1+u},
    \quad \quad
    X \in \Scal \setminus \Bcal
    .
\end{equation}
By Proposition~\ref{prop:change-of-variable-1},
\begin{equation}
\label{e:Koutbd1a}
    K^{(1)} - \McalnowM \in B_{\Fcal^{(1)}} (\textstyle{\frac{1}{2}}r^{(1)}\rhoFcal^{1+u})
    \quad \text{with} \quad u =\textstyle{\frac{1}{2}} ( a^{(0)} - a^{(1)}) >0,
\end{equation}
and then \eqref{e:Kout-bound2M} follows from \eqref{e:JCK0}, where we
remove the constant in $\prec$ by decreasing $u$ and
taking $\rhoFcal$ small.
\end{proof}

We also verify that $K^{(1)}$ obeys the remaining properties of
Definition~\ref{def:Kplusprops}, namely: $(V,K)$-analyticity, the
restriction property, the isometry property, and mass continuity.  The
isometry property holds because $\iota$ is an algebra homomorphism
and, for each polymer $X$, $K^{(1)} (X)$ is polynomial in $K$.  The
mass continuity is vacuous here, since $K^{(1)}$ does not depend on
the mass.

The restriction property holds in the sense that $K^{(1)}(X)$ is a function of $K(Y)$
for $Y \in X^\Box$.  This follows from the explicit formula for $\Kout$ given
in \refeq{Kdef-new}.  For example, for the case where $X$ is a single block $B$,
\refeq{Kdef-new-ex} gives
\begin{align}
    K^{(1)}(B)
    & = K(B) -J(B,B) + \sum_{U : (U,B) \in \Dcal(J)} I^U J(U,B).
\end{align}
It is due to the dependence of $J$ on $K(U)$ for $U \in B^\Box$
that $K^{(1)}(B)$ develops
its dependence on $K$ in the small set neighbourhood of $B$.

Finally, for the $(V,K)$-analyticity, by Lemma~\ref{lem:prod-analyticity}
it suffices to show that for each polymer $X \in \Pcal$, the map
$(V,K) \mapsto K^{(1)}(X)$ is an analytic function from $\DV \times
B_{\Fcal}(r^{(0)}\epdV)$ to $\Ncal(X^\Box)$, $\|\cdot\|_j$
(here $\DV$ is the domain for $V$ defined in \eqref{e:DV1}, not to be
confused with the domain $\Dcal (J)$ for $J$ in
Proposition~\ref{prop:change-of-variable-1}).
We know from Proposition~\ref{prop:change-of-variable-1}
that $K^{(1)}(X)$ is a polynomial in $I,\bar J,K$.
By Lemma~\ref{lem:multilinear-analyticity}, it suffices to show that
each of the maps $(V,K) \mapsto I(B)$ and $(V,K) \mapsto \bar J(U,B)$
is an analytic map from
$\DV$ to $\Ncal(B^\Box)$, $\|\cdot\|_j$.  For $I(B)$, this follows from
Proposition~\ref{prop:Ianalytic1}.  It therefore suffices to show that
the map $(V,K) \mapsto \bar J$ is analytic, with $\bar J$ defined
by \eqref{e:Kout-poly} and \eqref{e:Map1JXB}--\eqref{e:Map1JBB}.  This map is linear in
$K$, $\LT$ is a bounded map on $T_0$ by \refeq{LTXY5-first},
and $I^{-X}$ is an analytic map into $\Ncal(B^\Box)$, $\|\cdot\|_{T_0}$
by Proposition~\ref{prop:Ianalytic1}.  Therefore $J(U,B)$ is an analytic function
of $(V,K)$ taking values in $\Vcal(B)$, $\|\cdot\|_{T_0}$.
The bilinear map $(I,J) \mapsto \bar J$ is bounded with domain norms
$\|\cdot\|_j, \|\cdot\|_{T_0}$ and range norm $\|\cdot\|_j$, by
Proposition~\ref{prop:Ianalytic1}, and the desired analyticity of $\bar J$ then
follows from Lemma~\ref{lem:multilinear-analyticity}.

\subsection{Map~2: Transfer from block to \texorpdfstring{$V$}{V}}
\label{sec:Map2}

Map~2 transfers relevant parts from $K^{(1)} (B)$ into $V$.  It provides
the rationale for the formula \eqref{e:Vhat} for $\Vhat$.  We define
\begin{gather}
\label{e:single-block0}
    I^{(2)} = \hat{I} = I_{j} (\Vhat),
\quad\quad
    \delta I^{(2)} = I - \hat{I},
\quad\quad
    K^{(2)} = K^{(1)} \circ \delta I^{(2)}
    .
\end{gather}

\begin{lemma}
\label{lem:K2} There exists $\mu_2 \ge 1$ such that for $u>0$ and
$\rhoFcal$ both sufficiently small,
for $(V,K)$ obeying
\eqref{e:VKhyp},
\begin{align}
(i)&\quad
    (I \circ K^{(1)})(\Lambda)
    =
    (\Ihat \circ K^{(2)})(\Lambda),
\nonumber
\\
(ii)&\quad
    K^{(2)} \in B_{\Fcal^{(2)}} (r^{(2)} \rhoFcal)
    ,
\nonumber
\\
(iii)&\quad
    \text{$K^{(2)} = 0$ when $K=0$}
    ,
\nonumber
\\
(iv)&\quad
    \| \LT_{X}I^{-X}K^{(2)} (X)\|_{T_0}
    \le
    r^{(2)}
    \rhoFcal^{1+u}
    \quad \quad
    X \in \Scal
    .
\nonumber
\end{align}
\end{lemma}

\begin{proof}
(i) By Lemma~\ref{lem:bin},
\begin{align}
    I \circ K^{(1)}
    &=
    (\hat{I} + \delta I^{(2)}) \circ K^{(1)}
    =
    (\hat{I} \circ \delta I^{(2)}) \circ K^{(1)}
=
    \hat{I} \circ (\delta I^{(2)} \circ K^{(1)})
    =
    I^{(2)} \circ K^{(2)}
    .
\end{align}
(ii) By \eqref{e:JCK1} and the $K$ hypothesis of Theorem~\ref{thm:1}
we have $\|\delta I^{(2)} (B) \|_{j} \prec
\rhoFcal$.  By
choosing $r^{(2)}=\mu_2 r^{(1)}$ with $\mu_2$ sufficiently
large, this implies
\begin{equation}\label{e:deltaIhat-bound}
    \delta I^{(2)} \in B_{\Fcal^{(1)}} (2^{-2^{d}}r^{(2)} \rhoFcal).
\end{equation}
By Lemma~\ref{lem:K1}, $K^{(1)} \in B_{\Fcal^{(1)}} (2^{-2^{d}}r^{(2)}
\rhoFcal)$.  The desired estimate then follows from Lemma~\ref{lem:FcircG}.
The fact that $K^{(2)}$ vanishes at weighted infinity is a consequence of the fact
that $K^{(1)}$ has this property, and that both $I$ and $\Ihat$ vanish
at weighted infinity by Proposition~\ref{prop:Ivan}.
The field locality, symmetry and
component factorisation properties can be verified by inspection.
\\
(iii) For $K=0$,
it follows from \eqref{e:Q-def}--\eqref{e:Vhat} that $\Vhat = V$.  Then
\eqref{e:single-block0} gives $I^{(2)}=I$
and hence $K^{(2)}=0$.
\\
(iv) We first prove (iv) for the case $X \in \Scal \setminus
\Bcal$. By Lemma~\ref{lem:K1}(iv) (and increasing $\mu_2$, it is
sufficient to prove the result when $K^{(2)}(X)$ is replaced by
$K^{(2)}(X)-K^{(1)}(X)$.  By the triangle inequality,
\begin{equation}
     \|K^{(2)}(X)-K^{(1)}(X)\|
     \le
     \sum_{Y\in\Pcal :Y \subsetneqq X} \|K^{(1)}(Y)\|
     \|\delta I^{(2)}\|^{X\setminus Y}
.
\end{equation}
There are at most $2^{2^{d}}$ terms in this sum. By
\eqref{e:deltaIhat-bound} and
Lemma~\ref{lem:K1}(ii), together with the
exclusion of the term $Y=X$ on the right-hand side, we obtain
\begin{equation}
    \|K^{(2)}(X)-K^{(1)}(X)\| \prec (r^{(2)})^{2}\rhoFcal^{2}
.
\end{equation}
Then we obtain the desired bound by using \eqref{e:JCK0}, $r^{(2)}<1$,
$u<1$, and choosing $\rhoFcal  $ small.

Finally, we prove (iv) for the case $X = B \in \Bcal$.  By definition,
$K^{(2)} (B) = K^{(1)} (B) + \delta I^{(2)} (B)$.  Therefore, by
\eqref{e:MQB},
\begin{align}
    \LT_{B}I^{-B}K^{(2)} (B)
    &=
    \LT_{B}I^{-B}K^{(1)} (B) +
    \LT_{B}I^{-B} \delta I^{(2)} (B)
    \nnb
    &=
    \LT_{B}I^{-B}\left( K^{(1)} (B)- \McalnowM (B)\right)
     + \LT_{B}\left[ I^{-B} \delta I^{(2)} (B) + Q(B) \right]
     ,
\end{align}
where we used the fact that $\LT_BQ(B)=Q(B)$, which is
a consequence of the fact that $\LT_X \circ \LT_X=\LT_X$ by
\cite[\eqref{loc-e:oLT4}]{BS-rg-loc}.
By \eqref{e:I-b}, \eqref{e:LTXY5-first}, \eqref{e:JCK0},
\eqref{e:JCK2}, and \eqref{e:Koutbd1a}, it follows that
\begin{equation}
    \| \LT_{B}I^{-B}K^{(2)} (B)\|_{T_0}
    \le
    r^{(2)}\rhoFcal^{1+u}
    ,
\end{equation}
as required.
\end{proof}

We also verify that $K^{(2)}$ obeys the $(V,K)$-analyticity property,
the restriction property, the isometry property, and mass continuity.
The restriction property is evident from the definition of $K^{(2)}$,
and the mass continuity is again vacuous.

From the above construction and the explicit formula
\eqref{e:Kdef-new} for the change of variables in
Appendix~\ref{sec:change-of-variable} it follows that, for each
polymer $X$, $K^{(2)} (X)$ is a polynomial in $\hat{I}$ and
$K^{(1)}$. Since $\iota$ is a homomorphism the isometry property holds
provided for each block $B$ the function $\hat{I} (K^{(1)},B)$ of
$K^{(1)}|_{B^{\Box}}$ satisfies $\iota \hat{I} (K^{(1)},B) = \hat{I}
(\iota K^{(1)},\iota B)$. We omit this mechanical step.

For the analyticity, we observe that $K^{(2)}$ is multilinear in
$I,\Ihat,K^{(1)}$.  We have already verified the analyticity of
$K^{(1)}$ in the previous section, and the analyticity of $I$ and
$\Ihat$ is given by Proposition~\ref{prop:Ianalytic1}.  The desired
analyticity of $K^{(2)}$ then follows from
Lemmas~\ref{lem:prod-analyticity}--\ref{lem:multilinear-analyticity}.

\section{Maps~3--4: Expectation and change of scale}

Map~3 expresses the action of the expectation $\Ex_{+}$ in terms of
$K\circ I$.  A reblocking takes place in the process, yielding
$K^{(3)}\in \Kspace_{j+1}$.  Thus we measure the size of $K^{(3)}$ in
scale $j+1$ norms---the change of norm is an important ingredient in
establishing contractivity.

In Section~\ref{sec:polymers}, we define $I^{(3)}$ and $K^{(3)}$ and
summarise the principal facts about Map~3.  The proof of estimates on
$K^{(3)}$ is deferred to Section~\ref{sec:Map3estimates}, which relies
heavily on results from \cite{BS-rg-IE} that were designed expressly
for this analysis.

There are two types of potentially dangerous contributions to
$K^{(3)}$.  One type consists of the leading contributions in
$K^{(3)}$ which form a part of the \emph{perturbative} contributions
that arise in $K^{(3)}$ even when $K^{(2)}=0$.  These
perturbative contributions are larger than what is permitted in $K_{+}$
when $K=0$.
The second type consists of the contribution to $K^{(3)}$ which is
linear in $K^{(2)}$.  The latter contribution will be shown to be
contractive due to our having removed the relevant and marginal parts
of $K^{(2)}$ in Map~2, as expressed by Lemma~\ref{lem:K2}(iv); large
$L$ plays an important role in this step.  The leading perturbative
contributions will be redistributed by Map~4, via a second application
of the change of variables implemented by
Proposition~\ref{prop:change-of-variable-1}, leading to $K^{(4)}$
which obeys the better estimate of Lemma~\ref{lem:K5a}(iii), which is
the principal achievement of this section.

\subsection{Map~3:  Expansion, expectation, change of scale}
\label{sec:polymers}

The next proposition gives a formula for $K^{(3)}$.  This is the only
place where the factorisation property \eqref{e:Efaczz} of $\Ex_{+}$
is used: it ensures that $K^{(3)}$ obeys the component factorisation
property demanded by the space $\Kcal_{j+1}$ in
Definition~\ref{def:Kspace}.

Recall that $\Itilde$ was defined in \eqref{e:Itildef}.  We set
\begin{equation}
    I^{(3)} = \Itilde_{j+1}(\Vpt) = \Ipttil,
\end{equation}
and define $\delta I \in \BKspace_j$ (the space $\BKspace_j$ is
augmented here by the fluctuation fields introduced by $\theta$) by
\begin{equation}
    \label{e:deltaI-def}
    \delta I = \theta I^{(2)} - I^{(3)} = \theta \hat{I} - \Ipttil
    .
\end{equation}
Thus, for $U\in\Pcal_{j}$,
\begin{equation}
    \delta I^{U}
=
    \prod_{b\in\Bcal_{j} (U)}\big(
    \theta \hat{I} (b)-\Ipttil (b)
    \big).
\end{equation}

\begin{prop}
\label{prop:EIK}
Given $K^{(2)} \in \Kspace_{j}$,
\begin{equation}
\label{e:EIK1}
    \Ex_{+}\theta (\Ihat \circ K^{(2)})(\Lambda)
    = (\Ipttil \circ K^{(3)})(\Lambda),
\end{equation}
where, for $U \in \Pcal_{j+1}$,
\begin{equation}
\label{e:EIK2}
    K^{(3)} (U)
=
    \sum_{X \in\overline{\Pcal}_{j}(U)}
    \Ipttil^{U\setminus X}
    \Ex_{+}
    (\delta I
    \circ
    \theta K^{(2)}
    )(X).
\end{equation}
Also, $K^{(3)} \in \Fcal_{j+1}$.
\end{prop}

\begin{proof}
We use $\theta\Ihat (B) = \Ipttil(B)+\delta I (B)$
and Lemma~\ref{lem:bin} to see that
\begin{align}
    \Ex_{+}\theta
    (
    \Ihat \circ K^{(2)} )(\Lambda)
    &   =
        \Ex_{+}(
    \Ipttil   \circ \delta I \circ \theta K^{(2)})(\Lambda)
    .
\end{align}
There is no $\theta$ operating on $\Ipttil$, and this factor contains
no fluctuation fields upon which $\Ex_{+}$ can act.  Therefore, with
sums over disjoint $X_K, X_{\delta I}$, we have
\begin{align}
    \Ex_{+}\theta
    (
     \Ihat \circ K^{(2)})(\Lambda)
    & = \sum_{X_K,X_{\delta I} \in \Pcal_j(\Lambda)}
    \Ex_{+}\big(\delta I^{X_{\delta I}} \theta K^{(2)}(X_K) \big)
    \Ipttil ^{\Lambda \setminus (X_K \cup X_{\delta I})}
    \nnb
    & = \sum_{X_K,X_{\delta I} \in \Pcal_j(\Lambda)}
    \Ex_{+}\big(\delta I^{X_{\delta I}} \theta K^{(2)}(X_K)  \big)
    \Ipttil ^{\overline{X_K \cup X_{\delta I}} \setminus
    (X_K \cup X_{\delta I})}
    \Ipttil ^{\Lambda \setminus \overline{X_K \cup X_{\delta I}}}
    .
\end{align}
We write $X_{I} =\overline{X_K \cup X_{\delta I}} \setminus (X_K \cup
X_{\delta I})$ and $U = X_K \cup X_{\delta I}\cup X_{I}= \overline{X_K
\cup X_{\delta I}}$, to obtain
\begin{equation}
\label{e:Itil0}
    \Ex_{+}\theta (\Ihat \circ K^{(2)})(\Lambda)
    = \sum_{X_K,X_{\delta I},X_{I} \in \Pcal_j(\Lambda)}
    \Ex_{+}\big(\delta I^{X_{\delta I}} \theta K^{(2)}(X_K) \big)
    \Ipttil ^{X_{I}}
    \Ipttil ^{\Lambda \setminus U},
\end{equation}
where the sum is over disjoint $X_K,X_{\delta I},X_{I}$ with
$\overline{X_K \cup X_{\delta I}}= X_K \cup X_{\delta I}\cup X_{I}=U$
(see Figure~\reffg{reblock}).  With $K^{(3)}$ defined as in
\eqref{e:EIK2} this becomes the conclusion \eqref{e:EIK1}.

To see that $K^{(3)}$ is in $\Kspace_{j+1}$, the field locality is
straightforward.  Component factorisation is an immediate consequence
of the factorisation properties for $K^{(2)}$ and the finite-range
property \eqref{e:Efaczz} of $\Ex_{+}$, and the symmetry properties
required by Definition~\ref{def:Kspace} follow for $K^{(3)}$ from
Lemma~\ref{lem:Ex-sym}.

Finally, we show that $K^{(3)}$ vanishes at weighted infinity, as
required by the definition of the space $\Fcal^{(3)}$.  For this, we
rewrite \refeq{EIK2} as
\begin{equation}
\label{e:K3van}
    K^{(3)} (U)
=
    \sum_{X \in\overline{\Pcal}_{j}(U)}
    \Ipttil^{U\setminus X}
    \Ex_{+}\theta
    (\Ihat^X K^{(2)} (X))
    -
    \sum_{X \in\overline{\Pcal}_{j}(U)}
    \Ipttil^{U}
    \Ex_{+}\theta K^{(2)}(X).
\end{equation}
By Proposition~\ref{prop:Ivan}, each of
$\Ipttil^{U},\Ipttil^{U\setminus X}, \Ihat^X$ vanishes as
$\|\phi\|_\Phi \to \infty$.  So does $K^{(2)}(X)$, by
Lemma~\ref{lem:K2}.  By Proposition~\ref{prop:vanish-at-infinity}, the
property of vanishing at weighted infinity is preserved by the
operator $\Ex_{+}\theta$, and the proof is complete.
\end{proof}

\begin{figure}
\begin{center}
\includegraphics[scale = 0.2]{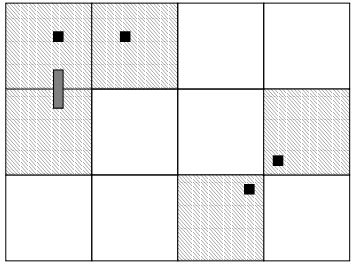}
\end{center}
\caption{\lbfg{reblock}
The black squares represent
$X_{\delta I}$, the small shaded polymer represents $X_K$ and the
five large shaded squares comprise $X_{I}$ in \eqref{e:Itil0}.}
\end{figure}

For connected sets $U\in \Ccal_{j+1}$, we define
\begin{align}
    h (U)  &=
    \sum_{X \in\overline{\Pcal}_{j}(U)}
    \Ipttil^{-X}\Ex_{j+1} \delta I^X,
\label{e:hdef}
\\
\label{e:kdef}
    k (U) &=
    \sum_{X \in\overline{\Ccal}_{j}(U)}
    \Ipttil^{-X} \Ex_{j+1} \theta   K^{(2)}(X)
    ,
\end{align}
where $\overline{\Pcal}_{j}(U)$ is
defined in Definition~\ref{def:blocks1} and
\begin{equation}
\lbeq{Cbardef}
    \overline \Ccal_{j}(U)=\{X \in\Ccal_j \mid\overline X = U\}.
\end{equation}
Then we rewrite the formula
\eqref{e:EIK2} for $K^{(3)}$ as
\begin{equation}
    \label{e:decomposeK3}
    K^{(3)} = \Ipttil h + \Ipttil k + R.
\end{equation}
Here $\Ipttil h$ is the \emph{perturbative} contribution resulting
from the terms in \eqref{e:EIK2} in which no explicit $K^{(2)}$
appears, $\Ipttil k$ is the contribution to \eqref{e:EIK2} that
contains exactly one factor $K^{(2)}$ supported on a connected set,
and $R$ consists of the remaining terms in $K^{(3)}$ which are not
included in $\Ipttil h+ \Ipttil k$.
In Section~\ref{sec:Map3estimates} below
(see \eqref{e:Ih0} and Lemmas~\ref{lem:contraction}--\ref{lem:Ktil}),
we prove that there exist $\kappa^*$ and $u>0$ such that
for all $U \in \Pcal_{j+1}$,
\begin{align}
\label{e:Map3h}
    \|\Ipttil(U) h(U)\|_{j+1}
    &
    \prec_L \rhoFcal^{2+f_{j+1} (a^{(3)},U)},
    \\
\label{e:Map3k}
    \|\Ipttil(U) k(U)\|_{j+1}
    &
    \le \kappa^* r^{(2)} \rhoFcal^{1+f_{j+1} (a^{(3)},U)},
    \\
\label{e:Map3R}
    \|R(U)\|_{j+1}
    &
    \prec_L \rhoFcal^{1+u+f_{j+1} (a^{(3)},U)}.
\end{align}
Crucially, $\kappa^*$ can be made as small as desired by taking
$L$ sufficiently large.  In fact, as we shall see, $\kappa^*$
is an $L$-independent multiple of $L^{-1}$.

The following lemma provides a summary of Map~3.
Parts~(i-ii) identify what this map
achieves---it performs the expectation with the resulting $K^{(3)}$
properly bounded at scale $j+1$.
There are two important improvements in the lemma:
the radius of the ball for $K^{(3)}$ has decreased by the factor
$\kappa_3$ compared to the ball for $K^{(2)}$, and the value
of $a^{(3)}$ has increased as in \eqref{e:achoice}.

\begin{lemma}
\label{lem:K3}
Let $\kappa_3=2\kappa^*$.  For $L$ sufficiently large
and
$\rhoFcal$ sufficiently small
depending on $L$, there
exists $K^{(3)}$ such that, for $(V,K)$ obeying \eqref{e:VKhyp},
\begin{align}
(i)&\quad
    \Ex_{+}(\Ihat \circ K^{(2)})(\Lambda)
    =
    (\Ipttil \circ K^{(3)})(\Lambda),
\nonumber
\\
(ii)&\quad
    K^{(3)} \in B_{\Fcal^{(3)}} (r^{(3)} \rhoFcal)
    ,
\nonumber
\\
(iii)&\quad
    \text{$K^{(3)} = \Ipttil h$
    (with $\Vpt(\hat V)=\Vpt(V)$) if $K=0$}
    .
\nonumber
\end{align}
\end{lemma}

\begin{proof}
(i) This follows immediately from Proposition~\ref{prop:EIK}.
\\
(ii) A decomposition of $K^{(3)}$ into three terms is given in
\eqref{e:decomposeK3}.  According to \eqref{e:Map3h}--\eqref{e:Map3R},
once we set $\kappa_3=2\kappa^*$ and
choose $\rhoFcal$ to be sufficiently small
depending on $L$, we find that $K^{(3)}$ obeys the estimate implied
by the statement that $K^{(3)} \in
B_{\Fcal^{(3)}} (\kappa_3 r^{(2)} \rhoFcal)$.  The fact that
$K^{(3)}$ has the properties required of membership in the space $\Fcal^{(3)}$
was established already in
Proposition~\ref{prop:EIK}.
\\
(iii) This follows from \eqref{e:decomposeK3} and the fact that
$K^{(2)}=0$ if $K=0$ by
Lemma~\ref{lem:K2}(iii), together with the fact that $\Vhat =V$ when
$K=0$ by \eqref{e:Vhat}.
\end{proof}

The restriction and isometry properties of
Definition~\ref{def:Kplusprops} can be verified from the definition of
$K^{(3)}$ as in maps 1,2.  For the mass continuity and
$(V,K)$-analyticity, we argue as follows.

\begin{proof}[Mass continuity of Map~3]
The mass continuity is needed in the setting of Theorem~\ref{thm:Kmcont}.
At this stage of the proof,
we consider continuity of $m^2 \mapsto K^{(3)}(U)$
for each polymer $U \in \Pcal_{j+1}$, as a map into
$\Ncal(U^\Box)$ with either norm $\|\cdot\|_j$ or $\|\cdot\|_{j+1}$.
(Further discussion occurs in Section~\ref{sec:pfthm1}.)
The dependence of $K^{(3)}(U)$ on the mass $m$
in \refeq{K3van} arises from the mass dependence
of the covariance $C_{j+1}$ in the decomposition of $(-\Delta +m^2)^{-1}$,
and occurs in in two ways.

One occurrence is via the dependence of $\Vpt$,
and hence of $I^{(3)}=\Ipttil$, on $C_{j+1}$ and thus on the mass.
The continuity of the coefficients of $\Vpt$ in small non-negative $m^2$
is established in
\cite[Proposition~\ref{pt-prop:rg-pt-flow}]{BBS-rg-flow}.
With the analyticity of $I$ in $\Vpt$ given by Proposition~\ref{prop:Ianalytic1},
the continuity of $\Ipttil^X$
(as an element of $\Ncal(X^\Box)$, $\|\cdot\|_{j+1}$)
in $m$ follows.

The second occurrence is in the covariance $C_{j+1}$ of the expectation in the
definition of $K^{(3)}$ in \refeq{EIK2}.  According to Proposition~\ref{prop:masscont},
given $F(X) \in \Ncal(X^\Box)$,
the linear map $m^2 \mapsto \Ex_{j+1}\theta F(X)$ from
the interval $\Iint_{j+1}$ of \refeq{massint} to $\Ncal(X^\Box)$, $\|\cdot\|_{j+1}$
is a continuous function.
Therefore $K^{(3)}(U)$ is a continuous function of $m^2 \in \Iint_{j+1}$.
\end{proof}

\begin{proof}[Analyticity of Map~3] By
Proposition~\ref{prop:Ianalytic1}, for $b \in \Bcal_j$, the map $(V,K)
\mapsto \Ihat(b)$ is an analytic map into $\Ncal(b^\Box),
\|\cdot\|_{j}$, and the map $(V,K) \mapsto \Ipttil(b)$ is an analytic
map into $\Ncal(b^\Box), \|\cdot\|_{j+1}$.  In Section~\ref{sec:Map2},
it is shown that the map $(V,K) \mapsto K^{(2)}(X)$ is also analytic
into $\Ncal(X^\Box), \|\cdot\|_{j}$.  By Proposition~\ref{prop:ip}
below, the map $\Ex_{+}\theta$ is a bounded linear map from
$\Ncal(X^\Box), \|\cdot\|_{j}$ to $\Ncal(X^\Box), \|\cdot\|_{j+1}$.
Therefore, the formula \refeq{K3van} expresses $K^{(3)} (U)$ as a
finite sum of bounded multilinear maps evaluated on factors which are
themselves analytic in $(V,K)$. By
Lemmas~\ref{lem:prod-analyticity}--\ref{lem:multilinear-analyticity},
$(V, K) \mapsto K^{(3)}$ is an analytic map into $\Fcal^{(3)}$.
\end{proof}

\subsection{Map~3:  Estimates}
\label{sec:Map3estimates}

Throughout this section, we work exclusively with the norm pairs
\eqref{e:np1}--\eqref{e:np2}.  Norms with subscripts are used to
denote these pairs of norms: when the scale-$j$ norm is the $G_j$ norm
then the scale-$(j+1)$ norm is the $T_{0,j+1}$ norm, and when the
scale-$j$ norm is the $\tilde G_j$ norm then the scale-$(j+1)$ norm is
the $\tilde G_{j+1}$ norm.  All the estimates given in terms of norms
$\|\cdot\|_j$ and $\|\cdot\|_{j+1}$ apply for each of these two
choices of norm pairs.  Our goal is to prove
\eqref{e:Map3h}--\eqref{e:Map3R}.  We begin with the bound on $\Ipttil
h$.  For this, we recall the following integration property, which is
\cite[Proposition~\ref{IE-prop:ip}]{BS-rg-IE}.

\begin{prop}
\label{prop:ip} Let $\epdV$ be sufficiently small (depending on $L$).
There is an $\Econst >0$ (independent of $L$) such that for disjoint
$X,Y \in \Pcal_j$ and for $F(Y) \in \Ncal(Y^\Box)$,
\begin{equation}
\label{e:integration-property}
    \|\Ex_{j+1} \delta I^X \theta F(Y)  \|_{j+1}
\le
    \Econst^{|X|_j+|Y|_j} \epdV^{|X|_j}\|F(Y)\|_j,
\end{equation}
where the pair of norms is given by either choice of \eqref{e:np1} or
\eqref{e:np2}, and where $|X|_j$ denotes the number of scale-$j$
blocks in $X$.
\end{prop}

The following proposition is overkill for our needs in Map~3,
but we will use its full power in Map~4 and it is convenient to
state it here in this form.
The leading part of $h$, denoted $\hldg$,
was defined in
\cite[\eqref{IE-e:hptdefqq}]{BS-rg-IE} by
\begin{equation}
\label{e:hptdefqq}
    \hldg (U,B)
    =
    \begin{cases}
    -\frac{1}{2}\Ex_{\pi ,j+1} \theta ( V(B); V(\Lambda \setminus B))
    & U=B
    \\
    \;\;\;
    \frac{1}{2}
    \Ex_{\pi ,j+1}\theta  ( V(B);  V(U\setminus B)) & U \supset B, |U|_{j+1}=2
    \\
    \;\;\;
    0 &\text{otherwise}
    .
    \end{cases}
\end{equation}
The subscript $\pi$ above corresponds to a bookkeeping device (see
\cite[\eqref{IE-e:Epi}]{BS-rg-IE}) that does not play an explicit role
in what follows.
It is shown in \cite[\eqref{IE-e:hpt0bis}]{BS-rg-IE} that,
given $B \in \Bcal_{j+1}$,
\begin{equation}
\label{e:hpt0}
    \sum_{U : (U,B) \in \Dcal(\hldg)} \hldg (U,B) = 0
    ,
\end{equation}
and this property is essential in Map~4 where it is used
in conjunction with \eqref{e:kprimecondition}.
We
define
\begin{equation}
    \hldg (U) = \sum_{B \in \Bcal(U)}\hldg(U,B).
\end{equation}
We now extend the definition of $f_j(a,X)$ in \eqref{e:f0def} as follows.
Given $z \ge 0$, for
$\amain \in (0,2^{-d}]$ and $X \in \Pcal_{j}$ we define
\begin{equation}
    \label{e:fdef}
    f_{j} (z, \amain ,X)
    =
    \begin{cases}
    z + f_{j} (\amain, X)
    &X\not = \varnothing
    \\
    0&X = \varnothing.
    \end{cases}
\end{equation}

\begin{prop}
\label{prop:Ih}
There exists a constant $\cldg = \cldg (L)$ such that 
\begin{align}
\label{e:Ih1}
    \|\Ipttil(U) \hldg (U,B)\|_{j+1}
    &
    \le \cldg  \epdV^{\zldg}
    ,
\quad\quad
    \zldg =2
    .
\end{align}
Also, for $U \in\Ccal_{j+1}$,
\begin{align}
\label{e:Ih2}
    \|\Ipttil(U) [h(U)-\hldg (U)]\|_{j+1}
    &
    \le
    c_{\pt} \epdV^{f_{j+1} (\zh,a^{(3)},U)},
\quad\quad
    \zh=3
    .
\end{align}
The constants $\cldg,c_{\pt}$ may depend on $L$, and the norms are either
of \eqref{e:np1} or \eqref{e:np2}.
\end{prop}

\begin{proof}
The inequality
\eqref{e:Ih1} is proved in \cite[Proposition~\ref{IE-prop:hldg}]{BS-rg-IE}.
For \eqref{e:Ih2}, given $U \in \Ccal_{j+1}$, we define
\begin{align}
\label{e:hreddef}
    \hred(U) & = \sum_{X \in \overline\Pcal_j(U): |X|_j \le 2}
    \Ipttil^{- X} \Ex_{+}\delta I^X,
\\
    \hrem(U)
    &=
    \sum_{X \in \overline\Pcal_j(U): |X|_j \ge 3}
    \Ipttil^{- X} \Ex_{+}\delta I^X,
\end{align}
so that
\begin{equation}
\label{e:hhh}
    h(U) =  \hred(U) + \hrem(U)
.
\end{equation}
The bound \eqref{e:Ih2} with $h$ replaced by $\hred$ was proved in
\cite[Proposition~\ref{IE-prop:h}]{BS-rg-IE},
and hence it
suffices to prove that, for $U \in \Ccal_{j+1}$,
\begin{align}
\label{e:Icalbd}
    \|\Ipttil^U  \hrem(U) \|_{j+1}
    &\le
    c_\pt \epdV^{f_{j+1} (\zh,\amain ,U)}
    .
\end{align}
But by definition, the fact that $|\overline X|_{j+1} \le |X|_j$,
 and Proposition~\ref{prop:ip},
\begin{align}
    \|\Ipttil^U  \hrem(U) \|_{j+1}
    &\le
    \sum_{X \in \overline\Pcal_j(U): |X|_j \ge 3}
    \Econst^{|U|_j} \epdV^{|X|_j}
    \le
    (C_{L,d}\epdV)^{3 \vee |U|_{j+1}}.
\end{align}
If $|U|_{j+1} \le 2^d$ then the right-hand side is less than $C_{L,d}^{3}
\epdV^{3}$, and \eqref{e:Icalbd} holds
in this case.
On the other hand, if $|U|_{j+1}\ge 2^d+1$, then, since $3 \le 2^d$,
\begin{equation}
    |U|_{j+1} \ge 3 + (|U|_{j+1} - 2^d)
    =
     3 + a(|U|_{j+1} - 2^d) + (1-a)(|U|_{j+1} - 2^d),
\end{equation}
where we take $a=a^{(3)}$ (though in fact any larger $a<1$ would also
work here).
Thus we can choose $t>0$ depending on $a$ such that
\begin{equation}
    |U|_{j+1} \ge 3 + (|U|_{j+1} - 2^d)
    =
     3 + a(|U|_{j+1} - 2^d) + t|U|_{j+1} .
\end{equation}
The resulting factor $\epdV^{t|U|_{j+1}}$ can be used to control
$C_{L,d}^{|U|_{j+1}}$ and the desired result follows.
\end{proof}

Since $U$ is a small set, $\epdV^{f_{j+1} (\zh,a^{(3)},U)} =
\epdV^{\zh}$ is much smaller than $\epdV^{f_{j+1} (\zldg,a^{(3)},U)} =
\epdV^{\zldg}$.  Therefore, it is an immediate consequence of
Proposition~\ref{prop:Ih}, via the triangle inequality, that
\begin{align}
\label{e:Ih0}
    \|\Ipttil(U) h(U)\|_{j+1}
    &
    \le
    c_{\pt} \epdV^{f_{j+1} (\zldg,a^{(3)},U)}.
\end{align}
This proves \eqref{e:Map3h}.
The constants $c_{\pt}$ in \eqref{e:Ih0} and \eqref{e:Ih2}
may be larger than the constant
with the same name in
\cite[Proposition~\ref{IE-prop:h}]{BS-rg-IE},
but this is of no importance.
The inequality \eqref{e:Ih2} shows that
$\hldg(U)$ is in fact the leading part of $h(U)$.

Next we estimate the term $\Ipttil k$ and prove \eqref{e:Map3k}.
For this we will apply the following crucial estimate,
which is \cite[Proposition~\ref{IE-prop:cl}]{BS-rg-IE}.
Given $X \subset \Lambda$, we define
\begin{equation}
\label{e:gamLdef}
    \gamma
    =
    \gamma(X)
    =
    L^{-d -1} +  L^{-1}\1_{X \cap \{a,b\} \not = \varnothing}.
\end{equation}

\begin{prop}
\label{prop:cl} Let $X \in \Scal_j$ and $U = \overline X$.  Let $F(X)
\in \Ncal(X^\Box)$ be such that $\pi_\alpha F(X) =0$ when
$X(\alpha)=\varnothing$, and such that $\pi_{ab}F(X)=0$ unless $j \ge
j_{ab}$ (recall \eqref{e:Phi-def-jc}).  Let $\kappa_F=\|F (X)\|_{j}$
and let $\kappa_{\LT F} =\|\tilde I^X \LT_X \tilde I^{-X} F(X) \|_j$.
Then
\begin{align}
    \label{e:contraction3z}
    \|\tilde I^{U\setminus X} \Ex_{+}\theta F (X) \|_{j+1}
    &
    \prec
    \cgam (X)
    \kappa_F
    + \kappa_{\LT F}
    ,
\end{align}
where the pair of norms is given by
either choice of \eqref{e:np1} or \eqref{e:np2}.
\end{prop}

The following result is at the heart of our method. It
establishes the contractivity of the linear part of the map $K^{(2)}
\mapsto K^{(3)}$ via two different and essential principles: for
small sets $X$ we have arranged in Lemma~\ref{lem:K2}(iii) that
$K^{(2)}$ has a small relevant/marginal local part
and we can apply Proposition~\ref{prop:cl}, while for large
sets we apply the geometric fact in Lemma~\ref{lem:small}
to exploit the decay of $K(X)$ in the size of $X$.

\begin{lemma}
\label{lem:contraction} For $L$ sufficiently large,
$\rhoFcal$ sufficiently small
depending on $L$, and $U \in \Ccal_{j+1}$,
\begin{equation}
    \label{e:contraction3}
    \|\Ipttil(U) k(U) \|_{j+1}
    \le \kappa^* r^{(2)} \rhoFcal^{1+f_{j+1} (a^{(3)},U)}
    ,
\end{equation}
where $\kappa^*$ is an $L$-independent multiple of
$L^{-1}$.
\end{lemma}

\begin{proof}
By Lemma~\ref{lem:K2},
\begin{align}
\label{e:contraction6}
    \|K^{(2)} (X)\|_{j}
    &\le
    r^{(2)}\rhoFcal
    .
\end{align}

In the definition of $k (U)$ in \eqref{e:kdef}, we first consider those terms
in the sum over $X$ where $X \in \Scal_j$ and we prove that the
contribution from these terms is bounded by the right-hand side of
\eqref{e:contraction3}.
By
Lemma~\ref{lem:K2} and \eqref{e:JCK3} (reducing $u$ slightly
to absorb the $L$-dependence in \eqref{e:JCK3}),
\begin{align}
    \label{e:contraction4}
    \|\LT_X \Ipttil^{-X} K^{(2)}(X) \|_{T_0}
    &\prec
    r^{(2)}\rhoFcal^{1+u}
    ,
\end{align}
so it follows from the first inequality of \eqref{e:JCK0} that
\begin{align}
    \label{e:contraction4-I}
    \|\Ipttil^X \LT_X \Ipttil^{-X} K^{(2)}(X) \|_j
    &\prec
    r^{(2)}\rhoFcal^{1+u}
    .
\end{align}
We have already shown in Proposition~\ref{prop:Rcc} that
$\pi_{ab}\Vpt=0$ unless $j \ge j_{ab}$.  That proof used the
observation that no small set can contain both points $a,b$ when
$j<j_{ab}$.  By taking $L$ larger if necessary, it is similarly the
case that the small set neighbourhood $X^\Box$ of a small set $X$
cannot contain $a,b$ when $j<j_{ab}$.  By the assumption in
Definition~\ref{def:Kspace} that $\pi_{ab}K(X)=0$ unless $a,b \in
X^\Box$, and by the definition of $K^{(2)}$, we conclude that
$\pi_{ab}K^{(2)}(X)=0$ when $j<j_{ab}$.  It then follows from
Proposition~\ref{prop:cl} that
\begin{align}
    \|\Ipttil^{U\setminus X} \Ex_{+}\theta K^{(2)} (X) \|_{j+1}
    &
    \prec
    \cgam (X)
    r^{(2)}\rhoFcal   +
    r^{(2)}\rhoFcal^{1+u}
    .
\end{align}
We drop the second term because it is small compared with the first
term. After summation over the $O(L^d)$ small sets whose closure is
$U$, and by taking $\rhoFcal$ small depending on $L$, the resulting
right-hand side is bounded above by a multiple of $L^{-1}
r^{(2)}\rhoFcal$, which is the correct size for
\eqref{e:contraction3}.  This completes the analysis for $X\in
\Scal_j$.

We now consider those terms in the sum over $X$ in \eqref{e:kdef} due
to $X \not \in \Scal_j$, and prove that the contribution from these
terms is bounded by the right-hand side of \eqref{e:contraction3}.  In
this sum over $X$ there are fewer than $2^{|U|_{j}}$ terms.
Therefore, by Proposition~\ref{prop:Ianalytic1},
\eqref{e:integration-property}, \eqref{e:contraction6}, and
Lemma~\ref{lem:small}, there is a constant $C=C(d,L)$ such that
\begin{align}
    \|\sum_{X \in \overline\Ccal_j(U): X\not\in \Scal_j}
    \Ipttil^{U\setminus X} \Ex_{j+1}  \theta   K^{(2)}(X)
    \|_{j+1}
    &\le
    r^{(2)}C^{|U|_j}
    \sup_{X \in \overline\Ccal_j(U): X\not\in \Scal_j}
    \rhoFcal^{1+f_{j} (a^{(2)},X)}
    \nnb
    & \le
    r^{(2)}C^{|U|_j}
    \rhoFcal^{1+a^{(2)} \eta|U|_{j+1} - 2^d }
    \nnb
    & \le
    r^{(2)}C^{|U|_j} \rhoFcal^{(a^{(2)} \eta - a^{(3)})|U|_{j+1}}
    \rhoFcal^{1+f_{j} (a^{(3)},U)},
\end{align}
where $\overline{\Ccal}_j(U)$ is defined by \refeq{Cbardef}.
By taking $\rhoFcal$ small,
this contribution is negligible
compared to the contribution due to $X \in \Scal_j$.
\end{proof}

Finally, we show that the remainder term $R$ of \eqref{e:decomposeK3}
is negligible compared with the term that contains $k$.

\begin{lemma}
\label{lem:Ktil}
For $\rhoFcal$ small depending on $L$, and for
all $U \in \Ccal_{j+1}$,
\begin{equation}
\label{e:Kbdnok}
    \|R (U)\|_{j+1}
    \prec
    \rhoFcal^{u+1+f_{j+1} (a^{(3)} ,U)}
    .
\end{equation}
\end{lemma}

\begin{proof}
By \eqref{e:decomposeK3}
\begin{equation}
    \label{e:Rformula}
    R(U) = K^{(3)} (U) - \Ipttil^U [h(U)+k(U)]
    .
\end{equation}
To obtain a convenient expression for
\eqref{e:Rformula}, we
introduce $\vec{X} = \big(X_{K},X_{\delta I},X_{\Ipttil}\big)$, and
say that $\vec{X} \in \Xcal(U)$ if $X_{K},X_{\delta I},X_{\Ipttil}$
are disjoint sets in $\Pcal_{j}$ whose union is $U$ and also
$X_{K}\cup X_{\delta I} \in \overline{\Pcal}_{j}(U)$.  We set
\begin{align}
    \label{e:nKdefz} n_{K} & =
    \text{number of components of $X_{K}$},
    \\
\label{e:ndeldefz}
    n_{\delta I} & = |X_{\delta I}|_j,
    \\
    n & =(n_{K},n_{\delta I}).
\end{align}
Then
\begin{align}
\label{e:R1defz}
    R(U) &=
    \sum_{\vec{X}  \in \Xcal(X)}
    \1_{n \in N_0^c}
    \Ex_{+}\left( \delta I^{X_{\delta I}} K^{(2)}(X_K) \right)
    (I^{(3)} )^{X_{\Ipttil}},
\end{align}
where $ N_0^c$ is the complement of $N_0 = \{(0,\N_0),(1,0)\}$.

With the value $2$ arising in \eqref{e:Iass1}, we write
\begin{equation}
\label{e:stabilityC}
    \EGconst = \max \{\Econst ,2 \}.
\end{equation}
By the product property \cite[\eqref{IE-e:norm-fac}]{BS-rg-IE} of the
norm, \eqref{e:stabilityC}, \eqref{e:integration-property}, and
Lemma~\ref{lem:K2},
\begin{align}
    \|R(U)\|_{j+1}
&\le
    \sum_{\vec{X}  \in \Xcal(X)}
    \mathbb{I}_{n \in N_0^c}
    \EGconst^{|X_K|_j+|X_{\delta I}|_j}
    \epdV^{|X_{\delta I}|_j}
    \|K^{(2)}(X_K)\|_j
    \EGconst^{|X_{\Ipttil}|_j}
    \nn\\
&\le
    \EGconst^{|X|_{j}}
    \sum_{\vec{X} \in \Xcal(X)}
    \mathbb{I}_{n \in N_0^c}
    \epdV^{|X_{\delta I}|_j}
    \prod_i
    (r^{(2)}\rhoFcal^{1+f_{j}( a^{(2)},X_{K,i})})
    \nnb
&\le
    (c\EGconst)^{L^d|X|_{j+1}}
    \sum_{\vec{X} \in \Xcal(X)}
    \mathbb{I}_{n \in N_0^c}
    \epdV^{|X_{\delta I}|_j}
    \rhoFcal^{\sum_i f_{j}( a^{(2)}, X_{K,i})},
\end{align}
where the product over $i$ is a product over the connected components
$X_{K,i}$ of $X_K$.  By Lemma~\ref{lem:czf} this gives
\begin{equation}
    \|R(U)\|_{j+1}
    \le
    (c\EGconst C (L))^{L^d|X|_{j+1}}
    \rhoFcal^{\delta |X|_{j+1} + u+1+f_{j+1} (a^{(3)} ,X)}
    \sum_{\vec{X} \in \Xcal(X)}
    \mathbb{I}_{n \in N_0^c}
    .
\end{equation}
Since the number of terms in the sum over $\vec{X} \in \Xcal(U)$ is at
most the number of ways of assigning each $j$-block in $U$ to either
$X_K$, $X_{\delta I}$, or $X_{\Ipttil}$, which is
$3^{|U|_j}=3^{L^d|U|_{j+1}}$, this gives
\begin{equation}
\label{e:KLIbd}
    \|R(U)\|_{j+1}
    \le
    b^{|U|_{j+1}}
    \rhoFcal^{u+1+f_{j+1} (a^{(3)} ,U)}
    ,
\end{equation}
with $b=\rhoFcal  (3c\EGconst C (L))^{L^d}$. This completes the proof
since $b \le 1$ for $\rhoFcal$ sufficiently small.
\end{proof}

\subsection{Map~4: Reapportionment of \texorpdfstring{$K^{(3)}$}{K-3}}
\label{sec:IKnext}

Map~4 removes the second-order perturbative contribution $\hldg$ from
$K^{(3)}$ when $K=0$.
Part~(iii) is the purpose of this map---the leading part
of $K^{(3)}$ has been reapportioned and is now absent in $K^{(4)}$.
All norms in this section are scale $j+1$ norms, either the
$T_{0,j+1}$ or the $\tilde G_{j+1}$ norms, with their corresponding
$\Fcal$ norms.  We drop the label $j+1$ on the norms, to simplify the
notation.

\begin{lemma}\label{lem:K5a}
There exist $\mu_4 \ge 1$, $K^{(4)}$, and a constant $C=C (L)$ such that,
for $(V,K)$ obeying \eqref{e:VKhyp},
\begin{align}
(i)\quad&
    \Ipttil  \circ   K^{(3)}
    =
    \Ipttil \circ K^{(4)} ,
\nonumber
\\
(ii)\quad&
    K^{(4)} \in B_{\Fcal^{(4)}} (r^{(4)} \rhoFcal)
    ,
\nonumber
\\
(iii)\quad&
    \text{$\|K^{(4)}(U)  \|
    \le
    (C\epdV)^{f_{j+1} (3
    ,a^{(4)},U)}$ \;
    for $U \in \Pcal$ \;  if \; $K=0$}
    .
\nonumber
\end{align}
\end{lemma}

\begin{proof}
(i)
Let $\mu_4=2$, so that $r^{(4)}=2r^{(3)}$.
By Proposition~\ref{prop:Ih} and by taking $\ggen$ sufficiently small,
there is a constant $C$ such that
\begin{equation}
    \label{e:Jbd2}
    \|\Ipttil  (U) \hldg (U,B)\|
    \le
    (C\epdV)^2
    \le {\textstyle{\frac 12}}
    r^{(4)}  \rhoFcal
.
\end{equation}
We apply Proposition~\ref{prop:change-of-variable-1} with
$J (U,B) = \hldg (U,B)$ and
\begin{gather}
    \Iin = \Ipttil  , \quad
    \Kin = K^{(3)}, \quad
    \rhogen = \rhoFcal, \quad
    \ain = a^{(3)}, \nnb
    \aout = a^{(4)},\quad
    \epsilon = {\textstyle{\frac 12}} r^{(4)} ,\quad
    z = z' = 1
    ,
\end{gather}
and we set $K^{(4)} = \Kout$.  Hypothesis \eqref{e:kprimecondition} is
provided by \eqref{e:hpt0}. Hypothesis \eqref{e:J-localisation} holds
by Lemma~\ref{lem:Ex-sym}, and by the use of $\Ex_\pi$ in
\refeq{hptdefqq} to localise observables properly.  The stability
hypothesis \eqref{e:Iin-stability} holds by
Proposition~\ref{prop:Ianalytic1}.  Hypothesis \eqref{e:Jbd} holds by
\eqref{e:Jbd2}.  Hypothesis \eqref{e:Mcalbd} is obtained from
\eqref{e:Jbd2} and Lemma~\ref{lem:K3}(ii) by using the triangle
inequality in the definition \eqref{e:McalnowM-def} of $\McalnowM$.
Thus all hypotheses hold and
Proposition~\ref{prop:change-of-variable-1} implies (i).
\\
(ii) The relevant estimate follows from Proposition~\ref{prop:change-of-variable-1} and
\eqref{e:Koutbd-K}.
The fact that
$K^{(4)}$ has the properties required of membership in the space $\Fcal^{(3)}$
can be established similarly to what was done in
Proposition~\ref{prop:EIK} for $K^{(3)}$.
\\
(iii) We again apply Proposition~\ref{prop:change-of-variable-1} but
with different choices for parameters.  We continue to take $J (U,B) =
\hldg (U,B)$, and use the first inequality of \eqref{e:Jbd2}.  With
Lemma~\ref{lem:K3}(iii), \eqref{e:McalnowM-def} now gives
\begin{equation}
    \McalnowM(U)
    =
    \Ipttil^U  (h - \hldg)(U)
    .
\end{equation}
By Proposition~\ref{prop:Ih},
\begin{equation}
\label{e:Ih-hldg}
    \|M(U)\|
    =
    \|\Ipttil (U) (h - \hldg) (U)\|
    \le
    (C\epdV)^{f_{j+1} (3,a^{(3)},U)} 
    .
\end{equation}
We apply Proposition~\ref{prop:change-of-variable-1} with $\rhogen
=C\epdV$, $\epsilon =1$, and with $z=1$ replaced by $z=3$.  
We have the required
inequality $z'=2>\frac 12 z = \frac 32$.  All hypotheses of
Proposition~\ref{prop:change-of-variable-1} then hold with the new
parameter values.  The conclusion of
Proposition~\ref{prop:change-of-variable-1} then implies that
\begin{equation}
\label{e:newKnullbd}
    \|K^{(4)}(U)\|
    \le
    2(C\epdV)^{f_{j+1} (3,a^{(4)},U)},
\end{equation}
and by absorbing the factor $2$ into $C$ we complete the proof of (iii).
\end{proof}

The restriction and isometry properties of $K^{(4)}$ are verified as
in maps 1 and 2.  In particular, the restriction property is a
consequence of the change of variables formula \refeq{Kdef-new}.  For
the mass continuity and the $(V,K)$-analyticity, the new ingredient
compared to what we have seen in previous Maps is to establish the
mass continuity and analyticity of $\Ipttil^U J(U,B)= \Ipttil^U
\hldg(U,B)$, with $\hldg(U,B)$ the degree-6 polynomial in the fields
defined by \eqref{e:hptdefqq}.  The subscript $\pi$ in
\eqref{e:hptdefqq} is not relevant for the continuity or analyticity
properties, and it suffices to verify mass continuity and
$(V,K)$-analyticity for the case where $\Ex_\pi$ is replaced by
$\Ex_{+}$ in \eqref{e:hptdefqq}.  As in the proof of analyticity in
Map~1, what is needed is to establish mass continuity and
$V$-analyticity (for there is no $K$-dependence) of $\hldg(U,B)$ as a
map into $\Ncal(U^\Box)$ with the $T_0$ norm.  The analyticity follows
from the fact that $\hldg$ is a bilinear function of $V$, by
Lemma~\ref{lem:multilinear-analyticity}.  The mass continuity of
$K^{(4)}(U)$ for each $U$ follows from Proposition~\ref{prop:masscont}
and the continuity of $K^{(3)}(U)$ for each $U$.

\section{Maps~5--6: Final adjustments}
\label{sec:Map5-6}

The interaction output by Map~4 is $\Ipttil(b) =
e^{-\Vpt(b)}\big(1+W_{j+1}(\Vpt,b)\big)$, which involves blocks $b$ of
scale $j$.  Also, the polynomial $\Vpt$ contains monomials
$\tau_{\nabla\nabla}$ and $\sigma\bar\sigma$ arising in $Q$, and hence does not
lie in $\Qcalnabla^{(0)}$.  The purpose of Maps~5--6
is to perform the bookkeeping tasks of replacing $\Ipttil$ by a scale
$j+1$ interaction of the form \eqref{e:IVB}, and replacing $\Vpt$ by
$V_{+} = \Vpt^{(0)} \in \Qcalnabla^{(0)}$.
As in \eqref{e:Vjplus1}, $\Vpt^{(0)}$
is the polynomial obtained by dropping the $\sigma\bar\sigma$ term, and
by replacing $z\tau_\Delta +
y\tau_{\nabla\nabla}$ in $\Vpt$ by $(z-y)\tau_\Delta$ as a formal
summation by parts would suggest.  Then $V_+$ and $I_{+}=I_{j+1}(V_+)$
have the same form as the initial $V$ and $I=I_j(V)$.

To accomplish this we use two steps. First, in Map~5 we eliminate
blocks $b\in\Bcal_j$ in favour of blocks $B\in\Bcal_{j+1}$, and we
simultaneously adjust $W(\Vpt)$ to $W(V_+)$.  Second, in Map~6 we
replace $e^{- \Vpt(B)}$ by $e^{- V_{+}(B)}$ to obtain
$I_{+}=I_{j+1}(V_+)$ with corresponding $K_{+}$.

Norms in this section are scale $j+1$ norms, either the
$T_{0,j+1}$ or the $\tilde G_{j+1}$ norms, with their corresponding
$\Fcal$ norms.  We drop the label $j+1$ on the norms, to simplify the
notation.

\subsection{Map~5: Adjustment to \texorpdfstring{$W$}{W}}

For $B\in \Bcal_{j+1}$, we define
\begin{equation}
\label{e:Iptplusdef}
    \Ipt^+(B) = e^{-\Vpt(B)}\big(1+W_{j+1}(V_{+},B)\big)
    ,
\end{equation}
and then define $\delta^+ I(B)$ by
\begin{equation}
\label{e:DeltaI}
    \Ipttil (B)
    =
    \Ipt^+ (B) + \delta^+ I(B)
    .
\end{equation}
We write
\begin{align}
\label{e:delIptplusdef}
    \delta \Ipt^{+}(B)
    &=
    e^{-\Vpt(B)}\big(W_{j+1}(\Vpt,B) -  W_{j+1}(V_{+},B)\big),
    \\
    \Delta I(B) &= e^{-\Vpt(B)}
    \left(
    \prod_{b \in \Bcal_j(B)}(1+W_{j+1}(\Vpt,b)) - \left(1+W_{j+1}(\Vpt,B) \right)
    \right),
\end{align}
so that
\begin{equation}
\label{e:delIplusB}
    \delta^+I(B) = \Delta I(B) + \delta \Ipt^+(B).
\end{equation}
It is proved in \cite[Lemma~\ref{IE-lem:DelIbd}]{BS-rg-IE} (with
$j$ replaced by $j+1$) that
\begin{equation}
\label{e:DelIbd}
    \|\Delta I(B)\|  \prec_L \, \epdV^{4}
.
\end{equation}

To estimate \refeq{delIptplusdef}, we first recall from \refeq{Qbound} that
$\|Q(b)\|_{T_{0,j}} \prec r^{(0)}
\epdV$
for $b \in \Bcal_j$.  Since the $\tau_{\nabla\nabla}$ term in $\Vpt$ arises solely
from a contribution due to $Q$, it then follows from
\cite[Lemma~\ref{IE-lem:T0ep}]{BS-rg-IE}  that
$\|y\tau_{\nabla\nabla}(b)\|_{T_{0,j}} \prec \epdV$.
Thus we can apply
\cite[Lemma~\ref{IE-lem:sbp-bds}]{BS-rg-IE} to conclude that
\begin{gather}
    \label{e:map6-bd}
    \|\delta \Ipt^+(B)\|  \prec_L \, \epdV^{2}
\end{gather}
(note that in \cite[Lemma~\ref{IE-lem:sbp-bds}]{BS-rg-IE} the $\sigma\bar\sigma$ term
is not eliminated when creating $V_+$, but such a term in $V$ does not contribute to $W(V)$
by definition, so \cite[Lemma~\ref{IE-lem:sbp-bds}]{BS-rg-IE} does apply).

\begin{lemma}\label{lem:K4}
There exist $\mu_5 \ge 1$, $K^{(5)}$, and a constant $C=C (L)$ such that,
for $(V,K)$ obeying \eqref{e:VKhyp},
\begin{align}
(i)\quad&
    (\Ipttil \circ K^{(4)})(\Lambda)
    =
    (\Ipt^+ \circ K^{(5)})(\Lambda),
\nonumber
\\
(ii)\quad&
    K^{(5)} \in B_{\Fcal^{(5)}} (r^{(5)} \rhoFcal),
\nonumber
\\
(iii)\quad&
    \text{$\|K^{(5)}(U)  \|
    \le
    (C\epdV)^{f_{j+1} (\zh,a^{(5)},U)}$ \;
    for $U \in \Pcal$ \;  if \; $K=0$}
    .
\nonumber
\end{align}
\end{lemma}

\begin{proof}
(i)
Let
\begin{equation}
    K^{(5)} = K^{(4)} \circ \delta^+ I.
\end{equation}
By \eqref{e:DeltaI} and Lemma~\ref{lem:bin},
\begin{equation}
    K^{(4)}  \circ \Ipttil
    =
    K^{(4)}  \circ (\delta^+ I \circ \Ipt )
    = K^{(5)} \circ \Ipt^+
    .
\end{equation}
(ii) By \eqref{e:delIplusB}--\eqref{e:map6-bd}, for
$\rhoFcal$ sufficiently small we have
\begin{equation}
\label{e:dIplusbd}
    \|\delta I^+(B) \|
    \le
    C \epdV^{2}
    \le
    r^{(4)}\rhoFcal
    ,
\end{equation}
i.e., $\delta I^+ \in B_{\Fcal^{(4)}}(r^{(4)}\rhoFcal)$.
The desired estimate is then a consequence of Lemmas~\ref{lem:K5a} and
\ref{lem:FcircG},
once we set $r^{(5)}=\mu_5 r^{(4)}$ with $\mu_5 = 2^{2^d}$.
The vanishing at weighted infinity,
field locality, symmetry, and component factorisation properties inherent
in the statement that $K^{(5)}\in \Fcal^{(5)}$ can be verified using the
fact that $K^{(4)}$ has
these properties.
(In particular, $\delta^+I$ vanishes at weighted infinity since each of $\Ipttil$ and $\Ipt^+$
do by Proposition~\ref{prop:Ivan}.)

\smallskip \noindent (iii) This follows from the first inequality of
\eqref{e:dIplusbd} and from Lemma~\ref{lem:K5a}(iii), by
Lemma~\ref{lem:FcircG}.
\end{proof}

The restriction and isometry properties of $K^{(5)}$ can be verified
as in maps 1 and 2 using its definition, as can the mass continuity
$m^2 \mapsto K^{(5)}(U) \in \Ncal(U^\Box), \|\cdot\|_{j+1}$ and the
$(V,K)$-analyticity, using the corresponding properties of $K^{(4)}$.
For the analyticity, slight modifications to the proof of
\cite[Proposition~\ref{IE-prop:Ianalytic1:5}]{BS-rg-IE} show that the
map $(V,K) \mapsto \Ipt^+$ is analytic from $\DV_j \times B_{\Fcal_j}
(r^{(0)}\rhoFcal)$ into $\Ncal(B^\Box)$, $\|\cdot\|_{j+1}$ (and so is
$(V,K) \mapsto I_{+}$).

\subsection{Map~6:  Adjustment to \texorpdfstring{$V$}{V}}
\label{sec:int-by-parts2}

As in \eqref{e:Vjplus1}, we define
\begin{equation}
    I^{(6)} = I_{+}
    =
    e^{-V_{+}}
    \big(1+W_{j+1}(V_{+})\big)
\end{equation}
Map~6 performs two tasks.  First, it removes the monomial
$y\tau_{\nabla\nabla}$ from the exponent of $\Ipt^+$ by converting it
to $y\tau_{\Delta}$ by summation by parts.  Second, it extracts
$\dq\sigma\bar\sigma$ from $\Ipt^+$ to bring it out of the circle
product; here $\dq = \frac{1}{2}(\dq^a+\dq^b)$ as in \refeq{qqq}.  The
boundary term resulting from the first task, and an adjustment to
achieve the second, convert $K^{(5)}$ to $K_{+}=K^{(6)}$.  All objects
in this section are at scale $j+1$.

In more detail, for $Z \in \Pcal_{j+1}$ we define
\begin{gather}
\label{e:int-by-parts}
    V_{\partial, Z}
    = \sum_{z \in Z} \ypt (\tau_{\nabla\nabla,z} - \tau_{\Delta,z}).
\end{gather}
With the definition of $\Ipt^+$ in \eqref{e:Iptplusdef}, this gives
\begin{equation}
    \Ipt^+ (Z)
    =
    e^{v(Z)}
    I_{+} (Z)
    e^{-V_{\partial,Z} }
    ,
\end{equation}
where $v(Z) = \sigma\bar\sigma\frac 12 (\dq^a\1_{a\in Z}+\dq^b\1_{b\in
Z})$ (cf.~$V_{ab}$ in \refeq{Vx}).  By performing summation by parts
on the right-hand side of \eqref{e:int-by-parts}, we find that there
exists $V_{\partial, Z,B}$, which depends only on fields that are in
the intersection of $B$ and the boundary of $Z$, such that
\begin{equation}
    V_{\partial, Z}
    =
    \sum_{B \in \Bcal} V_{\partial, Z,B}.
\end{equation}
Here $V_{\partial,Z,B}=0$ if $B$ is not a block
in $Z$ which is on the boundary of $Z$ in the sense that it has
a neighbour not in $Z$ (in particular, $V_{\partial,\Lambda,B}=0$).

We therefore have
\begin{gather}
    \Ipt^+ (\Lambda \setminus X)
    =
    e^{v(\Lambda \setminus X)}
    \!\!\!\!\!\!
    \prod_{B \in \Bcal_{j+1} (\Lambda \setminus X)}
    I_{+} (B)
    \left(1 +  R_X  (B)\right)
\quad
\text{with}
\quad
    R_X  (B)
    =
    e^{- V_{\partial,\Lambda \setminus X,B}}-1
    .
\end{gather}
By Lemma~\ref{lem:bin},
\begin{equation}
\label{e:Iptplus}
    \Ipt^+ (\Lambda \setminus X)
    =
    e^{v(\Lambda \setminus X)}
    (\delta I^{(6)}_X \circ I_{+})(\Lambda \setminus X)
    =
    e^{v(\Lambda \setminus X)}
    \!\!\!\!\!\!
    \sum_{Y \in \Pcal_{j+1} (\Lambda \setminus X)}
    (\delta I^{(6)}_X)^Y
    \,
    I_{+}^{\Lambda \setminus (X \cup Y)}
    ,
\end{equation}
where
\begin{equation}
\label{e:delI6def}
    (\delta I^{(6)}_X)^Y
    =
    \prod_{B \in \Pcal_{j+1} (Y)}
    R_X(B) I_{+} (B)
.
\end{equation}
Note that $\delta I^{(6)}_\varnothing =0$ by definition.  It follows
from \cite[Lemma~\ref{IE-lem:sbp-bds}]{BS-rg-IE}
(together with the verification of its assumption as above \refeq{map6-bd}) that there is
a constant $c$ such that
\begin{equation}
\label{e:delIX6}
    \|\delta I^{(6)}_X (B)\|_{j+1}
    \le c \rhoFcal
   .
\end{equation}

\begin{lemma}
\label{lem:K7a}
There exist $\mu_6 \ge 1$ and $K^{(6)}=K_{+}$ such that,
for $(V,K)$ obeying \eqref{e:VKhyp},
\begin{align}
(i)\quad&
    \Ipt^+  \circ K^{(5)}
    =
    I_{+} \circ K_{+},
\nonumber
\\
(ii)\quad&
    K_{+} \in B_{\Fcal^{(6)}} ( r^{(6)} \rhoFcal  )
    ,
\nonumber
\\
(iii) \quad&
    \text{$\|K_{+}(U)  \|
    \le
    (C\epdV)^{f_{j+1} (\zh,a^{(6)},U)}$ \;
    for $U \in \Pcal$ \;  if \; $K=0$}
    .
\end{align}
\end{lemma}

\begin{proof}
(i)
It follows from \eqref{e:Iptplus}, and from the formula for $\dq$ in
\refeq{qqq}, that
\begin{align}
    (\Ipt^+ \circ K^{(5)}) (\Lambda)
    &=
    e^{\dq \sigma\bar\sigma}
    \sum_{X \in \Pcal (\Lambda)}
    e^{-v(X)}K^{(5)} (X)
    (\delta I^{(6)}_X  \circ
    I_{+})(\Lambda \setminus X )
\nnb &
    =
    e^{\dq \sigma\bar\sigma}
    \sum_{Z \in \Pcal (\Lambda)}
    \sum_{X \in \Pcal (Z) }
    e^{-v(X)} K^{(5)} (X)  (\delta I_{X}^{(6)}  )^{Z\setminus X}
    I_{+}^{\Lambda \setminus Z}
    \nnb &
    =
    e^{\dq \sigma\bar\sigma}
     (I_{+} \circ K_{+}) (\Lambda)
    ,
\end{align}
where,
for $Z \in \Pcal$, we define
\begin{equation}
    \label{e:K7-def}
    K_{+} (Z)
    =
    \sum_{X \in \Pcal (Z) }
     e^{-v(X)} K^{(5)} (X)  (\delta I^{(6)}_X )^{Z\setminus X}.
\end{equation}

\noindent (ii) It follows from the product property of the $T_\phi$
norm that $\|e^{-v(X)} \| \le e^{\|v(X)\|}$ (for a proof, see
\cite[\eqref{norm-e:apos-e}]{BS-rg-norm}).  By definition, $\|v(X)\|
\le |\dq|\h_{\sigma}^2$.  Moreover, $\dq$ is non-zero only when $j$ is
at least the coalescence scale $j_{ab}$, and in this case $Q$ no
longer contains a $\lambda$ term since the corresponding monomials are
no longer in the range of $\LT$ above coalescence (see
\cite[Section~\ref{pt-sec:loc-specs}]{BBS-rg-pt}).  Therefore, by
\cite[Proposition~\ref{pt-prop:Vptg}]{BBS-rg-pt} and
\cite[\eqref{pt-e:qpt2}]{BBS-rg-pt}, $|\dq| \prec \lambda^2 L^{-2j}
\prec L^{-2j}$.  This implies that $\|v(X)\| \prec L^{-2j}\h_\sigma^2
$, and from the definitions of $\h_\sigma$ in \eqref{e:newhsig}, this
shows that $\|e^{-v(X)}\| \le 2$.  With Lemmas~\ref{lem:K4} and
\ref{lem:Kstar-comb2}, and by \eqref{e:delIX6}, this gives
\begin{align}
    \|K_{+} (Z) \|
    &\le
    \sum_{X \in \Pcal (Z): X \neq \varnothing}
     \|e^{-v(X)}\|
     \|K^{(5)} (X)\|
    \|\delta I_{X}^{(6)}\|^{|Z\setminus X|}
\nnb
    &\le
    2r^{(5)}
    \sum_{X \in \Pcal (Z): X \neq \varnothing}
    \rhoFcal^{1+f (a^{(5)},X)}
    (c\rhoFcal)^{|Z\setminus X|}
\nnb
    &\le
    2r^{(5)}
    (2c)^{|Z|}
    \sup_{X \in \Pcal (Z): X \neq \varnothing}
    \rhoFcal^{1+f (a^{(5)},X) +  |Z\setminus X|}
\nnb
    &\le
    2r^{(5)}
    (2c)^{2^{d}}
    \rhoFcal^{1+f (a^{(6)},Z)}
    =
    r^{(6)}
    \rhoFcal^{1+f (a^{(6)},Z)}
    ,
\end{align}
where in the last step we set $r^{(6)}=\mu_6 r^{(5)}$ with
$\mu_6 = 2(2c)^{2^{d}}$, and used $a^{(6)}$ to cancel the exponential growth
of $(2c)^{|Z|}$ for large sets $Z$.
Specialising to the case where $Z$ is connected, we obtain the estimate of (ii).

To see that $K_{+}$ obeys the component factorisation property,
let $Z$ be the disjoint union of $Z_1$ and $Z_2$.
The sum over $X$ in \eqref{e:K7-def} can then be written as
the sum over $X_1\in \Pcal(Z_1)$ and $X_2\in \Pcal(Z_2)$,
and the component factorisation property of $K^{(5)}$ implies
that $K^{(5)} (X) = K^{(5)} (X_1)   K^{(5)} (X_2)$.
It suffices if
$(\delta I^{(6)}_{X_1 \cup X_2} )^{Z\setminus (X_1\cup X_2)}
=
(\delta I^{(6)}_{X_1} )^{Z_1\setminus X_1}
(\delta I^{(6)}_{X_2} )^{Z_2\setminus  X_2}$,
and this indeed holds because
\begin{equation}
    \prod_{B \in \Bcal(Z\setminus (X_1\cup X_2)}
    R_{X_1 \cup X_2}(B)
    =
    \prod_{B \in \Bcal(Z_1\setminus X_1)}
    R_{X_1}(B)
    \prod_{B \in \Bcal(Z_2\setminus X_2)}
    R_{ X_2}(B).
\end{equation}

The fact that $K^{(6)}$ obeys the field locality, symmetry and component
factorisation properties can be seen from its definition.
The fact that $K^{(6)}$ vanishes at weighted infinity follows from the fact that
$K^{(5)}$ does, together with the fact that $\delta I_X^{(6)}$ vanishes
at weighted infinity by an extension of Proposition~\ref{prop:Ivan}.
\\
(iii) When $K=0$, by Lemma~\ref{lem:K4}(iii) we can replace $r^{(5)}$
in the proof of part~(ii) by $C\epdV^{z_h}$, and this immediately
gives the result.
\end{proof}

The restriction and isometry properties are straightforward to verify
as in maps 1 and 2.  We omit the tedious details which justify the
mass continuity $m^2 \mapsto K^{(6)}(U) \in \Ncal(U^\Box),
\|\cdot\|_{j+1}$, and the $(V,K)$-analyticity of $K^{(6)}$.

\begin{rk}
\label{rk:uphi4}
Recall the discussion of the
4-dimensional $n$-component $|\varphi|^4$ model in Section~\ref{sec:gmr}.
We now sketch how Lemma~\ref{lem:K7a} can be adapted so that the results of
the present paper can be applied in \cite{BBS-phi4-log}
to the $|\varphi|^4$ model.  The new ingredient is that $\Vpt(V-Q)$ contains a constant
term $u$, even when $V$ does not, because $Q$ will contain such a term and also
$\Vpt$ will produce one.
Let $|X|_j$ denote the number of scale-$j$ blocks in $X\in \Pcal_j$.
In particular, $|X|_0$ is the number of points in $X$.
The term $u$ in $V$ occurs in $I(X,V)$ only as
an overall factor $e^{\delta u|X|_0}$, since a constant term in $V$ cannot contribute to $W$.
In the scale-$(j+1)$ circle product considered in Lemma~\ref{lem:K7a},
we wish to replace the factor $e^{\delta u|X|_0}$ multiplying $\Ipt^+(X)$ by
$e^{\delta u|\Lambda|_0}$.  For this, we use
\begin{equation}
\lbeq{removeu}
    ((e^{\delta u}\Ipt^+)\circ K^{(5)})(\Lambda)
    =
    e^{\delta u|\Lambda|_0}(\Ipt^+ \circ (e^{-\delta u}K^{(5)})(\Lambda).
\end{equation}
The multiplication of $K^{(5)}$ on the right-hand side is controlled by the
estimate
\begin{equation}
\lbeq{uKbd}
    \|e^{-\delta u|X|_0}K^{(5)}(X)\|
    \le
    e^{|\delta u|\, |X|_{0}} \|K^{(5)}(X)\|
    =
    e^{|\delta u|L^{d(j+1)}|X|_{j+1}} \|K^{(5)}(X)\|
    .
\end{equation}
By definition,
\begin{equation}
    \delta u = \Vpt(V-Q)|_{\varphi=0}  =
    \Vpt(V)|_{\varphi=0} + \big(\Vpt(V-Q)|_{\varphi=0} - \Vpt(V)|_{\varphi=0}\big).
\end{equation}
As in the proof of Theorem~\ref{thm:mr-R}, we find that
\begin{equation}
    \big|\Vpt(V-Q)|_{\varphi=0} - \Vpt(V)|_{\varphi=0}\big| \le O(L^{-4d}\chicCov \ggen^3).
\end{equation}
Since $\Vpt(V)|_{\varphi=0}=O(L^{-dj}\chi\ggen)$ by
\cite[Lemma~\ref{phi4-log-lem:upt}]{BBS-phi4-log},
this gives
\begin{equation}
\lbeq{du-bis}
    |\delta u| = O(L^{-dj}\chicCov_j\ggen_j).
\end{equation}
Therefore,
\begin{equation}
\lbeq{uKbd2}
    \|e^{-\delta u|X|_0}K^{(5)}(X)\|
    \le
    e^{O(\chicCov_j\gbar_j)|X|_{j+1}} \|K^{(5)}(X)\|
    .
\end{equation}
The small amount of exponential
growth on the right-hand side
is handled by the increase from $a^{(5)}$ to $a^{(6)}$
which already performs a similar task in the proof of
Lemma~\ref{lem:K7a}.  Other aspects of the proof of our main results are
unchanged, and we apply this extension in \cite{BBS-phi4-log}.
\end{rk}

\section{Completion of proof of Theorem~\ref{thm:1}}
\label{sec:pfthm1}

We now assemble the conclusions obtained in the analysis of the six
Maps, to complete the proof of Theorem~\ref{thm:1}.

\begin{proof}[Proof of Theorem~\ref{thm:1}(i)] Since $K_{+}$ is the
composition of the six maps, the domain of $K_{+}$ is the domain of
the first map $K^{(1)}$, which, as specified in the hypothesis of
Lemma~\ref{lem:K1}, is $\DV \times B_{\Fcal}(r \epdV)$ as desired.
The range of $K_{+}$ is the range of the sixth map $K^{(6)}$.
By Lemma~\ref{lem:K7a}(ii), $K^{(6)} \in
B_{\Fcal_{j+1}(a^{(6)})}(r^{(6)} \rhoFcal)$. From
Section~\ref{sec:thm1parameters}, we find that $r^{(6)} = \kappa \rD$,
so $K^{(6)} \in B_{\Fcal_{j+1}(a^{(6)})}(\kappa \rD\rhoFcal)$. By
\eqref{e:gbarmono} and \eqref{e:rhoFcaldef}, $\rhoFcal
/\rhoFcal_{j+1} $ is bounded by a constant, so we can replace
$\rhoFcal$ by $\rhoFcal_{+}$ by absorbing this constant into the
constant $\gamma^{*}$ in $\kappa = O (L^{-1})$. By
\refeq{achoice}, $ a^{(6)} > a^{(0)}=a$, so the output
space $\Fcal_{j+1}(a^{(6)})$ has $a_{+}=a^{(6)}>a$ as
claimed. Furthermore, by \cite[Remark~\ref{IE-rk:h++}]{BS-rg-IE}, all estimates
involving our norm pairs
in \cite{BS-rg-IE} remain valid for some choice of $\h_{++} > \h_{+}$.
For this reason,
$\h_{+}$ can be replaced by $\h_{++}$ as required.

The bound \eqref{e:KTay0} is obtained in Lemma~\ref{lem:K7a}(iii),
again with a larger constant to allow the replacement of $\rhoFcal$
by $\rhoFcal_{+}$ as explained above.  The fact that the circle
product is preserved in the sense of \refeq{rgmapdef-bis} is a
consequence of part~(i) of Lemmas~\ref{lem:K2}, \ref{lem:K3},
\ref{lem:K5a}, \ref{lem:K4} and \ref{lem:K7a}.  The desired $(V,K)$
analyticity is a consequence of the analyticity established for each
Map.
\end{proof}

\begin{proof}[Proof of Theorem~\ref{thm:1}(ii-iii)]
These have both been established for the six individual Maps and therefore hold
for $K_+$.
\end{proof}

\begin{proof}[Proof of Theorem~\ref{thm:1}(iv)] For mass continuity,
the mass $m^2$ which is being varied appears in the analysis via the
mass dependence of the covariance $C_+$, which is a member of the
decomposition of the covariance $(-\Delta +m^2)^{-1}$.  The mass
continuity established for the six Maps provides a statement of
continuity of $K_+$ as a map from $m^2 \in \Iint_+$ into the space
$\Ncal(U^\Box)$, $\|\cdot\|_{j+1}$, for each polymer $U\in \Pcal_+$.

We wish to transfer this into a statement of continuity of $K_+$ as a
map from $m^2 \in \Igen_+(\mgen^2)$ into $\Fcal_+$.  The value of
$\mgen^2$ fixes the space $\tilde\Fcal_+$ and fixes $\chigen$ which
determines the radius of balls in this space, so that neither the
space nor the ball varies with $m^2$.  By
\cite[\eqref{pt-e:jmjOmega}]{BBS-rg-pt}, $\chi_j=\Omega^{-(j-\jm)_+}
\asymp \Omega^{-(j-j_m)_+}$, where $j_m=\lfloor \log_{L^2} m^{-2}
\rfloor$.  The values of $m^2$ and $\mgen^2$ are comparable by
definition of $\Igen$, hence so are $j_m$ and $j_{\tilde m}$, and
hence so are $\chi(m^2)$ and $\chigen=\chi(\mgen^2)$.  Consequently,
$\epdV$ of \refeq{rhoFcaldef} differs by a constant factor when
computed using $\chi$ or $\chigen$.  The estimates of
Propositions~\ref{prop:ip}--\ref{prop:Ih} produce $\epdV$ constructed
from $m^2$, since these are estimates based on the covariance $C_+$.
On the other hand, estimates implied by membership in the space
$\tilde\Fcal_+$ are in terms of $\epdV$ constructed from $\mgen^2$, by
definition.  The fact that the two versions of $\epdV$ are comparable
means that it does not matter if different versions appear at
different steps of the analysis.  As there are only finitely many
polymers $U$ in $\Lambda$, the continuity of $K_+$ as a map into
$\Ncal(U)$, $\|\cdot\|_{j+1}$ therefore implies continuity into
$\Fcal_+$, as required.
\end{proof}

\begin{proof}[Proof of Theorem~\ref{thm:1}(v)] We consider the case
$x=a$, as the case $x=b$ is similar.  If $\pi_a V=0$ and $\pi_a
K(X)=0$ for all $X \in \Pcal$, then neither $I$ nor $K$ has a
component in $\pi_a\Ncal$, and the observable field $\sigma$ is not
present in either of $I$ or $K$.  It is possible that $\sigmab$ or
$\sigma\sigmab$ are present in $K$.  However, in our construction of
$K_+$ via the six Maps, the operations involving the observable fields
consist of multiplication of polynomials in the quotient space
discussed around \refeq{Ncaldecomp}.  Therefore no $\sigma$ term,
i.e. no term in $\pi_a\Ncal$, can be created in $K_+$ if it is not
present initially.
\end{proof}

\setcounter{section}{0}
\renewcommand{\thesection}{\Alph{section}}


\section{Proof of Proposition~\ref{prop:Wcal-completeness}}
\label{sec:Banach}

In this section we prove Proposition~\ref{prop:Wcal-completeness},
which states that several normed spaces are complete.

We fix the scale $j$ and suppress it in the notation.  Thus $\Ccal
(\volume)$ is the set of connected polymers at scale $j$.  For $X\in
\Ccal (\volume)$, let $W (X,\phi)$ be a continuous positive function
of $\phi$ in the normed space $\Phi (X^{\Box})$.  This means that $W
(X,\phi)$ is a function of $\phi$ in the space of fields in $\phi
:\volume\to\C$ but only depends on the restriction of $\phi$ to
$X^{\Box}$.  Let $\Scal (\volume)$ be the space of maps $F:\Ccal
(\volume) \rightarrow \Ncal (\volume)$ such that $F (X)$ is in
$\Ncal (X^{\Box})$ for $X$ in $\Ccal (\volume)$.  The following
proposition provides the first step in the proof of
Proposition~\ref{prop:Wcal-completeness}.

\begin{prop}
\label{prop:W-Banach} For $\volume=\Lambda$ or $\volume=\Zd$ the
space $\Scal (\volume)$ is complete in the norm
\begin{equation}\label{e:W-norm-Banach}
    \|F\|_{W}
    =
    \sup_{X \in \Ccal (\volume) , \phi \in \Phi (\volume)}
    \|F\|_{T_{\phi}}W^{-1} (X,\phi)
    .
\end{equation}
\end{prop}

\begin{proof}
We suppress the $\volume$ argument. For $X\in \Ccal$, $\phi\in
\Phi$, $g\in \Phi$, define the linear functional
\begin{equation}
    \lambda_{X,\phi,g}: \Scal \rightarrow \C
    \quad\text{by}\quad
    F \mapsto \pair{F (X),g}_{\phi} W^{-1} (X,\phi)
,
\end{equation}
with the pairing on the right-hand side defined in
\cite[Definition~\ref{norm-def:Tphi-norm}]{BS-rg-norm}.
Then
\begin{equation}
    \|F\|_{\Scal}
    =
    \sup_{X \in \Ccal,\phi\in \Phi,g \in B (\Phi)}
    \big|\lambda_{X,\phi ,g} (F)\big|
.
\end{equation}
Therefore a sequence $F_{n}$ in $\Scal$ is Cauchy if and only if
$\lambda_{X,\phi ,g} (F_{n})$ is Cauchy in $\C$, uniformly in the
parameters $( X,\phi,g)\in \Ccal \times \Phi \times B (\Phi)$. Let
$F_{n}$ be a Cauchy sequence in $\Scal$. By completeness of $\C$ the
sequence $\lambda_{X,\phi ,g} (F_{n})$ has a limit $F_{X,\phi ,g}$ in
$\C$. Since $F_{n}$ is uniformly Cauchy the convergence is uniform in
the parameters. Therefore, to prove that $\Scal$ is complete, it
suffices to prove that there exists $F$ in $\Scal$
such that $\lambda_{X,\phi ,g} (F) = F_{X,\phi ,g}$ for all values of
the parameters.
Thus we fix $X \in \Ccal$, assume that $F_{n}$ is a sequence in $\Ncal
(X^{\Box})$, and it suffices to prove that there exists $F\in \Ncal
(X^{\Box})$ such that $\lambda_{X,g,\phi}(F_{n})\rightarrow
\lambda_{X,g,\phi}(F)$.

It suffices to restrict the test function $g$ to a small class of test
functions, as follows.  Let $z$ be a sequence in $\Lambdabold^{*}$ and
let $x$ and $y$ be the boson and fermion subsequences of $z$.  Let the
length $p (x)$ of the sequence $x$ be at most $p_{\Ncal}+2$, where the
$2$ allows for observables.  We define a test function $\delta_z\in
\Phi$ by setting $\delta_{z} (z')=1$ when $z' = z$ and $\delta_{z}
(z')=0$ otherwise, for $z' \in \Lambdabold^{*}$.  By the definition of
$\Phi (X^{\Box})$ all elements of $\Phi (X^{\Box})$ are finite linear
combinations of these special test functions.  Thus it suffices to
prove that $\lambda_{X,\phi,\delta_z}(F_{n})\rightarrow
\lambda_{X,\phi,\delta_z}(F)$ since this gives the corresponding
results for all $g$ in $\Phi (X^{\Box})$ and therefore also for all
$g$ in $\Phi$.

Since $\delta_z$ can be normalised to be in $B (\Phi)$,
$\lambda_{X,\phi ,\delta_z} (F_{n})$ is Cauchy in $\C$, uniformly in
$\phi \in \Phi$.  By the definition of the pairing,
\begin{equation}
\lbeq{diffpair}
    \pair{F_{n},\delta_z}_{\phi}
    =
    F_{n,x,y}
    =
    \left(\prod_{i=1}^{p (x)} \frac{\partial}{\partial \phi_{x_{i}}} \right)
    F_{n,y} (\phi)
\end{equation}
is a partial derivative of $F_{n,y}$ with respect to $\phi$. By the
definition of $\Scal$ this partial derivative is continuous in $\phi$
and the pairing is well defined on the equivalence classes $\phi \in
\Phi (X^{\Box})$ and $g \in \Phi (X^{\Box})$. By hypothesis, $W
(X,\phi)$ is bounded below uniformly on compact subsets of $\Phi
(X^{\Box})$. Therefore the uniform convergence of
$\lambda_{X,\phi,\delta_z} (F_{n})$ implies that the partial
derivative $F_{n,x,y}$ converges uniformly in $\phi $ for $\phi$ in
compact subsets of $\Phi (X^{\Box})$. By the continuity of $F_{n,x,y}$
as a function of $\phi$ the limit of $F_{n,x,y}$ is continuous in
$\phi$.
By integration we find that the derivatives of the limit are the
limits of the derivatives. Therefore there exists $F_{y} \in \Ncal
(X^{\Box})$ such that $F_{n,x,y} (\phi)$ converges to $F_{x,y} (\phi)$
for all $\phi$. Letting $F = \sum_y \frac{1}{y!}F_y(\phi)\psi^y$ and
noting that this sum over $y$ is a finite sum because $X$ is a finite
set, we have $\lambda_{X,\phi,\delta_z}(F_{n})\rightarrow
\lambda_{X,\phi,\delta_z}(F)$, and the proof is complete.
\end{proof}

As in Section~\ref{sec:norms}, we denote by $\Ical (\volume)$ the set
of elements of $\Ncal$ whose $T_{0}$ semi-norm is zero. Define
$\Scal(T_0)$ to be the space of maps $F:\Ccal (\volume) \rightarrow
\Ncal (\volume)/\Ical(\volume)$ such that $F (X)\in \Ncal
(X^{\Box})/\Ical(\volume)$. Since we have factored out the ideal
$\Ical(\volume)$, the $T_{0}$ semi-norm becomes a norm on this space.

\begin{prop}
\label{prop:T-Banach} For $\volume=\Lambda$ or $\volume=\Zd$, the space
$\Scal (T_{0})$ is complete.
\end{prop}

\begin{proof}
Given  $F\in\Ncal$, we replace $\phi$ and $\psi$ by
$t\phi$ and $t\psi$ and construct a polynomial $T \in \Ncal$ of degree
$p_{\Ncal}$ by making a Taylor expansion in powers of $t$ to order
$p_{\Ncal}$ and setting $t=1$.  Then derivatives of $T$ at $\phi =0$
match derivatives of $F$ up to and including order $p_{\Ncal}$.
Therefore $F-T\in\Ical (\volume)$, and the map $F\mapsto
T$ identifies $\Ncal (\volume)/ \Ical (\volume)$ with polynomials of
degree $p_{\Ncal}$.  Then, for all $X$, $\Ncal (X) /\Ical (\volume)$
is a finite dimensional space and therefore $\Scal(T_0)$ is complete
in $T_{0}$ norm.
\end{proof}

\begin{prop}
\label{prop:closed-subspace} For either of the two choices $\volume =
\Zd$ or $\volume = \Lambda$, the spaces $\Fcal(G)$, $\Fcal(\tilde G)$,
are closed subspaces of $\Scal$ and are complete. Likewise,
$\Fcal(T_{0})$ is a closed subspace of $\Scal (T_{0})$ and is also
complete.
\end{prop}

\begin{proof}
The spaces $\Fcal (G)$ and $\Fcal (\tilde{G})$ are obtained when $W
(X,\phi)$ is chosen as in \eqref{e:Wdef}.  According to the
definitions of the regulators in
\cite[\eqref{IE-e:GPhidef},\eqref{IE-e:9Gdef}]{BS-rg-IE}, $W (X,\phi)$
is positive and continuous in $\phi$.  Therefore, by
Proposition~\ref{prop:W-Banach}, with either choice of $W$, the space
$\Scal$ is complete.  Also, according to
Proposition~\ref{prop:T-Banach}, $\Scal (T_{0})$ is
complete. Therefore it is sufficient to prove that $\Fcal(G)$,
$\Fcal(\tilde G)$ and $\Fcal(T_{0})$ are closed subspaces.  As
discussed in Section~\ref{sec:Knorms}, elements of $\Fcal (G)$, $\Fcal
(\tilde{G})$ and $\Fcal(T_{0})$ must satisfy the symmetry and field
locality conditions of Definition~\ref{def:Kspace}.  These conditions
define closed subspaces.

Therefore, it only remains to prove for the cases $\Fcal (G)$, $\Fcal
(\tilde{G})$ that the condition of vanishing at weighted infinity
defines a closed subspace of $\Scal$.  For this let
$F_{1},F_{2},\dots$ be a sequence of elements of $\Scal$ that vanish
at weighted infinity and are such that the sequence converges in
$\Scal$ to a limit $F$.  We must prove $F$ vanishes at weighted
infinity. Let $\epsilon >0$ and let $X$ be a polymer.  By definition,
there exists $N$ such that $\|F (X) - F_{N} (X)\|_{T_{\phi}}W^{-1}
(X,\phi) < \epsilon$ uniformly in $\phi$. Therefore
\begin{align}
    \|F (X)\|_{T_{\phi}}W^{-1} (X,\phi)
 &<
    \epsilon + \|F_{N}\|_{T_{\phi}}W^{-1} (X,\phi)
.
\end{align}
Since $F_{N} \in \Scal$, it follows that
\begin{equation}
    \limsup_{
    \|\phi\|_{\Phi(X)}
    \rightarrow \infty}
    \|F (X)\|_{T_{\phi}}W^{-1} (X,\phi)
    <
    \epsilon
,
\end{equation}
and since this holds for all $\epsilon$, $F$ vanishes at weighted
infinity, as was to be proved.
\end{proof}

\begin{proof}[Proof of Proposition~\ref{prop:Wcal-completeness}] A
Cauchy sequence $F_{n}$ in $\Wcal (\volume)$ is Cauchy in each of the
$\Fcal (G)$ and $\Fcal (\tilde{G})$ norms. Therefore it has limits
$F_{G}$ and $F_{\tilde{G}}$ in the $\Fcal (G)$ norm and the $\Fcal
(\tilde{G})$ norm. Both norms imply convergence pointwise in $X,\phi$
so $F_{G} = F_{\tilde{G}}$ and therefore $F_{n}$ is convergent in
$\Wcal (\volume)$.
\end{proof}

\section{Two properties of the expectation}
\label{sec:massconty}

In this section, we prove that the expectation is continuous in
the mass, and that the expectation preserves the property of vanishing at infinity.
We begin with the continuity statement.

\subsection{Mass continuity of the expectation}
\label{sec:masscont}

\subsubsection{Statement of continuity}

We consider the continuity properties of the expectation as a function
of the covariance, and of the mass which defines the covariance.
There are two fixed scales, $j$ and $j+1$, and the scale advances in
norms when the expectation is taken.  We omit the scale when it is $j$
and write $+$ to indicate scale $j+1$.  The covariance $C$ is
always considered to be a test function with two arguments and,
furthermore, is assumed to be in the unit ball $B_1(\Phi_+)$ of the
space $\Phi_+$ of test functions.
Let $X \in \Ccal$ be a connected polymer $X \in \Ccal$.  Recall
from \eqref{e:np1}--\eqref{e:np2} that two norm pairs $\|\cdot\|_j$,
$\|\cdot\|_{j+1}$ are defined on $\Ncal(X^\Box)$.
We write $\Xcal$ for either of the normed spaces defined by the
two choices of $\|\cdot\|_j$ and for each of these choices let
$\Xcal_+$ be the normed spaces defined by the accompanying choice of
$\|\cdot\|_{j+1}$.
The main continuity result is the following proposition, whose proof
is given in the remainder of Section~\ref{sec:masscont}.

\begin{prop}
\label{prop:continuity2}
For $F\in \Xcal$, the map $C \mapsto
\Ex_{C}\theta F$ from $B_{1} (\Phi_{+})$ to $\Xcal_{+}$ is continuous.
\end{prop}

Now we choose the covariance $C$ to be one of the $m^2$-dependent
covariances $C=C_{j+1}$ for $j<N(\volume)$, or $C=C_{N,N}$ for
$j+1=N(\Lambda)$, which arise in the finite-range decomposition of the
covariance $(-\Delta +m^2)^{-1}$ described in
\cite[Section~\ref{IE-sec:cd}]{BS-rg-IE}.
Proposition~\ref{prop:continuity2} then implies the continuity of the
expectation as a function of the mass $m^2$.

\begin{prop}
\label{prop:masscont} For $F\in \Xcal$, the map $m^2 \mapsto
\Ex_{C}\theta F$ from $\Xcal$
to $\Xcal_{+}$ is continuous.
\end{prop}

\begin{proof}
By Proposition~\ref{prop:continuity2}, it suffices to show that $m^2 \mapsto C$ is
a continuous function from $I_j$ to $B_1(\Phi_+)$.
This is a consequence of
\cite[Proposition~\ref{pt-prop:Cdecomp}(b)]{BBS-rg-pt}.
(In fact, \cite[Proposition~\ref{pt-prop:Cdecomp}]{BBS-rg-pt}
does not directly address mass continuity in the $\Phi_{j+1}$ norm, but it
does when augmented with the estimate \cite[(1.15)]{Baue13a}.)
\end{proof}

\subsubsection{Reduction to dense subset}

The following lemma is a standard result in functional
analysis.  We omit the proof, which is an $\epsilon/3$ argument.

\begin{lemma}
\label{lem:TC}
Let $\Bcal$ and $\Bcal_+$ be Banach spaces.
Suppose that the sequence of linear operators
$T_{n}:\Bcal \rightarrow \Bcal_{+}$
is uniformly bounded in operator norm, and suppose that
$T:\Bcal \rightarrow \Bcal_{+}$ is a bounded linear operator. If
$T_{n} F \rightarrow T F$ for all $F$ in a dense subset of $\Bcal$,
then $T_{n} F \rightarrow T F$ for all $F\in \Bcal$.
\end{lemma}

The dense subset we use is the subspace
$\Xcal_{0}$ of $\Xcal$ whose elements are
compactly supported in $\phi$, namely, given $X \in \Pcal$,
\begin{equation}
\lbeq{Scal0def}
    \Xcal_{0} = \{F\in\Xcal:
    \exists R \;\text{such that}\;
    \|F\|_{T_{\phi}} =0
    \;\text{if}\; \|\phi\|_{\Phi(X^\Box)} \ge R\}.
\end{equation}

\begin{lemma}
\label{lem:Scompact}
The set $\Xcal_{0}$ is dense in $\Xcal$.
\end{lemma}

\begin{proof}
Let $\chi:\Cbold \rightarrow
\Rbold$ be a smooth non-negative function of compact support that is
bounded by 1, equals 1 on a neighbourhood of the unit disk and has
support inside the disk of radius $2$. For $R \ge 1$, $x\in Y$, and
$\phi \in \Phi$, let $\chi_{R,x} (\phi) = \chi
(\phi_{x}/R)$.
Let $F \in \Xcal$ and let $\epsilon >0$.  We will show that $R$ can be chosen
so that $\|F-\chi_RF\|_\Xcal < \epsilon$, and this suffices since $\chi_RF\in \Xcal_0$.

By the definition of the $T_{\phi}$ norm  (see
\cite[Definition~\ref{norm-def:Tphi-norm}]{BS-rg-norm}),
for $R$ large depending on $\h$,
\begin{gather}
\lbeq{chiRbd}
    \| \chi_{R,x} \|_{T_{\phi}}
    \le
    \sum_{p=0}^{p_\Ncal} \frac{1}{p!} \chi^{(p)} (\phi_{x}/R )
    (\h/R)^p
\le
    1+O \left(\h/R\right)
\le
    2 .
\end{gather}
Also, since $\|1-\chi_{R,x}\|_{T_\phi}=0$ for $|\phi_{x}| < R$,
\begin{gather}
\lbeq{1-chi}
    \|1 - \chi_{R,x} \|_{T_{\phi}}
    =
    \1_{|\phi_{x}| \ge R}\,
    \|1 - \chi_{R,x} \|_{T_{\phi}}
\le
    3\1_{\|\phi\|_{\Phi} \ge R}
.
\end{gather}
Let $\chi_{R} (\phi) = \prod_{x\in Y} \chi_{R,x} (\phi)$.  Let $\succ$
be any total ordering of the points in $Y=X^\Box$. We apply
\refeq{chiRbd}--\refeq{1-chi} and the product property of the
$T_{\phi}$ semi-norm to obtain
\begin{align}
    \|1 - \chi_{R}\|_{T_{\phi}}
    &\le
    \sum_{y \in Y}
    \|1 - \chi_{R,y} \|_{T_{\phi}}
    \prod_{x\in Y, x \succ y}
    \| \chi_{R,x} \|_{T_{\phi}}
    \le
    \1_{\|\phi\|_{\Phi (Y)}\ge R}\,
    3|Y| 2^{|Y|}
.
\end{align}
By hypothesis, $\|F \|_{T_{\phi}}\Gcal^{-1} (X,\phi) \to 0$ as
$\|\phi\|_{\Phi (Y)}\rightarrow \infty$.  By the product property, for
$R$ sufficiently large depending on $\epsilon ,\h, Y$, this gives
\begin{gather}
    \|F - \chi_{R} F\|_{\Xcal}
    \le
    3|Y| 2^{|Y|}
    \sup_{\phi:\|\phi\|_{\Phi (Y)}\ge R}
    \|F \|_{T_{\phi}}\Gcal^{-1} (X,\phi)
    <
    \epsilon
    .
\end{gather}
This completes the proof.
\end{proof}

\subsubsection{Continuity of expectation in covariance}

Before proving Proposition~\ref{prop:continuity2}, we first prove the
following lemma concerning a norm equivalence.  We write $Y=X^\Box$
below, to simplify the notation. The normed space $\Xcal$ is
defined above Proposition~\ref{prop:continuity2}.

\begin{lemma}\label{lem:Cp}
Let $S\subset \C^{Y}$ and let $F_{n}\in \Xcal$ for $n\in\N$. Then
$F_{n}$ is convergent in $T_{\phi}$ semi-norm uniformly in $\phi\in S$
if and only if $F_{n,y}$ and its derivatives up to order $p_{\Ncal}$
converge uniformly in $\phi\in S$ for the finitely many possible
sequences $y$ arising from $Y$.  This is the same as $F_{n,y}$ being
convergent in the $C^{p_{\Ncal}} (S)$ topology for each such $y$.
\end{lemma}

\begin{proof}
The proof is closely related to that of Proposition~\ref{prop:W-Banach}.
Given $\phi\in S$ and $g\in \Phi$, we define the linear
functional $\lambda_{\phi,g}: \Xcal \to \C$ by $F \mapsto \pair{F,g}_{\phi}$.
Then
\begin{equation}
    \sup_{\phi \in S}\|F\|_{T_{\phi}}
    =
    \sup_{\phi \in S,g \in B (\Phi)}
    \big|\lambda_{\phi ,g} (F)\big|
.
\end{equation}
Therefore the sequence $F_{n}$ in $\Scal$ is convergent in $T_{\phi}$
uniformly in $\phi\in S$ if and only if $\lambda_{\phi ,g} (F_{n})$
is convergent in $\C$, uniformly in the parameters $(\phi,g)\in S
\times B (\Phi)$.

Let $z \in \Lambdabold^{*}$.  We define a test function $\delta_z\in
\Phi$ by setting $\delta_{z} (z')=1$ when $z' = z$ and $\delta_{z}
(z')=0$ otherwise, for $z' \in \Lambdabold^{*}$.  All test functions
are finite linear combinations of these special test functions and
they comprise a finite basis for the test functions in $\Phi (Y)$.
Therefore $\lambda_{\phi ,g} (F_{n})$ is convergent uniformly in
$(\phi,g)$ if and only if $\lambda_{\phi ,g} (F_{n})$ is convergent
uniformly in $\phi \in S$ when $g$ is a basis test function.
Exactly as in \refeq{diffpair},
$\pair{F_{n},\delta_{x,y}}_{\phi}$ is a partial
derivative of $F_{n,y}$ with respect to $\phi$. Therefore,
uniform convergence of $\lambda_{\phi,g} (F_{n})$ is equivalent to
uniform convergence of partial derivatives, as claimed.
\end{proof}

\begin{proof}[Proof of Proposition~\ref{prop:continuity2}] By
Proposition~\ref{prop:ip}, the linear map $T_C: F \mapsto
\Ex_{C}\theta F$ from $\Xcal$ to $\Xcal_{+}$ is a bounded operator.
The proof of Proposition~\ref{prop:ip} shows that $\|T_C\|$ is bounded
uniformly in $C\in B_{1} (\Phi_{+})$.  Let $C_n$ be a sequence of
covariances in $B_1(\Phi)$ that converges to $C$, and let $T_{C_n}: F
\mapsto \Ex_{C_n}\theta F$.  By
Lemmas~\ref{lem:TC}--\ref{lem:Scompact}, it suffices to show that
$T_{C_n}F \to T_CF$ for all $F\in \Xcal_0$, with $\Xcal_0$ the dense
subset of $\Xcal$ defined by \refeq{Scal0def}.  An element
$F\in\Xcal_{0}$ has the form $F = \sum_{y}\frac{1}{y!}F_{y} \psi^{y}$,
and this is a finite sum because there are finitely many fermion
fields with labels in $Y$. Therefore it suffices to show that, for
each finite sequence $y$, $\Ex_{C_n}\theta F_{y}\psi^{y}$ converges to
$\Ex_{C}\theta F_{y}\psi^{y}$.  By
\cite[\eqref{norm-e:ECbfssz}]{BS-rg-norm}, $\Ex_{C}\theta F =
(\Ex_{C}\theta F_{y}) (\Ex_{C}\theta \psi^{y})$, so it suffices to
prove that $\Ex_{C_n}\theta F_{y} \to \Ex_{C}\theta F_{y}$ and
$\Ex_{C_n}\theta \psi^{y} \to \Ex_{C}\theta \psi^{y}$.

Since $F_{y}\psi^{y}\in\Ncal (Y)$ we can regard $C$ as an element of
the finite-dimensional vector space $\Phi_{+} (Y)$, on which all norms
are equivalent.  In particular a sequence $C_{n}$ of covariances
converges in $\Phi_{+} (Y)$ if and only if the sequence converges in
the sense of convergence of matrix elements of $Y \times Y$ matrices.
By definition, $\Ex_{C}\theta \psi^{y}$ is a polynomial in $\psi_{x}$
for $x \in Y$, with coefficients that are polynomials in matrix
elements of $C$.  The space of such polynomials is finite-dimensional,
so the map $C \mapsto \Ex_{C}\theta \psi^{y}$ is continuous.

It remains to prove that $C \mapsto \Ex_{C} \theta F_{y}$ is
continuous as a map from a domain of fixed-size matrices to
$\Xcal_{+}$, for $F_{y}$ a compactly supported $p_{\Ncal}$ times
continuously differentiable function of $\phi \in \C^{Y}$.  However,
the map $\Ex_{C} \theta$ represents convolution by a Gaussian, and
from this it can be seen that $\Ex_{C} \theta F_{y}$ is jointly
continuous in $\phi$ and $C$.  Derivatives commute with convolution so
the same is true for derivatives.  Therefore, $\Ex_{C} \theta F_{y}$
is continuous in $C$, as a map into $C^{p_{\Ncal}} (\C^{Y})$.  By
Lemma~\ref{lem:Cp}, it is therefore also continuous in the topology of
convergence in $T_{\phi}$ norm uniformly in $\phi$.  This is a
stronger topology than the norm on $\Xcal_{+}$, so the proof is
complete.
\end{proof}

\subsection{Expectation and vanishing at weighted infinity}

We now prove that the property of vanishing at weighted
infinity is preserved by the expectation.  Since we only take
expectations in finite volume we consider the vector space
$\Kcal(\Lambda)$ with the $\Fcal(\Gcal)$ norm defined in
\eqref{e:KFcal} in terms of the weight
\begin{equation}
    \label{e:Wdef-copy}
    W (X,\phi)
=
    \rhogen^{f (a ,X)}\Gcal(X,\phi)
\end{equation}
of \eqref{e:Wdef}, with $\rho$ given by \refeq{rhoFcaldef}.
There is also the space $\Fcal_+(\Gcal_+)$ defined in terms
of the weight
\begin{equation}
    \label{e:Wplusdef}
    W_+ (X,\phi)
=
    \rhogen_+^{f_+ (a_+ ,X)}\Gcal_{+}(X,\phi),
\end{equation}
where we assume that $a<a_+$.  Our norm pairs
\refeq{np1}--\refeq{np2}
are such that $\Gcal=G_j$ is paired
with $\Gcal_+=T_{0,j+1}$, and $\Gcal_j=\tilde G_j$ is paired with
$\Gcal_+=\tilde G_{j+1}^{\Gtilp}$.
As a first step, we prove the following lemma.

\begin{lemma}
\label{lem:W}
Let
$f_{R}=\1_{\{\|\xi\|_{\Phi (X)}\le R\}}$.  Then for $X\in\Pcal$ and $\phi\in \C^{X^\Box}$,
\begin{equation}
\lbeq{W1}
    \Ex_{+} W (X,\phi + \xi)
    \le
    W_{+}(X,\phi),
\end{equation}
\begin{equation}
\lbeq{W2}
    \lim_{R\rightarrow \infty}
    \sup_{\phi \in \Phi}
    \frac{1}{W_{+}(X,\phi)}
    \Ex_{+} \left[ (1-f_{R}(\xi)) W (X,\phi + \xi) \right]
    =
    0
.
\end{equation}
\end{lemma}

\begin{proof}
By definition of the regulators,
and by the inequality $\|\phi+\xi\|^2
\le 2(\|\phi\|^2+\|\xi\|^2)$,
\begin{equation}
    \Gcal (X,\phi + \xi) \le \Gcal^2(X,\phi) \Gcal^2(X,\xi)
    \le \Gcal^2(X,\phi) G^2(X,\xi).
\end{equation}
Using \cite[Lemma~\ref{IE-lem:mart}]{BS-rg-IE},
we obtain
\begin{equation}
    \label{e:Wmart1}
    \Gcal (X,\phi + \xi)
    \le
        \Gcal_{+} (X,\phi) G^{2}(X,\xi)
.
\end{equation}
By \cite[\eqref{IE-e:EG}]{BS-rg-IE},
\begin{equation}
    \label{e:Gintegrable}
    \Ex_{+} G^{2}(X,\xi) \le 2^{|X|_j}
,
\end{equation}
and hence
\begin{equation}
    \label{e:EW}
    \Ex_{+} \Gcal (X,\phi + \xi)
\le
    2^{|X|_{j}}\Gcal_{+} (X,\phi)
.
\end{equation}
By \eqref{e:Wdef-copy}, \eqref{e:EW}, and the fact that
$\rho^{f(a,X)} 2^{|X|_j} \le \rho_+^{f_+(a_+,X)}$ by definition (for small $\ggen$),
\begin{equation}
    \Ex_{+} W (X,\phi + \xi)
\le
    \rhogen^{f (\amain ,X)} 2^{|X|_{j}}\Gcal_{+} (X,\phi)
\le
    W_{+} (X,\phi)
.
\end{equation}
This completes the proof of \refeq{W1}.

To prove \refeq{W2}, we repeat the steps
in the proof of \refeq{W1} but with the factor $1-f_{R}$ included.  This
factor then appears under the expectation in \eqref{e:Gintegrable}, and
\refeq{W2} then follows by dominated convergence.
\end{proof}

A second ingredient we need is that for a function $f=f(\xi)$ of the
fluctuation field $\xi$,
\begin{equation}
\lbeq{EfFbd}
    \|\Ex_{+}fF(X)\|_{T_{\phi,+}(\h_+)}
    \le
     \Ex_{+}\left[ |f(\xi)|\, \|F(X)\|_{T_{\phi+\xi}(\h)}  \right].
\end{equation}
This follows from a slight adaptation of
\cite[\eqref{IE-e:ip1}--\eqref{IE-e:ip1.1}]{BS-rg-IE},
with the improved version of \cite[Proposition~\ref{norm-prop:EK}]{BS-rg-norm}
provided by
\cite[\eqref{norm-e:EKzh}]{BS-rg-norm} to include the function $h=f_R$.
These give
the inequality
\begin{equation}
    \|\Ex_{+}fF(X)\|_{T_\phi(\h/2)}
    \le
    \Ex_{+}\left[ |f(\xi)|\, \|F(X)\|_{T_{\phi+\xi}(\h)}  \right].
\end{equation}
With the monotonicity in $\h$ of the $T_\phi$ norm provided by
\cite[Lemma~\ref{IE-lem:Imono}]{BS-rg-IE},
\refeq{EfFbd} then follows from $ \h_+ \le \frac 12 \h$.
This application of \cite[Lemma~\ref{IE-lem:Imono}]{BS-rg-IE} requires that
$F \in \Ncal$ is gauge invariant and such
that $\pi_{ab}F=0$ when $j<j_{ab}$, so we make this assumption throughout the
rest of the section without further mention.

\begin{prop}\label{prop:vanish-at-infinity}
Suppose $F\in \Kcal$ vanishes at $W$-weighted infinity.  Then $\Ex_{+}\theta F$ vanishes at
$W_+$-weighted infinity.
\end{prop}

\begin{proof}
Let $C_{F} = \|F \|_{W}$ and let $X$ be a polymer.  Given $R>0$, let
$f_{R}=\1_{\{\|\xi\|_{\Phi (X)}\le R\}}$.  We write $I_{R} = \Ex_{+}
f_{R}\theta F (X)$ and $I_{\not R} = \Ex_{+} (1-f_{R})\theta F (X)$ so
that
\begin{equation}
    \label{e:F-decompose}
    \Ex_{+}\theta F (X) = I_{R} + I_{\not R}.
\end{equation}
By \refeq{W2}, we can
choose $R$ large such that
\begin{equation}
    \label{e:notR}
    \Ex_{+} \left[ |1-f_{R}(\xi)|\, W (X,\phi + \xi) \right]
\le
    C_{F}^{-1}\epsilon W_{+}(X,\phi)
.
\end{equation}
Therefore, by \refeq{EfFbd},
\begin{align}
    \|I_{\not R}\|_{T_{\phi} (\h_{+})}
&\le
    \Ex_{+}  |1-f_{R}(\xi)|\, \|F (X)\|_{\Ttimes_{\phi+\xi}(\h)}
\nnb &
\le
    \Ex_{+}  |1-f_{R}(\xi)|\, C_{F} W (X,\phi +\xi)
\le
    \epsilon W_{+}(X,\phi)
,
\end{align}
and hence
\begin{gather}
    \label{e:large-xi}
    \limsup_{\|\phi\|_\Phi \rightarrow \infty}
    \frac{1}{W_{+} (X,\phi)}
    \|I_{\not R}\|_{T_{\phi} (\h_{+})}
\le
    \epsilon
    .
\end{gather}
Let
\begin{equation}
    M (\phi)
=
    \sup_{\xi}  |f_{R}(\xi)|
    \frac{1}{W (X,\phi+\xi)}
    \|F (X)\|_{T_{\phi+\xi} (\h)}
.
\end{equation}
By \refeq{EfFbd} and \refeq{W1},
\begin{gather}
    \|I_{R}\|_{T_{\phi} (\h)}
\le
        \Ex_{+} |f_{R}(\xi)|\, \|F (X)\|_{\Ttimes_{\phi+\xi}(\h)}
\le
    M(\phi)
    \Ex_{+} W(\phi+\xi)
\le
    M(\phi)
    W_{+} (X,\phi)
.
\end{gather}
When $\|\xi\|_{\Phi}\le R$, if $\|\phi\|_\Phi \to \infty$ then also
$\|\phi+\xi\|_\Phi \to\infty$.  Since $F$ vanishes at weighted
infinity, it follows that $M (\phi) \rightarrow 0$ as
$\|\phi\|_\Phi \rightarrow \infty$, and hence
\begin{equation}
    \lim_{\|\phi\|_\Phi \rightarrow \infty}
    \|I_{R}\|_{T_{\phi} (\h)}
    W_{+}^{-1} (X,\phi)
=
    0
\end{equation}
With \eqref{e:large-xi} and \eqref{e:F-decompose}, this
concludes the proof.
\end{proof}

\section{Polymer geometry}
\label{sec:gl}

We now prove some geometric lemmas used in our analysis.  They concern
$f_{j} (z,\amain ,X)$, which
is defined for $z \ge 0$ and $a\in (0,2^{-d}]$ and $X
\in \Pcal_{j}$ by \eqref{e:fdef}.
We begin with the following elementary but useful observation.
We claim that for  $X \in \Pcal$, $0 \le a' < a$, $C \ge 1$,
and for $\epsilon$ sufficiently small,
\begin{align}
\label{e:Khatcombiii}
    C^{|X|_{j}}\epsilon^{f (z,a , X)}
    &
    \le
    C^{2^d} \epsilon^{f (z,a',X)}
    .
\end{align}
This follows from
\begin{align}
    C^{|X|_{j}} \epsilon^{f (z,a ,X)}
    &
    \le
    \epsilon^z C^{2^d} (C\epsilon^a)^{(|X|_{j}-2^d)_+}
    \le
    \epsilon^z C^{2^d} (\epsilon^{a'})^{(|X|_{j}-2^d)_+}
    =
    C^{2^d} \epsilon^{f (z,a',X)},
\end{align}
when $\epsilon$ is small enough that $C\epsilon^a \le \epsilon^{a'}$.

The following is a subadditivity property of $f_j$.
Fix any $a \in [0,2^{-d}z]$, and let
$X = \bigcup_i X_{i}$ be  a nonempty union  of disjoint sets
$X_{1},\dotsc X_{n} \in \Pcal_{j}$.  Then
\begin{equation}
\label{e:fsubadditive}
    f_{j} (z, a ,X)
    \le
    \sum_{i} f_{j} (z, a ,X_{i})
    .
\end{equation}
To prove this, we observe that
for $|X|_{j} \le 2^{d}$ the inequality reduces to $z \le
\sum_{i}z$, and otherwise the left-hand side equals
\begin{equation}
    z - a 2^{d} + \sum_{i} a |X_{i}|_{j}
    \le
    \sum_{i}
    \left(
    z - a 2^{d} + a |X_{i}|_{j}
    \right)
    \le
    \sum_{i}
    f_{j} (z, a ,X_{i})
    .
\end{equation}

For $F,G \in \Kspace_{j}$ it is straightforward to check that $F\circ
G$ is in $\Kspace_{j}$.
We use the following estimate for the circle product several times.

\begin{lemma}
\label{lem:FcircG} Fix $0 < \aout < \ain \in (0,2^{-d}]$ and let
$\rhoFcal$ be sufficiently small
depending on $\aout,\ain$.
Let $\epsilon
\in (0,1)$ and $\delta =2^{-2^{d}}\epsilon$.
If $F,G \in B_{\mathrm{in}}
(\delta)$ then $F \circ G \in B_{\mathrm{out}} (\epsilon)$.
\end{lemma}

\begin{proof}
By the triangle inequality, the product property of the norm, the
hypotheses and the subadditivity \eqref{e:fsubadditive} of $f_{j}$, we
have, for $Z$ connected and $\rhoFcal  \le 1$,
\begin{gather}
    \|(F \circ G ) (Z)\|_{j}
    \le
    \sum_{X \in \Pcal (Z)}
    \|F (X)\|_{j}\,
    \|G (Z\setminus X)\|_{j}
    \le
    \delta \sum_{X \in \Pcal (Z)}
    \rhoFcal^{f_{j} (\ain ,X)}
    \rhoFcal^{f_{j} (\ain ,Z\setminus X)}
\\
    \le
    \delta 2^{|Z|}
    \rhoFcal^{f_{j} (\ain ,Z)}
    \le
    \delta 2^{2^{d}}
    \rhoFcal^{f_{j} (\aout ,Z)}
    =
    \epsilon
    \rhoFcal^{f_{j} (\aout ,Z)}
.
\end{gather}
The last inequality is obtained from \eqref{e:Khatcombiii} and
requires $\rhoFcal$ to be small.
This completes the
proof.
\end{proof}

\begin{lemma}
\label{lem:Kstar-comb2}
For $z \ge 0$, $\zldg \ge a \ge 0$, and $X, Y$ disjoint with $X  \neq \varnothing$,
\begin{equation}
    f (z, a,X ) + \zldg |Y|_{j} \ge  f (z,a,X \cup Y).
\end{equation}
\end{lemma}

\begin{proof}
\emph{Case} $|X |_{j}\ge 2^{d}$. The left-hand side equals
\begin{gather}
    z + a \big(|X |_{j}-2^{d}\big) + \zldg |Y|_{j}
    \ge
    z + a \big(|X \cup Y|_{j}-2^{d}\big),
\end{gather}
which equals the desired right-hand side.

\smallskip \noindent
\emph{Case} $|X |_{j} < 2^{d}, \, |X\cup Y|_{j} \ge 2^{d}$.
Since $X$
is not empty the left-hand side equals
\begin{align}
    z + \zldg |Y|_{j}
    & >
    z + a |Y|_{j} + a \big(|X|_{j}-2^{d}\big)
    =
    z + a \big(|X\cup Y|_{j}-2^{d}\big) ,
\end{align}
which equals the desired right-hand side.

\smallskip \noindent
\emph{Remaining case} $|X\cup Y|_{j} < 2^{d}$. Since $X$ is not empty, the
left-hand side equals
\begin{gather}
    z + \zldg |Y|_{j}
    \ge
    z
    =
    z+ a \big(|X\cup Y|_{j}-2^{d}\big)_{+} ,
\end{gather}
which equals the desired right-hand side.
\end{proof}

The following lemma is stated (but not proved) above
\cite[Lemma~2]{DH92}.
A consequence of the lemma is that if $X\in
\Scal_j$ then $\overline{X} \in \Scal_{j+1}$.
The important
geometrical constant $\eta = \eta (d)>1$ used in
Lemma~\ref{lem:contraction} is introduced in
Lemma~\ref{lem:small}.

\begin{lemma}
\label{lem:small}
There is an $\eta = \eta(d)>1$ such that for all $L \geq L_0(d)=2^d+1$
and for all large
sets $X \in \Ccal_{j}$,
\begin{equation}
\label{e:etaineq}
    |X|_j
    \geq
    \eta |\overline{X}|_{j+1}.
\end{equation}
In addition, \eqref{e:etaineq} holds with
$\eta =1$ for all $X\in \Pcal_{j}$ (not necessarily connected,
and possibly small).
\end{lemma}

\proof Fix $L \geq L_0(d)=2^d+1$ (this restriction enters only in the
third paragraph of the proof).  It is clear that for any $m \geq 1$
the closure of any set of $m$ $j$-blocks contains at most $m$
$(j+1)$-blocks, and hence \eqref{e:etaineq} always holds with $\eta =1$.

Assume that $X$ is a large connected set.  Let $\Delta = \Delta(d)$
denote the maximum possible number of blocks that touch a connected
set of $2^d+1$ blocks.
We will prove \eqref{e:etaineq} by induction on $|\overline{X}|_{j+1}$,
with $\eta = 1+1/(2^d+1+2^d\Delta)$.

To begin the induction, we claim that if $|\overline{X}|_{j+1}=2^d+1$
then $|X|_j \geq 2^d+2$, and hence
\begin{equation}
    \frac{|X|_j}{|\overline{X}|_{j+1}}
    \geq
    \frac{2^d+2}{2^d+1} = 1+ \frac{1}{2^d+1} \geq \eta.
\end{equation}
To prove the claim, we proceed as follows.
The maximum possible value of $|\overline{X}|_{j+1}$ is
$|X|_j$, so we only need to rule out the case $|X|_j
= |\overline{X}|_{j+1} = 2^d +1$, which we now assume.
Let $D(X)$ be the integer part of
$L^{-j}\max_{x,y \in X}|x -y |_\infty$; this is a measure of
the diameter of $X$ counted
in number of $j$-blocks.  Then $D(X) \leq 2^d+1\leq L$.
Also, every $j$-block
in $X$ lies in a different $(j+1)$-block in $\overline X$.
However, any set of $2^d+1$ $(j+1)$-blocks contains a pair of blocks
$B_1,B_2$ that do not touch.  Therefore $D(b_1\cup b_2) >L$
for every pair of $j$-blocks $b_1\in B_1$
and $b_2\in B_2$,  so that $b_1 \cup b_2 \subset X$ is
not possible.  This contradiction proves the claim.

To advance the induction, suppose that \eqref{e:etaineq} holds when
$2^d+1 \leq |\overline{X}|_{j+1} \leq n$, and suppose now that
$|\overline{X}|_{j+1}=n+1$.
We remove from $\overline{X}$ a connected subset of $2^d+1$ blocks.
The complement of this connected subset  consists of
no more than $\Delta$ connected components (since
if there were more then
one of these components is not adjacent to the removed subset nor to
any of the at most $\Delta$ components adjacent to the removed subset,
and hence $X$ would be disconnected).
We list these components
as $\overline{X}_1, \ldots, \overline{X}_\Delta$, and choose $k \in \{0,1,\ldots,\Delta\}$
such that $|\overline{X}_i|_{j+1} \geq 2^d+1$ for $i \leq k$
and $|\overline{X}_i|_{j+1} \leq 2^d$ for $i >k$ (some of the latter components
may be empty).
Let $M =\sum_{i=1}^{k}|\overline{X}_i|_{j+1}$ and $m=
\sum_{i=k+1}^{\Delta}|\overline{X}_i|_{j+1}$.
By the induction hypothesis applied to $\overline{X}_i$
for $i \leq k$, and by \eqref{e:etaineq} with $\eta =1$ for $i > k$,
\eqalign
    \frac{|X|_j}{|\overline{X}|_{j+1}}
    & \geq
    \frac{2^d+2+ \eta M + m}
    {2^d+1+ M + m}
    = 1 +
    \frac{1+ (\eta -1) M}
    {2^d+1+ M + m}
    \nnb
    & \geq 1 +
    \frac{1+ (\eta -1) M}
    {2^d+1 + M + \Delta 2^d
    }
    = 1 +
    \frac{1+ (\eta -1) M}
    {\frac{1}{\eta -1} + M}
    =
    \eta,
\enalign
where we used our specific choice for the value of $\eta$ in the
penultimate step (note that the last equality is true no matter what the
value of $M$).
This advances the induction and completes the proof.
\qed

\begin{lemma}
\label{lem:czf}
Suppose that either $X_K$ has at least two components,
or $X_K$ has at least one component and $X_{\delta I} \neq
\varnothing$.
Let $n_{\delta I} =|X_{\delta I}|_j$
and write $X_{K_i}$ for the connected components of $X_K$.
Let $z \ge z_0 >0$.  Let $0<a \le 1$
and let $\abig \in (a ,\eta a)$.  There exist
positive $\delta ,v$, depending on $d,z_0,\abig,a$, such that
\begin{equation}
    n_{\delta I} + \sum_i f_{j} (z,a , X_{K_i})
    \ge
    v + \delta |\overline{X_{\delta I} \cup X_K}|_{j+1} +
    f_{j+1} (z, \abig , \overline{X_{\delta I} \cup X_K})
    .
\end{equation}
\end{lemma}

\begin{proof}
Suppose first that $\overline{X_{\delta I} \cup X_K} \in \Scal_{j+1}$.
Then the right-hand side is at most $v + \delta 2^{d} + z$.  In the two
cases listed at the beginning of the statement of the lemma, the
left-hand side is at least $2z$, $1+z$.
There exist
$v,\delta$ positive so that each of these is greater than $v +\delta
2^{d} + z$.

So suppose now that $\overline{X_{\delta I} \cup X_K} \not\in
\Scal_{j+1}$.  For non-empty $X_{K}$ we let $\sum_i$ denote the sum
over components $X_{K_{i}}$.
We reduce $v,\delta$, if necessary,
so that $\abig + \delta \le \eta a$ and
$ v - \abig 2^{d} \le - a 2^{d}$. By
Lemma~\ref{lem:small}, using $a \le 1$ and
\eqref{e:fsubadditive}, we have
\begin{align}
    v + \delta |\overline{X_{\delta I} \cup X_K}|_{j+1} +
    f_{j+1} (z, \abig , \overline{X_{\delta I} \cup X_K})
    & \le
    v + z - \abig 2^{d} +
    \eta a |\overline{X_{\delta I} \cup X_K}|_{j+1}
    \nnb
    & \le
    v + z - \abig 2^{d} + a n_{\delta I} +
    a |X_K|_{j}
    \nnb
    & \le
    z - a 2^{d} +
    n_{\delta I} + a |X_K|_{j}
    \nnb
    & \le
    n_{\delta I} + f_{j} (z, a , X_K)
    \nnb
    & \le
    n_{\delta I} + \sum_{i} f_{j} (z, a , X_{K_{i}})
    ,
\end{align}
as required.
\end{proof}

\begin{lemma}
\label{lem:Y0}
Let $0<z<2z'$.
Recall the definition of $\Ycal_0(W)$ below \eqref{e:K-M-new}.
There exists $c = c (d)$ such that for $\ain \in
(0,c)$, $\aout \in [0,\ain]$,
and for $(X, \{(U_B,B)\}, U_{\McalnowM}) \in \Ycal_0(W)$,
\begin{equation}
\label{e:Y0eq}
    z'|X|_j + \sum_i f_j (z,\ain ,U_{\McalnowM,i})
    \ge (\ain - \aout) |W|_j + f_j (z,\aout ,W)
    .
\end{equation}
\end{lemma}

\begin{proof}
Let $U_{\McalnowM ,i},\, i = 1,\dotsc ,n_{\McalnowM}$, be the components of
$U_{\McalnowM}$.  Let $S$ denote the number of small sets $U$ that can
contain a given block $B$.  Then $|X^{\Box}|_j\le 2^d S|X|_j$, and hence, since
$W=X^{\Box}\cup U_\McalnowM$,
\begin{equation}
    |W|_j
    \le
    2^d S|X|_j + \sum_i |U_{\McalnowM,i}|_j
    .
\end{equation}
Letting $u = \ain - \aout$ we rewrite this as
\begin{equation}
    u|W|_j
    +
    \aout |W|_j
    +
    z-\aout 2^{d}
    \le
    \ain 2^d S|X|_j
    +
    \sum_i \ain |U_{\McalnowM,i}|_j
    +
    z-\aout 2^{d}
    .
\end{equation}
The definition of $\Ycal_{0}$ excludes the case $X = \varnothing$ so
we assume $X \not = \varnothing$ and we can also assume $W \not \in
\Scal$, because $W=X^{\Box}\cup U_\McalnowM$ and $X^{\Box} \not \in \Scal$. Then $
f_j (z,\aout ,W) = z-\aout 2^{d} + \aout |W|_j$.  Therefore the left-hand side
is $u|W|_j + f_j (z,\aout ,W)$.  Let $v = z-\ain 2^{d}$ so that $v + \ain
|U|_j \le f_j (z,\ain ,U)$.  Then we can rewrite the inequality as
\begin{align}
    u|W|_j + f_j (z,\aout ,W)
    &\le
    \ain 2^d S|X|_j
    +
    \sum_i \ain |U_{\McalnowM,i}|_j
    +
    v
    +
    u 2^{d}
    \nnb
    &=
    \ain 2^d S|X|_j
    +
    \sum_i \left( v + \ain |U_{\McalnowM,i}|_j \right)
    +
    (1 - n_{\McalnowM}) v
    +
    u 2^{d}
    \nnb
    &\le
    \ain 2^d S|X|_j
    +
    \sum_i f_j (z,\ain ,U_{\McalnowM,i})
    +
    (1 - n_{\McalnowM}) v
    +
    u 2^{d}
    .
\end{align}
We choose $\ain > 0$ sufficiently small that $v = z - \ain 2^{d}\ge
0$. Decreasing $\ain$ if necessary we have $\ain 2^dS + u 2^{d} \le
z'$. If $n_{\McalnowM} \ge 1$ then we use $\ain 2^dS|X|_j + u 2^{d}
\le z' |X|_j$ to obtain the desired result.

Now we consider the case $n_{\McalnowM}=0$, which is the same as
$U_{\McalnowM} = \varnothing $.  Decreasing $\ain$ if necessary,
and using $z<2z'$, we have
$\ain 2^d S + \frac 12 z + u 2^{d} \le z'$.  The definition of
$\Ycal_{0}$ requires $|X|_j\ge 2$ when $U_{\McalnowM} = \varnothing $ so
\begin{align}
    \ain 2^d S|X|_j
    +
    (1-n_{\McalnowM})v
    +
    u 2^{d}
    &=
    \ain 2^d S|X|_j
    +
    v
    +
    u 2^{d}
    \nnb
    &\le
    \ain 2^d S|X|_j
    +
    z
    +
    u 2^{d}
    \nnb
    &\le
    \big(
    \ain 2^d S
    +
    {\textstyle{\frac{1}{2}}}z
    +
    u 2^{d}
    \big)|X|_j
    \nnb
    &\le
    z' |X|_j
    .
\end{align}
This completes the proof.
\end{proof}

\section{
Change of variables}
\label{sec:change-of-variable}

In this section, we prove Proposition~\ref{prop:change-of-variable-1},
which for convenience we restate here as
Proposition~\ref{prop:change-of-variable-app}.  For further
discussion of this proposition, see \cite[Section~5]{Bryd09}.
This section applies for \emph{any} norm
$\|\cdot\|$ on $\Ncal$ which obeys the product property
\cite[\eqref{IE-e:norm-fac}]{BS-rg-IE}.  We make use of
\refeq{DcalJ}--\refeq{McalnowM-def}, and in particular recall that $M$
is defined in \refeq{McalnowM-def}, for $\Kin \in \Kspace$ and $U \in
\Ccal$ and with $\bar J (U,B)=\Iin^UJ(U,B)$, by
\begin{equation}
    \label{e:McalnowM-def-bis}
    \McalnowM (U) = \Kin (U) -  \sum_{B \in \Bcal (U)} \bar J (U,B)
    .
\end{equation}

\begin{prop}
\label{prop:change-of-variable-app} Let $\ain$ be small as specified
in Lemma~\ref{lem:Y0}.  Let $\aout<\ain$ and $z' > \frac 12 z$. Let
$\rhogen$ be sufficiently small depending on the difference
$\aout-\ain$.  Let $\epsilon \in (0,1)$.  Let $J,\Iin$ be as
in \eqref{e:DcalJ} -- \eqref{e:Iin-stability}.  Suppose
that $\Kin \in \Kspace$ and $J$ satisfy
\begin{align}
\label{e:Jbd-app}
    &
    \sup_{\Dcal (J)}
    \|\Iin^{U} J (U,B) \|
    \leq
    \epsilon \rhogen^{z'},
    \\
\label{e:Mcalbd-app}
    &
    \McalnowM \in B_{\Fcal_{\rm in}} (\epsilon \rhogen^{z})
.
\end{align}
Then there exists $\Kout \in \Kspace$ such that
\begin{align}
\label{e:Kout-IK-app}
    &
    (\Iin \circ \Kout) (\Lambda)
=
    (\Iin \circ \Kin)(\Lambda),
\\
\label{e:Kout-poly-app}
    &
    \text{$\Kout$ is polynomial in $\Iin,\bar J,\Kin$},
\\
\label{e:Koutbd-KM-app}
    &
    \text{$\Kout = \McalnowM +E$
    with $E\in  B_{\Fcal_{\rm out}} (\epsilon \rhogen^{z+(\ain-\aout)/2})$}.
\end{align}
If $\Kin =0$ and
$J=0$, then $\Kout =0$.
\end{prop}

\begin{proof}
For $U \in \Ccal$ and $B \in \Bcal$, with $\bar J (U,B)=\Iin^UJ(U,B)$,
let
\begin{equation}
    \label{e:JprimeUdef}
    \bar J(U) = \sum_{B \in \Bcal(U)} \bar J(U,B)
    .
\end{equation}
For $U_J\in \Pcal$, let
\begin{equation}
    \bar J(U_J) = \prod_{U \in {\rm Comp}(U_J)} \bar J(U).
\end{equation}
By the definition of $\McalnowM$ and the component factorisation
property of $\Kin$,
\begin{align}
    (\Iin \circ \Kin)(\Lambda)
    &=
    \sum_{\Uin \in \Pcal}
    \Iin^{\Lambda \setminus \Uin} \Kin(\Uin)
    \nnb
    &=
    \sum_{\Uin \in \Pcal}
    \Iin^{\Lambda \setminus \Uin}
    \prod_{U \in {\rm Comp}(\Uin)}\big(\bar  J (U)  + \McalnowM(U)\big)
    \nnb
    & =
    \sum_{\Uin \in \Pcal}
    \Iin^{\Lambda \setminus \Uin}
    \sum_{\hat{U}_M \subset {\rm Comp}(\Uin)}
    \bar J (\Uin \setminus U_M)  \McalnowM(U_M)
    ,
\end{align}
where $U_{M}$ is the union of components in $\hat{U}_{M}$.
\begin{figure}
\begin{center}
\includegraphics[scale = 0.8]{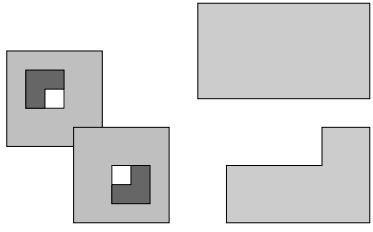}
\end{center}
\caption{\lbfg{reblockY} This figure illustrates an element of $\Ycal(W)$.  The white
squares are the blocks of $X$, and are centred in two larger squares whose union is
$X^\Box$.  Each white square $B$ is contained in a
small set $U_B$, which is itself contained in $B^{\Box}$. The
components of $U_M$ are the two shaded
components without white squares, and $W$ is
the total area.}
\end{figure}

Given $X \in \Pcal$, let $B_1,\ldots, B_n$ be a list of
the blocks in $\Bcal(X)$, and let
\begin{align}\label{}
    \Ucal(X)
    =
    &\{
    \{(U_{B_1},B_1), \ldots, (U_{B_n},B_n)\} :
    \nnb
    &
    U_{B_i} \in \Pcal, \; U_{B_i} \supset B_i,\;
    \text{
    $U_{B_i}$ does not touch $U_{B_j}$ for $i \neq j$
    }
    \}
.
\end{align}
(In particular, $\Ucal(X)$ is empty if any two blocks of $X$ touch
each other.)  Given an element of $\Ucal(X)$, we write $U_J = \cup_{B
\in \Bcal(X)}U_B$, and write $\Pcal'(\Lambda \setminus U_J)$ for the
set of polymers that do not touch $U_J$.  By interchanging the
sums over blocks $B$ in \eqref{e:JprimeUdef} and polymers $U_B$, we
obtain
\begin{align}
    (\Iin \circ \Kin)(\Lambda) &
    =
    \sum_{X\in \Pcal(\Lambda)}
    \sum_{\{(U_B,B)\}\in \Ucal(X)}
    \Big( \prod_{B \in \Bcal(X)}
    \bar J (U_B,B) \Big)
    \nnb & \quad\quad\quad \times
    \sum_{U_{\McalnowM} \in \Pcal'(\Lambda \setminus U_J)}
    \McalnowM (U_{\McalnowM})
    \Iin^{\Lambda \setminus (U_{\McalnowM} \cup U_J)}.
\end{align}
Recall the definition of the small set neighbourhood of $X$ in
Definition~\ref{def:blocks2}.
For $W=X^{\Box} \cup U_M$, we write
\begin{equation}
    \Iin^{\Lambda \setminus (U_{\McalnowM} \cup U_J)}
    = \Iin^{\Lambda \setminus W}
    \Iin^{W \setminus (U_{\McalnowM} \cup U_J)}.
\end{equation}
With this notation the claim \eqref{e:Kout-IK-app} holds with $\Kout$
given by
\begin{equation}
\label{e:Kdef-new}
    \Kout(W) =
    \sum_{(X, \{(U_B,B)\}, U_{\McalnowM}) \in \Ycal( W)}
    \left( \prod_{B \in \Bcal(X)}
    \bar J (U_B,B) \right)
    \McalnowM (U_{\McalnowM})
    \Iin^{W \setminus(U_{\McalnowM} \cup U_J)}
    \quad
    (W \in \Pcal)
    .
\end{equation}
Here $\Ycal(W)$ denotes the set of triples $(X, \{(U_B,B)\},
U_{\McalnowM})$, with $X \in \Pcal (W)$, $\{(U_B,B)\} \in \Ucal(X)$,
$U_{\McalnowM} \in \Pcal'(\Lambda\setminus U_J)$, and $X^{\Box} \cup
U_{\McalnowM} = W$. Note that the small set neighbourhood $X^\Box$
contains all possible unions $U_J$ of small sets in the summation over
the $U_B$.

Note that the above formula implies that if $J = 0$ then $\Kout
=\Kin$.  In particular, as claimed, $\Kout =0$ when $\Kin =0$ and
$J=0$.  Also, it shows that $\Kout$ is polynomial in $\Iin,\bar
J,\Kin$ as claimed in \eqref{e:Kout-poly-app}.  As an explicit
example, for the case where $W$ is a single block $B$,
\refeq{Kdef-new} gives
\begin{align}
\lbeq{Kdef-new-ex}
    \Kout
    (B) &= M(B) + \sum_{U : (U,B) \in \Dcal(J)} \bar J(U,B)
    .
\end{align}

The properties that define $\Ycal(W)$, together with the hypothesis
that $\Kin \in \Kspace$ and that $J$ obeys \refeq{J-localisation}, can
be used to verify the claim that $\Kout \in \Kspace$. In particular,
$\Kout$ obeys the factorisation property of
Definition~\ref{def:Kspace} by construction, and the field
locality property holds because we have constructed $\Kout (W)$ as a
polynomial in the local objects $\bar{J}$, $M$, $\Iin$ evaluated on
sets contained in $W$.

For the bound claimed in \eqref{e:Koutbd-KM-app}, we first show that the
contribution to \eqref{e:Kdef-new} due to triples with $|X|=1$ and
$U_{\McalnowM} = \varnothing$ vanishes.  This feature is a crucial
ingredient.  In this case, $X$ is a single block $B$, $\Kout (W)=0$
unless $W=X^{\Box}=B^{\Box}$ and thus $B$ is uniquely determined by
$W$, and by \eqref{e:kprimecondition}, the contribution to
\eqref{e:Kdef-new} is
\begin{equation}
\label{e:Jcancel}
    \sum_{U\in \Scal : U \supset B}
    J (U,B)
    \Iin^{B^{\Box}}
    =
    0
    .
\end{equation}

Let $W\in\Ccal$.
As in \eqref{e:fdef}, we write
\begin{equation}
    \label{e:fdef-bis}
    f_{j} (z, \amain ,X)
    =
    \begin{cases}
    z + f_{j} (\amain, X)
    &X\not = \varnothing
    \\
    0&X = \varnothing,
    \end{cases}
\end{equation}
for $z \ge 0$, $\amain \in (0,2^{-d}]$ and $X \in \Pcal_{j}$.  We
apply the triangle inequality, product property
\cite[\eqref{IE-e:norm-fac}]{BS-rg-IE}, and the hypotheses to
\eqref{e:Kdef-new}, to obtain
\begin{equation}
\label{e:K-M-new}
    \|\Kout(W) - \McalnowM (W)\|
    \leq
    \sum_{(X, \vec U, U_{\McalnowM}) \in \Ycal_0(W)}
    \epsilon \;
    \rhogen^{z'|X|}
    \prod_i
    \left(\rhogen^{f (z,\ain,U_{\McalnowM,i})}\right)
    \ItilstabC^{|W\setminus (U_J\cup U_{\McalnowM})|}
    ,
\end{equation}
where $\Ycal_0(W)$ imposes the constraints on $(X,\{(U_B,B)\},
U_{\McalnowM})$ required in \eqref{e:Kdef-new} with the additional
constraint that the terms with $X=\varnothing$, $U_{\McalnowM} = W$,
and with $|X|=1$, $U_{\McalnowM} = \varnothing$ are omitted (the first
omission is because $\McalnowM$ is subtracted in \eqref{e:K-M-new} and
the second is due to the cancellation in \eqref{e:Jcancel}).  Since
$\ItilstabC \geq 1$, and since $W\setminus (U_J \cup U_{\McalnowM})
\subset X^{\Box}$ and $|X^{\Box}| \leq 2^dS|X|$, the power of
$\IstabC$ above can be replaced by ${\rm const}^{|W|}$.  By
Lemma~\ref{lem:Y0}, for $(X, \{(U_B,B)\}, U_{\McalnowM}) \in
\Ycal_0(W)$, and with $2u = \ain - \aout$, we have
\begin{equation}
    z'|X| + \sum_i f (z,\ain , U_{\McalnowM,i})
    \ge 2u|W| + f (z,\aout ,W).
\end{equation}
Therefore,
\begin{equation}
    \|\Kout(W) - \McalnowM (W)\|
    \leq
    \epsilon
    \rhogen^{2u|W| + f (z,\aout ,W)}
    {\rm const}^{|W|}
    |\Ycal_0(W)|,
\end{equation}
where $|\Ycal_0(W)|$ denotes the cardinality of $\Ycal_0(W)$.  For
fixed $X$, there are at most $S^{|X|}$ possible choices of the small
sets $U_B$ specified in the definition of $\Ycal_0(W)$.  We use this,
and also relax the summation to disjoint $X$ and $U_\McalnowM$ in $W$.
Since there are at most $3^{|W|}$ ways to partition $W$ into
$X,U_{\McalnowM},W\setminus (X\cup U_{\McalnowM})$, we can absorb
$|\Ycal_0(W)|$ into ${\rm const}^{|W|}$.  Finally, we choose
$\rhogen$ sufficiently small depending on $u$ so that ${\rm
const}^{|W|} \le \frac{1}{2} \rhogen^{-u|W|}$. Then
\begin{equation}
    \label{e:Koutbd-2}
    \|\Kout(W) - \McalnowM (W)\|
    \leq
    {\textstyle{\frac 12}}
    \epsilon \,
    \rhogen^{u|W| + f (z,\aout ,W)}
    \le
    {\textstyle{\frac 12}}
    \epsilon \,
    \rhogen^{u + f (z,\aout ,W)},
\end{equation}
which implies \eqref{e:Koutbd-KM-app}.
\end{proof}

\section{Approximate isometry between finite and infinite volume}
\label{sec:iota}

The proof of Theorem~\ref{thm:1Zd} uses
Lemmas~\ref{lem:extendK}--\ref{lem:extendEuclideanK}, which are given
below.  Lemmas~\ref{lem:iota-adjoint}--\ref{lem:iotaisom} are used in
the proof of Lemmas~\ref{lem:extendK}--\ref{lem:extendK2}. In this
section $k$ is any scale, in particular it can be $j$ or $j+1$.  Let
$X\in \Pcal_{k} (\Zd)$.  A coordinate map $\iota$ from $X$ to a torus
$\Lambda$ exists for all $\Lambda$ with ${\rm diam}(\Lambda) \ge 2
{\rm diam}(X)$. Given a coordinate map $\iota:X\to \Lambda$ and a test
function $g\in \Phi (\iota X)$, we define a test function $g_{\iota}
\in \Phi (X)$ by $(g_{\iota})_{z} = g_{\iota z}$, where $\iota z$ is
defined by letting $\iota$ act on the sequence $z$ componentwise.

For the first lemma, recall the pairing in the definition of the
$T_\phi$ semi-norm in
\cite[Definition~\ref{norm-def:Tphi-norm}]{BS-rg-norm}.

\begin{lemma}\label{lem:iota-adjoint}
Let $X \in \Pcal_{k} (\Zd)$. For a coordinate map $\iota:X \to
\Lambda$, $F \in \Ncal (X)$, $g \in \Phi (\Lambda)$, and $\phi \in
\C^{\Lambda}$,
\begin{equation}
    \langle \iota F, g \rangle_\phi
    =
    \langle F , g_{\iota} \rangle_{\phi_{\iota}}
\end{equation}
\end{lemma}

\begin{proof}
Both sides are linear in $F$.  By \eqref{e:Fform}, it suffices to
consider $F=F_y \psi^y$.  Then $\iota F = \iota (F_y) \psi_{\iota}^y$.
According to \eqref{e:iotaNcal},
\begin{gather}
\lbeq{nearisometry}
    \langle \iota F, g \rangle_\phi
=
    \sum_{x \in (\iota X)^{*}}\frac{1}{x!} \Big(F_{y}(\phi_{\iota})\Big)_{x}g_{x,\iota y}
=
    \sum_{x \in X^{*}}\frac{1}{x!} F_{x,y}(\phi_{\iota}) g_{\iota x,\iota y}
    =
    \langle F , g_{\iota} \rangle_{\phi_{\iota}}
.
\end{gather}
\end{proof}

For real $t>0$ and a nonempty polymer $X\in \Pcal_k( \Zd)$, let
$X_{t}\subset \Zd$ be the smallest subset that contains $X$ and all
points in $\Zd$ that are within distance $tL^{k}$ of $X$.  The next
lemma expresses a sense in which coordinate maps are approximately
isometries as maps between spaces of test functions.  Our norms on
test functions (see \cite[Example~\ref{norm-ex:h}]{BS-rg-norm} and
\cite[\eqref{norm-e:PhiXdef}]{BS-rg-norm}) depend on a parameter $\h$.
For the next lemma, we exhibit this dependence by writing $\Phi_{k}
(\iota X,\h)$, etc.

\begin{lemma}
\label{lem:iota-Phi} Let $X \in \Pcal_{k}(\Zd)$, let $s>0$, and let
$\iota$ be a coordinate map from a polymer containing $X_{s}$
into a torus $\Lambda$.  There exists $\h_s>0$ and $c>0$ (independent
of $X,k,\iota,\Lambda$), with $1 \le\h/\h_s \le 1+cs^{-1}$, such that
for $g \in \Phi_{k} (\iota X,\h)$,
\begin{equation}
    \label{e:iota-phi-bounds}
    \|g_{\iota}\|_{\Phi_{k}(X,\h)}
\le
    \|g\|_{\Phi_{k}(\iota X,\h)}
    \le
    \|g_{\iota }\|_{\Phi_{k}(X,\h_{s})},
\end{equation}
and likewise for the $\tilde{\Phi}_{k}$ semi-norm.
\end{lemma}

\begin{proof}
We give the proof for the case $\Lambdabold = \Lambda$, because the
general case is merely an elaboration of notation.  We write $\Phi =
\Phi_{k}$.  By the definition of the norm on test functions, we see
that it is sufficient to fix an integer $p\ge 1$ and prove the lemma
for the case where the test function $g\in \Phi(\iota X,\h)$ is zero
except on sequences of length $p$.  The domain $\iota X$ is contained
in a torus $\Lambda$. By thinking of $\Lambda$ as a hypercube in a
lattice of hypercubes paving $\Zd $, we identify a test function on
$\Lambda$ with a function on $(\Zd)^{p}$ which is periodic in each
component.  The $\Phi (\iota X,\h)$ norm of $g$ is the infimum of
$\|g'\|_{\Phi (\Lambda)}$ over extensions $g'$ of $g$ to
$\Lambda^{p}$; by the identification and the definition of $g_{\iota}$
this is the infimum of $\|g'_{\iota}\|_{\Phi (\Zd,\h)}$ over
extensions $g'_{\iota}$ of $g_{\iota}$ to functions of $(\Zd)^{p}$
that are periodic in each component. The norm
$\|g_{\iota}\|_{\Phi(X,\h)}$ is the same but the extensions are not
constrained to be periodic.  Therefore
\begin{equation}
    \|g\|_{\Phi (\iota X,\h)}
\ge
    \|g_{\iota}\|_{\Phi(X,\h)}
\end{equation}
which is the lower bound claimed in \eqref{e:iota-phi-bounds}.

Let $r=\frac{s}{3}$.  By the definition of $\Phi(X,\h)$, there exists
an extension $\tilde{g}_{\iota} \in \Phi (\Zd,\h)$ of $g_{\iota}$ such
that
\begin{equation}
\lbeq{gZX}
    \|\tilde{g}_{\iota}\|_{\Phi (\Zd,\h)}
    \le
    (1+ r^{-1})
    \|g_{\iota}\|_{\Phi(X,\h)}
.
\end{equation}
By \cite[Lemma~\ref{loc-lem:gX}]{BS-rg-loc}, there exists a function
$\chi=\chi_r$, which is equal to $1$ on $X^{p}$ and $0$ on
$\Zd^{p}\setminus X_{2r}^{p}$, and a constant $c_0>0$ (independent of
$p$, $X$, and $L^j$), such that
\begin{equation}
\label{e:Phihnorm}
    \|\tilde{g}_{\iota}\chi\|_{\Phi (\Zd ,\h)}
\le
    \big(1 +  c_0 r^{-1} \big)^p\|\tilde{g}_{\iota}\|_{\Phi (\Zd,\h)}
.
\end{equation}
In combination with \eqref{e:gZX}, this gives the existence of $\h_s$
obeying the desired bound, such that
\begin{equation}
    \|\tilde{g}_{\iota}\chi\|_{\Phi (\Zd,\h)}
\le
    \|g_{\iota}\|_{\Phi(X,\h_{s})}
.
\end{equation}
By hypothesis, the domain of $\iota$ strictly contains $X_{2r}$, so we
can invert $\iota$ on $\iota X_{2r}$.  Therefore $(\tilde{g}_{\iota}
\chi )_{\iota^{-1}}$ is an extension of $g\rvert_{\iota X}$ to the
subset $\iota X_{2r}$ of $\Lambda$. Provided $L$ is large enough so
that $rL^{k}\ge p_{\Phi}$ for all $k$, the derivatives up to order
$p_{\Phi}$ in each argument of this extension are zero near the inner
boundary of $\iota X_{2r}$ so we can further extend by zero to all of
$\Lambda$. Call this extension $G$. Then, by definition of $\Phi
(\iota X,\h)$,
\begin{equation}
    \|g\|_{\Phi(\iota X,\h)}
    \le
    \|G\|_{\Phi (\Lambda,\h)}
    =
    \|\tilde{g}_{\iota}\chi\|_{\Phi (\Zd,\h)}
    \le
    \|g_{\iota}\|_{\Phi(X,\h_{s})}
\end{equation}
and this proves the upper bound of \refeq{iota-phi-bounds}.
\end{proof}

The next three lemmas express senses in which coordinate maps are
isometries, provided a small change is made in the parameter $\h$.

\begin{lemma}
\label{lem:iotaisom} Let $X \in \Pcal_{k}(\Zd)$, let $\iota : \tilde X
\to \Lambda$ be a coordinate map with $X_{N/4} \subset \tilde X\in
\Pcal_{k} (\Zd)$, and let $\phi\in\C^\Lambda$.  The induced map
$\iota: \Ncal(X)\to \Ncal(\iota X)$ is defined in \refeq{iotaNcal}.
This map is linear, and there exists $\h^-=\h^-(N) \le \h$, with $\h^-
\to \h$ as $N=N (\Lambda)\rightarrow \infty$, such that
$\|F\|_{T_{\phi_{\iota }}(\h^-)} \le \|\iota F\|_{T_{\phi}(\h)} \le
\|F\|_{T_{\phi_{\iota}}(\h)}$ for all $F \in \Ncal(X)$.
\end{lemma}

\begin{proof}
The linearity of the map is clear.  We write $\Phi =\Phi_{k}$.  Let
$F\in \Ncal(X)$ and $g\in\Phi (\iota X,\h)$.  By
Lemma~\ref{lem:iota-adjoint}, the definition of the
$T_{\phi}(\h)$ norm, and Lemma~\ref{lem:iota-Phi},
\begin{gather}
    |\pair{\iota F,g}_{\phi}|
=
    |\pair{F,g_{\iota}}_{\phi_{\iota}}|
\le
    \|F\|_{T_{\phi_{\iota}}(\h)}
    \|g_{\iota}\|_{\Phi (X,\h)}
\le
    \|F\|_{T_{\phi_{\iota}(\h)}}
    \|g\|_{\Phi (\iota X,\h)}
.
\end{gather}
Taking the supremum over $g$ with unit norm, we have $\|\iota
F\|_{T_{\phi}(\h)} \le \|F\|_{T_{\phi_{\iota}}(\h)}$ which is one of
the desired inequalities.  For the reverse estimate, we consider
$N(\Lambda) \to \infty$, assume $\diam{X} < \frac 14 \diam{\Lambda}$,
let $s= \frac 14 N(\Lambda)$ and write $h^-$ for $h_s$ of
Lemma~\ref{lem:iota-Phi}.  Note that $\h^- \uparrow \h$ as desired. By
the second bound in Lemma~\ref{lem:iota-Phi}, for a test function
$f\in\Phi (X,\h)$,
\begin{gather}
    |\pair{F,f}_{\phi_{\iota}}|
=
    |\pair{\iota F,f_{\iota^{-1}}}_{\phi}|
\le
    \|\iota F\|_{T_{\phi}(\h)}
    \|f_{\iota^{-1}}\|_{\Phi (\iota X,\h)}
\le
    \|\iota F\|_{T_{\phi}(\h)}
    \|f\|_{\Phi (X,\h^-)}
.
\end{gather}
Taking the supremum over $f$ with $\|f\|_{\Phi (X,\h^-)}=1$, we have
$\|F\|_{T_{\phi_{\iota }}(\h^-)} \le \|\iota F\|_{T_{\phi}(\h)}$,
which completes the proof.
\end{proof}

\begin{lemma}\label{lem:extendK}
Let $X\in \Pcal_{k} (\Zd)$, $F\in \Ncal (X^\Box)$, and let $\iota :
\tilde X \to \Lambda$ be a coordinate map with $X_{N/4} \subset \tilde
X\in \Pcal_{k} (\Zd)$.  The map $F \mapsto \iota F$ is a linear map
from $\Ncal (X^\Box)$ to $\Ncal (\iota X^\Box)$, and obeys
\begin{gather}
    \|\iota F \|_{\Gcal }
    \le
    \|F \|_{\Gcal }
\end{gather}
for either choice of the regulators $\Gcal =G$ or $\Gcal = \tilde G$
(recall \refeq{np1}--\refeq{np2}).
\end{lemma}

\begin{proof}
The linearity of $F \mapsto \iota F$ is clear.  By the definition of
the norm, followed by Lemma~\ref{lem:iotaisom} and then
Lemma~\ref{lem:iota-Phi} (to bound the norm in the
regulator),
\begin{gather}
    \|\iota F \|_{\Gcal}
    =
    \sup_{\phi\in\C^{\Lambda}}
    \|\iota F\|_{T_{\phi}}
    \Gcal^{-1} (\iota X,\phi)
    \le
    \sup_{\phi\in\C^{\Lambda}}
    \|F\|_{T_{\phi_{\iota}}}
    \Gcal^{-1} (\iota X,\phi)
    \nn\\
    \le
    \sup_{\phi\in\C^{\Lambda}}
    \|F\|_{T_{\phi_{\iota}}}
    \Gcal^{-1} (X,\phi_{\iota})
    \le
    \|F \|_{\Gcal}
,
\end{gather}
and the proof is complete.
\end{proof}

\begin{lemma}\label{lem:extendK2}
Let $\h^-$ be as in Lemma~\ref{lem:iotaisom}, and let $\Gtilp^+=
\Gtilp (\h/\h^-)^2$. For $X\in\Pcal_{k}(\Zd)$, $F\in \Ncal
(X^{\Box})$, $\Gtilp \in (0,1]$, and for a coordinate map $\iota :
\tilde X \to \Lambda$ with $X_{N/4} \subset \tilde X\in \Pcal_{k}
(\Zd)$,
\begin{align}
&
    \label{e:T0}
    \|F\|_{T_{0}(\h^-)}
    \le
    \|\iota F\|_{T_{0}(\h)}
\\
&
    \label{e:G-eta}
    \|F\|_{\tilde{G}^{\Gtilp^+}(\h^-)}
    \le
    \|\iota F\|_{\tilde{G}^{\Gtilp}(\h)}.
\end{align}
\end{lemma}

\begin{proof}
We only prove \eqref{e:G-eta}, because \eqref{e:T0} is a
specialisation of the same method to $\phi =0$.  Let $\phi \in
\C^{\Lambda}$.  By Lemma~\ref{lem:iota-Phi},
\begin{equation}
    \|\phi\|_{\Phi(\iota X,\h)} \le \|\phi_\iota\|_{\Phi(X,\h^-)}
    = (\h/\h^-)
    \|\phi_\iota\|_{\Phi(X,\h)},
\end{equation}
and hence, by definition of the regulator, $\tilde G^\Gtilp (\iota
X,\phi) \le \tilde G^{\Gtilp^+}(X,\phi_\iota)$.  Therefore, by
Lemma~\ref{lem:iotaisom},
\begin{gather}
    \|F\|_{T_{\phi_{\iota }}(\h^-)}
    \le
    \|\iota F\|_{T_{\phi}(\h)}
    \le
    \|\iota F\|_{T_{\phi}(\h)}
    \tilde{G}^{-\Gtilp}(\iota X,\phi)
    \tilde{G}^{\Gtilp^+}(X,\phi_{\iota})
.
\end{gather}
We divide by $\tilde{G}^{\Gtilp^+}$ and take the the supremum
over $\phi$ to complete the proof.
\end{proof}

Let $\Xcal\subset \Ccal_{k} (\volume)$, and let $F:\Xcal \rightarrow
\Ncal$ have the properties listed in Definition~\ref{def:Kspace}
except that Euclidean covariance is replaced by the restricted version
that if $X,Y \in \Xcal$ and $E$ is a Euclidean automorphism such that
$Y=EX$ then $E (F (X)) = F (EX)$. Let $W:\Ccal_{k}
(\volume)\rightarrow \R_{+}$ be a Euclidean invariant function such
that $\|F (X)\|_{k} \le W (X)$ for $X \in \Ccal (U)$.  The following
lemma, whose proof does not depend on the other lemmas in this
appendix, shows that $F$ has an extension to an element of
$\Kcal(\volume)$.

\begin{lemma}\label{lem:extendEuclideanK}
Any $F:\Xcal\to\Ncal$ as above has an extension to an element $\hat{F}
\in \Kcal (\volume)$ such that $\|\hat{F} (X)\|_{k} \le W (X)$ for $X
\in \Ccal (\volume)$. The map $F \mapsto \hat{F}$ is linear, and
if $F (X)$ satisfies \eqref{e:defnFcal} for $X$ in $\Xcal$, then the
same is true for $\hat{F} (X)$ for all polymers $X$.
\end{lemma}

\begin{proof}
For $X\in\Ccal_{k} (\volume)$ such that $X = EY$ for some automorphism
$E$ of $\volume$ and some $Y\in\Xcal$, define $\hat{F} (X)= EF (Y)$.
If there exists another $Y' \in \Xcal$ and an automorphism $E'$ such
that $X=E'Y'$, then $A=E^{-1}E'$ is a Euclidean automorphism such that
$AY'=Y$.  By hypothesis, $ E'F (Y') = EAF (Y') = EF (AY') = EF (Y) $,
so this definition of $\hat{F}$ is not dependent on choices. If there
is no pair $Y,E$ such that $X=EY$ then define $\hat{F} (X)=0$. By
construction $\hat{F}$ has the properties listed in
Definition~\ref{def:Kspace}, the extension is bounded by $W$, linear
in $F$, and preserves the property \eqref{e:defnFcal}.
\end{proof}

\section{Aspects of symmetry}
\label{sec:Qapp}

We now prove properties of the polynomial $Q$ of \refeq{Q-def},
and prove in particular that it lies in $\Qcal$ as claimed below \refeq{Q-def}.
In addition, we prove that Gaussian expectation preserves defining properties
of the space $\Kcal$ in Definition~\ref{def:Kspace};
this is used in the proof of Proposition~\ref{prop:EIK}.

We draw attention to a notational clash in this appendix:
$Q$ denotes the polynomial \refeq{Q-def}
in Lemma~\ref{lem:Qapp}, whereas $Q$ denotes the supersymmetry generator
(see \cite[Section~\ref{pt-sec:bulksym}]{BBS-rg-pt} or \cite[Section~6]{BIS09})
in Lemma~\ref{lem:Ex-sym}.

For the following lemma, we write $F|_0$ for the
\emph{constant part} of $F\in \Ncal$, which results from setting $\phi=0$ and
$\psi=0$ in $F$.

\begin{lemma}
\label{lem:LTcp}
For $F \in \Ncal$ and $X \subset \Lambda$, the constant monomial of $\LT_X F$ is $F|_0$.
\end{lemma}

\begin{proof}
It is the defining property of $\LT_X F$ in \cite[Definition~\ref{loc-def:LTsym}]{BS-rg-loc}
that $\pair{F,g}_0 = \pair{\LT_X F,g}_0$ for all test functions $g$ in the
space $\Phipol$ of polynomial test functions.  One such test function is
$g_\varnothing = 1$ (a test
function with no arguments).  By setting $g=g_\varnothing$ in the pairing, we obtain
$F|_0=(\LT_X F)|_0$.  Since $(\LT_X F)|_0$ is the constant
monomial of $\LT_X F$, the proof is complete.
\end{proof}

\begin{lemma}
\label{lem:Qapp}
The formula $Q (B) = \sum_{Y \in \Scal (\Lambda) : Y \supset B}
\LT_{Y,B} I^{-Y} K (Y)$
of \refeq{Q-def} defines an element $Q\in \Qcal$.
\end{lemma}

\begin{proof}
The operator $\LT$ preserves Euclidean covariance, gauge invariance,
and supersymmetry, according to
\cite[Proposition~\ref{loc-prop:9LTdef}]{BS-rg-loc} and
\cite[Proposition~\ref{loc-prop:Qcom}]{BS-rg-loc}.  Since $V$ and $K$
have these properties, therefore $Q$ also has them.  It is then a
consequence of \cite[Lemma~\ref{pt-lem:ss}]{BBS-rg-pt} that $Q$ lies
in $\Qcal$, once we prove that $Q$ cannot have a constant term.  But
by Lemma~\ref{lem:LTcp}, the constant monomial in $\LT_Y I^{-Y}K(Y)$
equals the constant part of $I^{-Y}K(Y)$, and this is zero by the
assumption that $K \in \Kcal$ and $V \in \Qcal$.  Therefore, the
constant monomial in $\LT_{Y,B} I^{-Y} K (Y)$ is also zero, and hence
so is the constant monomial in $Q(B)$, as desired.

To understand the role of the block $B$ in
more detail,  we first note that any choice of $B$
determines $\pi_{\varnothing}Q$, because by the Euclidean invariance of
$\pi_{\varnothing }K$ specified in Definition~\ref{def:Kspace},
\eqref{e:Q-def} assigns the same value to $\pi_{\varnothing }Q$ for
all choices of $B$.  For the observable terms, because
\eqref{e:lambda-defs} contains indicator functions, \eqref{e:Q-def}
does not determine the coupling constants in
$\pi_{\alpha}Q$ ($\alpha = \pp,\qq,\pp\qq$) unless $B$
contains $\pp$ or $\qq$. Taking all choices of $B$, \eqref{e:Q-def}
consistently determines a unique element $Q$ in $\Qcal$.
\end{proof}

\begin{lemma}
\label{lem:Ex-sym}
Let $F \in \Ncal$ and suppose that $\Ex_{j+1} \theta F$ exists.
\\
(i) If $F$ is
gauge invariant or Euclidean covariant, then so is
$\Ex_{j+1}\theta F$.
\\
(ii)
The supersymmetry generator $Q$ commutes with $\Ex_{j+1}\theta$, i.e.,
$Q\Ex_{j+1}\theta = \Ex_{j+1}\theta Q$.  In particular, if $F$ is supersymmetric
then so is $\Ex_{j+1}\theta F$.
\\
(iii)
If $F$ is
supersymmetric, then $\Ex_{j+1} F = F|_0$.  In particular, if $F$ has
zero constant part, then so does $\Ex_{j+1} \theta F$.
\end{lemma}

\begin{proof}
Throughout the proof, we write simply $\Ex$ for $\Ex_{j+1}$, and we omit
some details.  All forms in the proof have even degree.
\\
(i)
Let $A_{j+1}=C_{j+1}^{-1}$.  By definition,
\begin{equation}
    (\Ex \theta F)(\sigma,\bar\sigma,\phi,\bar\phi,\psi,\bar\psi)
    =
    \int e^{-S_{A_{j+1}}(\xi,\bar\xi,\eta,\bar\eta)}
    F(\sigma,\bar\sigma,\phi+\xi,\bar\phi+\bar\xi,\psi+\eta,\bar\psi+\bar\eta),
\end{equation}
where the action $S_{A_{j+1}}$ is Euclidean and gauge invariant.
The claim can be seen to follow from this.

\smallskip \noindent
(ii)
From \cite[\eqref{pt-e:Qcomm}--\eqref{pt-e:Qcomm2}]{BBS-rg-pt}, we
know that $\hat Q =(2\pi i)^{-1/2}Q$ commutes with $\Lcal$ and hence also with $e^{\Lcal}$.
Since the action of $\Ex \theta$ on polynomials is the
same as the action of $e^\Lcal$ by
\cite[Lemma~\ref{norm-lem:*heat-eq}]{BS-rg-norm},
this implies that $\Ex\theta Q P = Q\Ex\theta P$ for polynomials $P \in \Ncal$.

A proof for general integrable elements of $\Ncal$ can be based on the
argument of \cite[Lemma~A.2]{BI03d}, and we provide a sketch.  By definition,
$\theta F$ is a function of fluctuation and other fields, and the expectation
integrates out the fluctuation fields leaving dependence on the others.
We denote integration with respect to
the fluctuation fields by $\int_1$, with respect to the other fields by $\int_2$,
and with respect to all fields by $\int_{21}$.
Then for a form $K_{12}$ depending on both fields, and a form $K_2$ depending on
the other fields, since the integral of any $Q$-exact form is zero (see
\cite[p.58]{BIS09}), we have
$\int_{21}Q(K_2K_{12})=0$.  Therefore, since $Q$ is an antiderivation and $K_2$ has even degree,
\begin{equation}
    \int_2K_2\int_1QK_{12}=\int_{21}K_2(QK_{12})= - \int_{21}(QK_2)K_{12}
    = - \int_{2}(QK_2)\int_{1}K_{12} .
\end{equation}
Similarly, $\int_{2} Q(K_2\int_1K_{12})=0$, and hence
$\int_{2} (QK_2)\int_1K_{12}= -\int_{2} K_2Q\int_1K_{12}$.
Thus we have shown that
\begin{equation}
    \int_2K_2\int_1QK_{12}= \int_{2} K_2Q\int_1K_{12}.
\end{equation}
Since $K_2$ is arbitrary, this implies that $Q\int_1K_{12}=  \int_1QK_{12}$.

We set $K_{12}=e^{-S_A} \theta F$ and use $Q e^{-S_{A}}=0$ (for $A=C_{j+1}^{-1}$)
to conclude that
\begin{equation}
     Q\Ex \theta F= Q \int e^{-S_A}\theta F = \int Q(e^{-S_A} \theta F)
      = \int e^{-S_A} Q\theta F
    =  \Ex Q\theta F.
\end{equation}
In particular, $Q\Ex\theta F = \Ex Q\theta F$.
It suffices finally to show that $Q\theta F = \theta QF$.
Since $Q$
is an antiderivation and $\theta$ is a homomorphism, it
is enough to verify that $Q\theta F =\theta Q F$ for $F = f (\phi,\bar\phi)$, $F =
\psi_{x}$, and $F=\bar\psi_x$.  These are readily verified using $Q = d +\iota_{X}$
(see \cite[(6.4)]{BIS09}).

\smallskip \noindent
(iii)
Suppose that $F$ is supersymmetric.
Let $\tilde F = F - F|_0$, which is supersymmetric
 and has zero constant part.
For $m \geq 0$, let $\tilde F(m) = e^{-m\sum_{x\in \Lambda} \tau_x}\tilde F$.
We claim that $\Ex \tilde F(m)$ is independent of $m$. Indeed,
let $v_{x} = \phi_x\psi_x$.
Then $\tau_x = \hat{Q}v_x$ and since $\hat{Q}\tilde F=0$,
\begin{equation}
  \label{e:dmQ}
    \ddp{}{m} \Ex \tilde F(m)
    =
    \sum_{x\in\Lambda} \Ex (\tau_x \tilde F(m))
    = \sum_{x\in\Lambda} \Ex(\hat{Q}(v_{x} \tilde F(m))).
\end{equation}
The right-hand side of \refeq{dmQ} is zero,
since the integral of any $Q$-exact form vanishes (see
\cite[p.58]{BIS09}).  It follows that $\Ex \tilde F = \lim_{m\to\infty}
\Ex \tilde F(m)$, and this limit vanishes since $\tilde F$ has zero constant part.
Therefore, $\Ex F = F|_0$.  In particular,
since $(\Ex\theta F)|_0=\Ex F$, if $F$ has zero constant part then so does $\Ex\theta F$.
\end{proof}

\section*{Acknowledgements}

The work of both authors was supported in part by NSERC of Canada.
DB gratefully acknowledges the support and hospitality of the
Institute for Advanced Study at Princeton and of Eurandom during part
of this work.
GS gratefully acknowledges the support and hospitality of the Institut
Henri Poincar\'e, and of the Kyoto University Global COE Program in
Mathematics, during visits to Paris and Kyoto where part of this work
was done.
We thank Roland Bauerschmidt for numerous helpful discussions and
valuable advice, and Benjamin Wallace for corrections to a previous version.

\bibliography{../../bibdef/bib}
\bibliographystyle{plain}

\end{document}